\PassOptionsToPackage{english}{babel}

\documentclass[aps,pra,reprint,superscriptaddress,letterpaper,twocolumn,longbibliography,floatfix]{revtex4-1}
\usepackage{graphicx}
\usepackage{amssymb}     
\usepackage[english,ngerman]{babel}

\usepackage[english]{babel}

\usepackage{blindtext}
\usepackage{amsmath} 
\usepackage[T1]{fontenc}
\usepackage{dsfont}
\usepackage{esint}
\usepackage{mathtools}
\usepackage{amsmath}
\newmuskip\pFqmuskip

\newcommand*\pFq[6][8]{%
  \begingroup 
  \pFqmuskip=#1mu\relax
  \mathcode`\,=\string"8000
  \begingroup\lccode`\~=`\,
  \lowercase{\endgroup\let~}\pFqcomma
  {}_{#2}F_{#3}{\left[\genfrac..{0pt}{}{#4}{#5};#6\right]}%
  \endgroup`    `       `   
}
\newcommand{\pFqcomma}{\mskip\pFqmuskip}

\newcommand{\pp}{\phantom{\prime}}

\usepackage{setspace}
\usepackage[usenames, dvipsnames, x11names]{xcolor}

\def    \bse{\begin{subequations}}
\def    \ese{\end{subequations}}

\newcommand{\ket}[1]{\left| #1 \right>} 
\newcommand{\bra}[1]{\left< #1 \right|} 
\let\baraccent=\= 
\renewcommand{\=}[1]{\stackrel{#1}{=}} 

\begin{document}

\selectlanguage{english}

\title{Hidden time-reversal symmetry, quantum detailed balance and exact solutions of driven-dissipative quantum systems}

\author{David Roberts$^{1,2}$, Andrew Lingenfelter$^{1,2}$, A. A. Clerk}
\affiliation{Pritzker School of Molecular Engineering, University of Chicago, Chicago, IL, USA \\
$^2$Department of Physics, University of Chicago, Chicago, IL, USA}

\date{\today}

\begin{abstract} 
Driven-dissipative quantum systems generically do not satisfy simple notions of detailed balance based on the time symmetry of correlation functions.  We show that such systems can nonetheless exhibit a {\it hidden time-reversal symmetry}  which most directly manifests itself in a doubled version of the original system prepared in an appropriate entangled thermofield double state.  This hidden time-reversal symmetry has a direct operational utility:  it provides a general method for finding exact solutions of non-trivial steady states.   Special cases of this approach include the coherent quantum absorber and complex-$P$ function methods from quantum optics. We also show that hidden-TRS has observable consequences even in single-system experiments, and can be broken by the non-trivial combination of nonlinearity, thermal fluctuations, and driving.  To illustrate our ideas, we analyze concrete examples of driven qubits and nonlinear cavities.  These systems exhibit hidden time-reversal symmetry but not conventional detailed balance.  
\end{abstract}

\maketitle


\section{Introduction}

Time-reversal is a basic symmetry that plays a crucial role in a vast variety of physical systems.  For open classical systems subject to dissipation and driving, it manifests itself as detailed balance constraints on transition rates (or equivalently drift and diffusion functions).  It also places a strong symmetry constraint on steady-state two-time correlation functions $\overline{A(t) B(0)}$: they must be invariant when each quantity is replaced by its time-reversed version and $t \rightarrow -t$.  This symmetry is sometimes referred to as Onsager symmetry, as it plays a crucial role in the derivation of Onsager reciprocity relations.  In classical systems, this symmetry has a direct operational utility:  it provides a simple route for finding steady state probability distributions (i.e.~potential conditions that can be used to solve Fokker-Planck equations \cite{gardiner_stochastic_2009}).

 \begin{figure}[h!]
     \centering
    \includegraphics[width=0.95 \columnwidth]{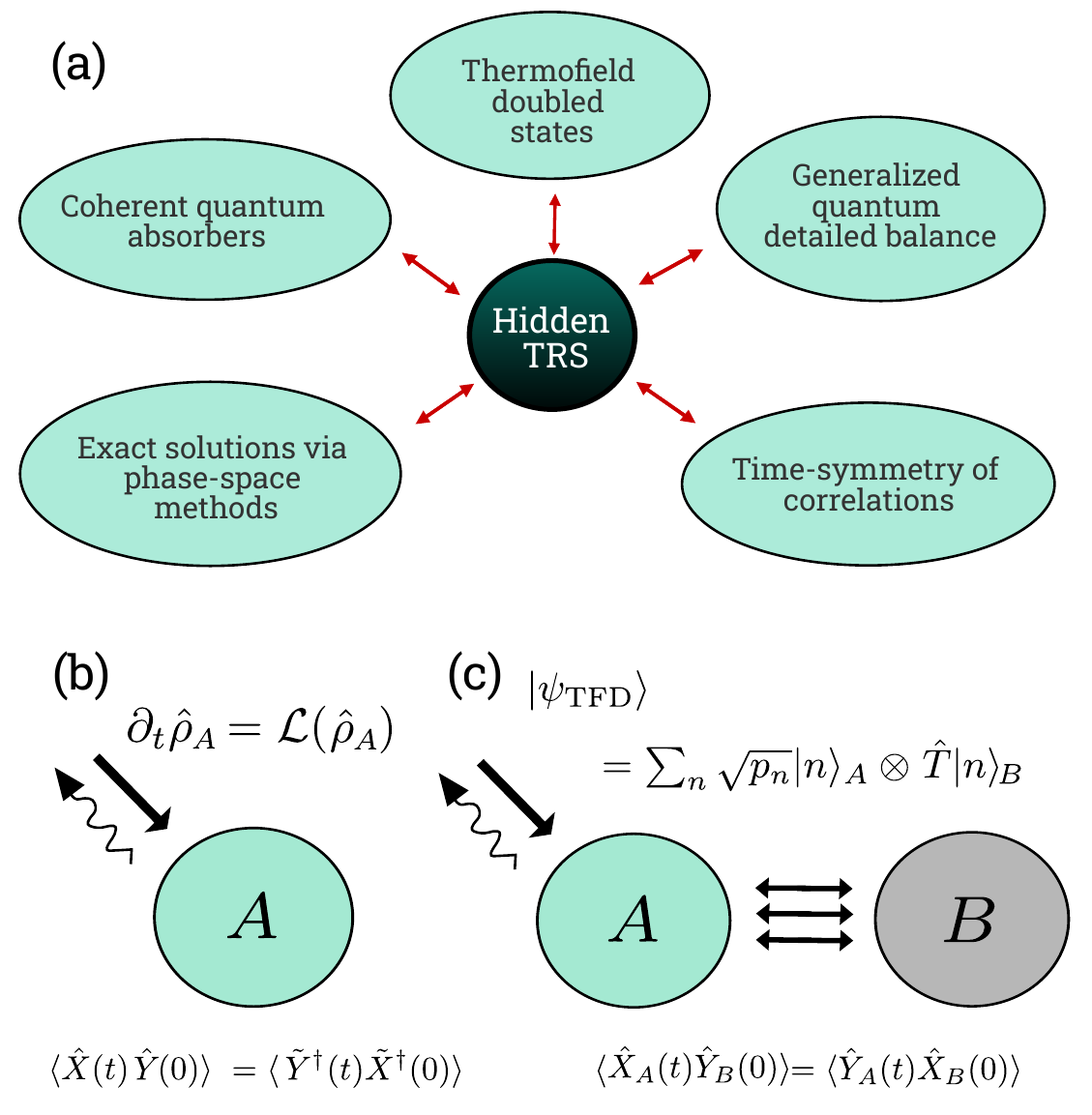}
     \caption{(a) The concept of hidden time-reversal symmetry helps unify disparate-seeming topics in the study of driven-dissipative quantum systems.  (b) A driven dissipative system described by a Lindblad master equation.  The conventional definition of  quantum detailed balance is formulated as a time symmetry of steady-state correlation functions. (c) Hidden TRS is a symmetry ensuring that correlation functions of a doubled version of the original system, prepared in a thermofield double state, are time symmetric.  It is defined by an anti-unitary operator $\hat{T}$.  Hidden TRS can hold even when the correlation function symmetry in (b) fails; it also enables a powerful method for finding exact solutions. }
\label{fig:intro_fig}
 \end{figure} 

There is a long history of works that extend notions of Onsager symmetry and detailed balance to quantum open systems described by a Markovian master equation in Lindblad form \cite{agarwal_open_1973,carmichael_detailed_1976,alicki_detailed_1976,kossakowski_quantum_1977,majewski_detailed_1984,Majewski1999,Denisov2002,fagnola_generators_2007,fagnola_generators_2010,duvenhage_balance_2018,Carlen2017,ramezani_quantum_2018}.
The most natural definition requires that steady-state correlation functions in the quantum theory obey an Onsager symmetry analogous to the classical case \cite{agarwal_open_1973,carmichael_detailed_1976,Denisov2002}; this condition necessarily holds if the microscopic system-bath dynamics obey time-reversal symmetry \cite{carmichael_detailed_1976}.  Later more formal works  considered generalized definitions of quantum detailed balance \cite{fagnola_generators_2007,fagnola_generators_2010}, framed in terms of quantities that are not directly measurable and whose physical interpretation is somewhat opaque.  The ultimate operational utility of all these quantum definitions of detailed balance are unclear.  Unlike classical detailed balance, these quantum symmetries are not known to enable a simple method for finding a non-trivial system's steady state density matrix\footnote{  The only attempt at such connections in the past were limited to systems that were easily solvable by other means (e.g.~linear bosonic systems, or systems that could be reduced to a classical master equation) \cite{agarwal_open_1973}.}.

In this work, we introduce a powerful, symmetry-based formulation of quantum detailed balance (QDB) that goes beyond the simple definition in Ref.~\cite{agarwal_open_1973}, and that {\it directly} enables an efficient way for finding non-trivial steady states.  Our work builds on Ref.~\cite{duvenhage_balance_2018}, which showed that a particular  generalized definition of 
QDB introduced in 
Ref.~\cite{goldstein_kms-symmetric_1995} can be formulated using an entangled, thermofield double state \cite{Takahashi1996}.  We use this to introduce the notion of ``hidden'' time reversal symmetry (TRS) in an open quantum system.  This anti-unitary symmetry need not reveal itself through some simple invariance of the original master equation, nor through a standard Onsager symmetry of two-time correlation functions.  Instead, this symmetry is directly tied to a time symmetry of correlation functions of a {\it doubled} version of the original system prepared in an thermofield double state whose form is directly tied to the symmetry operator (see Fig.~\ref{fig:intro_fig}(c)).  Crucially, we show that a system can possess hidden TRS even if it fails to have the conventional quantum detailed balance (CQDB) defined in Ref.~\cite{agarwal_open_1973} (though in the limit of infinitely weak dissipation, these notions coincide).

Hidden TRS is not just a formal curiosity: it provides a powerful tool for understanding complex non-thermal and non-classical steady states.  We show that {\it the existence of hidden TRS directly yields a simple and direct method for analytically finding the steady state density matrix of a Lindblad driven-dissipative quantum system}.  This method is not limited to situations of weak driving, interactions or dissipation.  It represents a generalization of the coherent quantum absorber (CQA) method introduced in Ref. \cite{stannigel_driven-dissipative_2012}, and extended in Ref.~\cite{roberts_driven_2019}. Hidden-TRS is also connected to well-known exact solution methods from quantum optics based on the complex-$P$ phase space quasiprobability \cite{drummond_generalised_1980,drummond_quantum_1980,bartolo_exact_2016,elliott_applications_2016}: these methods can be viewed as special cases of our more general approach.

While experiments on doubled quantum systems prepared in thermofield double states have recently been performed \cite{Monroe2020}, hidden TRS also has experimental consequences in experiments on just a single system.  Unlike CQDB, systems with hidden TRS will not exhibit Onsager time-symmetry of all correlators.  However, we show that there are always a class of special correlation functions that are guaranteed to have this time symmetry.  This provides a direct means for probing hidden TRS (and its possible breaking) in a variety of experimental platforms.  We explore in detail two classes of ubiquitous, experimentally-accessible systems (see Table \ref{tab:thermal-trs}): Rabi-driven qubits subject to dissipation, and driven-dissipative nonlinear quantum cavities.  These systems exhibit in general no correlation function time-symmetry, and hence do not possess CQDB as defined in \cite{agarwal_open_1973}.  They however do possess hidden-TRS in the low temperature or small nonlinearity limit.  This explains the surprising exact solvability of a variety of driven nonlinear cavity models \cite{drummond_generalised_1980,drummond_quantum_1980,bartolo_exact_2016,elliott_applications_2016,roberts_driven_2019}.
We explore how hidden TRS is broken in these models by the combination of non-zero temperature, driving and nonlinearity.  For nonlinear cavities, breaking of detailed balance was extensively studied in the semiclassical limit \cite{DykmanKrivoglaz1979,Dykman1988,marthaler_switching_2006,dykman_periodically_2012,Guo2013,LingzhenThesis,Yaxing2019}.  

The rest of this paper is organized as follows: in Sec.~\ref{sec:ClassicalDB} we review the doubled-system formulation of classical detailed balance and the definition of CQDB \cite{agarwal_open_1973,carmichael_detailed_1976}; we show that CQDB only holds for a limited class of systems that possess trivial steady states.  Sec.~\ref{subsec:HiddenTRS} introduces the notion of hidden TRS, and connects it to the definition of generalized QDB introduced in Ref.~\cite{goldstein_kms-symmetric_1995}.  In Secs.~\ref{sec:DynamicalEquiv} and \ref{sec:ExactSols} we demonstrate how the existence of hidden TRS enables an extremely direct method for finding exact solutions for steady states.  In Sec.~\ref{sec:CavityHiddenTRS} we discuss how a variety of driven quantum cavity models possess hidden TRS, while in Sec.~\ref{sec:ThermalHiddenTRSBreaking} we discuss how thermal fluctuations in some cases can break this symmetry.  Sec.~\ref{sec:complex_P} shows how hidden-TRS underlies the complex-$P$ function exact solution method.  We conclude in Sec.~\ref{sec:Conclusions}.

\begin{table}[]
    \centering
    \begin{tabular}{l|c|c|c}
        & & {\bf Hidden} & {\bf Hidden TRS:}\\ 
        {\bf System} & {\bf CQDB} & {\bf TRS} & {\bf unique?}\\ \hline
        Thermal qubit & Yes & Yes & $\hat{T}_\theta,~e^{i\theta}\equiv 1$\\
        Thermal linear cavity  & Yes & Yes &  $\hat{T}_\theta,~e^{i\theta}\equiv 1$ \\
        \it{Kerr cavity, 1-ph. drive} & {\it No} & {\it Yes} & {\it Yes}\\
        \it{Kerr cavity, 2-ph. drive} & {\it No} & {\it Yes} & No; $\hat{T}_+,\hat{T}_-$\\ 
        \,\,\,\, with nonzero temp. & No & No & N/A\\
        \it{Driven qubit} & {\it No} & {\it Yes} & \it{Yes}\\
        \,\,\,\, with non-zero temp. & No & No & N/A\\
    \end{tabular}
    \caption{Common driven-dissipative quantum systems 
    and their status both with respect to conventional quantum detailed balance (CQDB) (c.f.~Sec.~\ref{subsec:CQDB}), and our new notion of hidden-TRS.  Italics indicate systems with hidden-TRS that do not have CQDB.  Some of these systems possess multiple distinct hidden TRS (right-most column).}
    \label{tab:thermal-trs}
\end{table}

\section{Classical detailed balance and conventional quantum detailed balance}
\label{sec:ClassicalDB}

\subsection{Classical detailed balance}

Consider a classical stochastic system with a discrete set of microstates 
indexed by integers $n$, 
whose Markovian dynamics is fully described by a set of transition rates $\Gamma_{n\to m}$.  The time-dependent probability $p(n,t)$ for the system to be in a given state $n$ then obeys: \begin{align}
    \frac{d}{dt} p(n,t) = \sum_{m} p(m,t)\Gamma_{m\to n} - \sum_{m} p(n,t)\Gamma_{n\to m}.\label{eq:kinetic}
\end{align}

We assume that this equation admits a time-independent steady-state probability distribution $\bar{p}(n)$.  This steady state is said to have detailed balance if there is a balancing of probability fluxes between any given pair of states and their time-reversed partners.  Letting $\tilde{n}$ denote the time-reversed version of the microstate $n$, the condition is 
(see e.g.~\cite{gardiner_stochastic_2009}):
\begin{align}
    \bar{p}(n)\Gamma_{n\to m} = \bar{p}(\tilde{m})\Gamma_{\tilde{m}\to \tilde{n}}.
    \label{eq:DBRates}
\end{align}
This definition generalizes directly to systems with a continuous state space: Eq.~(\ref{eq:kinetic}) then becomes a Fokker-Planck equation, and Eq.~(\ref{eq:DBRates}) becomes a constraint on drift and diffusion matrices (so-called potential conditions).
As is well known, these ``zero probability flux'' conditions give a direct way to find the steady state distribution.  In the discrete case, one uses the zero-flux condition to iteratively find the probability of each microstate in terms of the rates.  The continuous version of this yields the steady state in terms of a potential function that is determined by drift and diffusion matrices (see e.g.~\cite{gardiner_stochastic_2009}).

One can equivalently define detailed balance by a time symmetry of correlation functions, what we term here an Onsager symmetry.  Suppose $X$ and $Y$ are arbitrary functions of the microstate $n$ of our system.  Steady state correlation functions can be defined in the usual manner in terms of conditional probabilities 
$p(m,t | n,0)$, which can be computed from Eq.~(\ref{eq:kinetic}).  For example:
\begin{equation}
    \overline{X(t)Y(0)} \equiv  
    \sum_{m,n} X(m) p(m,t | n,0) Y(n) \bar{p}(n)
\end{equation}
The detailed balance condition of Eq.~(\ref{eq:DBRates}) is then equivalent to requiring that the following symmetry hold for all steady state correlation functions:
\begin{align}
    \overline{X(t)Y(0)} = \overline{\tilde{Y}(t) \tilde{X}(0)}.\label{eq:cdb_wT}
\end{align}
Here, the time-reversed function $\tilde{X}$ is defined as $\tilde{X}(n) \equiv X(\tilde{n})$.

 \begin{figure}[t]
     \centering
    \includegraphics[width=0.99\columnwidth]{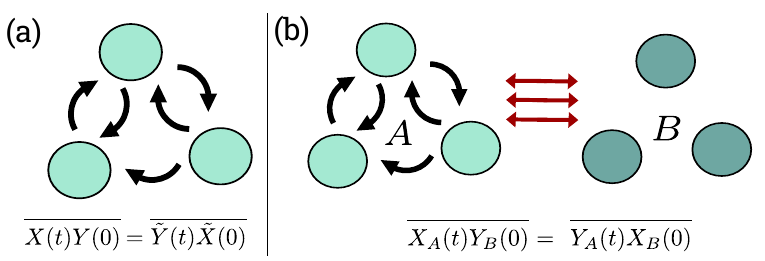}
     \caption{(a) System with discrete microstates,  described by a classical master equation.
     Detailed balance (Eq.~(\ref{eq:DBRates})) is equivalent to a time-symmetry of stationary correlation functions. (b)  Equivalent formulation of classical detailed balance, involving 
     a doubled system prepared in an initial correlated state given in Eq.~(\ref{eq:delta_correlated_state}); the auxiliary system $B$ has no dynamics.  The detailed balance condition in (a) is equivalent to requiring a time-symmetry of doubled-system correlators.}
\label{fig:classical_fig}
 \end{figure} 

\subsection{Doubled-system formulation of classical detailed balance}
\label{sec:classical-doubled-system}

Consider now an alternate but equivalent formulation of classical detailed balance \cite{duvenhage_balance_2018}.  We imagine making a copy of our original system that has exactly the same set of microstates as the original.  This auxiliary system (system $B$) is completely static, whereas the original system (system $A$) retains its transition rates and dynamics, see Fig.~\ref{fig:classical_fig}.  To be explicit, the doubled-system is described by a probability distribution $p_{AB}(n_A,n_B;t)$ which satisfies the master equation:
\begin{align}
    \frac{d}{dt} p_{AB}(n,n'; t)  &= 
        \sum_{m} p_{AB}(m,n'; t)\Gamma_{m\to n} \nonumber\\ 
        &~~~~~~~~~~~~~-\sum_{m} p_{AB}(n,n'; t)\Gamma_{n\to m}.
        \label{eq:kinetic_doubled}
\end{align}

We further assume that the doubled system is initially prepared in a correlated state described by the probability distribution:
\begin{align}
    &p_{AB}(n,m;0)= \bar{p}(n) \delta_{m,\tilde{n}}, \label{eq:delta_correlated_state}
\end{align}
where $\bar{p}(n)$ is the single-system steady state of interest.  Note that the correlations in this state partner each microstate $n$ in system $A$ with its time-reversed partner $\tilde{n}$ in system $B$.  It is easy to confirm that the marginal probability distribution describing each subsystem alone ($A$ or $B$) is time independent and equal to $\bar{p}(n)$ for system $A$, and $\bar{p}(\tilde{n})$ for system $B$.  In contrast, the total state of the two systems will evolve non-trivially.  The result is that while quantities involving only each subsystem are completely static, correlations between the two subsystems will evolve in time.

Given this dynamics and initial state, we now ask whether inter-system correlations at time $t>0$ are invariant if we ``exchange'' the two subsystems.  To be concrete, we consider two observable quantities $X$ and $Y$ that we could measure on either system $A$ or $B$; these are functions of microstates, i.e. $X_A(n,m) = X(n)$,  $X_B(n,m) = X(m)$.  We now define a doubled-system Onsager time-symmetry by the requirement:      
\begin{align}
    \overline{X_A(t) Y_B(0)} = \overline{Y_A(t)X_B(0)},
    \label{eq:doubledcdb}
\end{align}
for all possible observables $X, Y$.  We stress that system $B$ is not dynamical.

Superficially, this looks very different from the standard Onsager time-symmetry constraint in Eq.~(\ref{eq:cdb_wT}).  We are not exchanging observation times, but rather the subsystem in which each quantity is measured.  There is also no explicit time-reversal operation in this equation (it is instead hardwired into the initial correlated state).  Despite these differences, one can easily show (see Ref.~\cite{duvenhage_balance_2018} and App.~\ref{app:DoubledClassicalDB}) that the above symmetry relation is {\it completely} equivalent to standard Onsager time symmetry.  While this formulation thus provides nothing new in the classical context, we will see that it motivates an extremely powerful generalized notion for quantum systems, namely the notion of a hidden time-reversal symmetry.


\subsection{Markovian quantum open system: general setting}

Onsager-like correlation function time-symmetries can also be considered in the context of open quantum systems \cite{agarwal_open_1973,carmichael_detailed_1976,Denisov2002}.  Our general setting will be an open system whose dynamics is described by a Markovian Lindblad master equation.  The dynamics is completely specified by the Hermitian system Hamiltonian $\hat{H}$ and a set of jump operators $\hat{c}_l$ that describe the influence of external dissipative reservoirs.  
Defining $\hat{H}_\text{eff}= \hat{H}-\frac{i}{2}\sum_l \hat{c}_l^\dag \hat{c}_l$, our general Lindblad equation of motion for the system density matrix $\hat{\rho}(t)$ is:
\begin{align}
    \frac{d}{dt} \hat{\rho} = -i(\hat{H}_\text{eff}\hat{\rho}-\hat{\rho}\hat{H}_\text{eff}^\dag )  + \sum_{l=1}^M \hat{c}_l \hat{\rho}\hat{c}_l^\dag
    \equiv \mathcal{L}[\hat{\rho}],
    \label{eq:qme}
\end{align}
where we have introduced the Liouvillian superoperator $\mathcal{L}$.  Given a particular steady state of this equation (i.e.~a time-independent solution $\hat{\rho}_{\rm ss}$), steady-state correlators between two system operators $\hat{X}$ and $\hat{Y}$ obey time-translation symmetry and can be calculated from the master equation (see, e.g., \cite{Breuer2002}).  To state this compactly, we first introduce the adjoint Liouvillian superoperator $\bar{\mathcal{L}}$, determined by (see e.g.~Ref.~\cite{gardiner_quantum_2000})
\begin{align}
    \bar{\mathcal{L}}[\hat{A}] \equiv i(\hat{H}_\text{eff}^\dag \hat{A} - \hat{A}\hat{H}_\text{eff})+
        \sum_{l=1}^M \hat{c}_l^\dag \hat{A}\hat{c}_l. \label{eq:LAdjointEqn}
\end{align}
Correlation functions are then computed as
\begin{align}
    \langle \hat{X}(t) \hat{Y}(0) \rangle \equiv 
        \textrm{Tr} \left( \mathcal{E}_t[\hat{X}] \hat{Y} \hat{\rho}_{\rm ss} \right),
        \,\,\,\,\,
        \mathcal{E}_t \equiv \exp \left( \bar{\mathcal{L}} t \right).
        \label{eq:CorrFuncDefn}
\end{align}

\subsection{Conventional quantum detailed balance (CQDB)}
\label{subsec:CQDB}
 
As mentioned in the introduction, a variety of definitions of quantum detailed balance have been formulated for Lindblad Markovian master equations
\cite{duvenhage_balance_2018,agarwal_open_1973,cipriani_dirichlet_1997,kossakowski_quantum_1977}. The simplest and best-known definition of quantum detailed balance involves the same constraint on correlation functions that exists in the classical setting \cite{agarwal_open_1973,carmichael_detailed_1976}.  
With this definition (which we call conventional quantum detailed balance (CQDB)), detailed balance requires:
\begin{align}
    \langle \hat{X}(t)\hat{Y}(0)\rangle = \langle  \tilde{Y}^\dag(t)\tilde{X}^\dag(0)\rangle.\label{eq:qdb_wT}
\end{align}
Here $\hat{X}$ and $\hat{Y}$ are arbitrary system operators, and tilde is used to denote the time-reversed version of an operator, 
i.e.~$\tilde{B}\equiv \hat{T}\hat{B}\hat{T}^{-1}$,  where $\hat{T}$ is the anti-unitary time-reversal operator.  
Note that on the RHS of Eq.~(\ref{eq:qdb_wT}), we have 
$\tilde{Y}^\dagger(t) \equiv \mathcal{E}_t \left[ \tilde{Y}^\dagger \right] $.

Besides being simple to state, the CQDB correlation function symmetry also has a direct connection to a microscopic symmetry:  this correlation function time symmetry {\it necessarily} holds if the microscopic system-bath dynamics has time-reversal symmetry \cite{carmichael_detailed_1976}.  This requires both that the entire system-plus-bath Hamiltonian has time-reversal, and that the bath state also relaxes to a time-independent state (not just the system state). This connection to microscopics (and the fact that it involves measurable correlation functions) has led many to consider CQDB the most natural generalization of classical detailed balance to quantum systems \cite{Denisov2002}.  Note that CQDB is also referred to as ``GNS detailed balance'' \cite{Carlen2017}.

\subsection{CQDB implies a trivial steady state}
\label{subsec:CQDBtrivial}

As has been noted previously \cite{alicki_detailed_1976, fagnola_generators_2007, fagnola_generators_2010}, the CQDB condition is extremely restrictive: it {\it only}
holds for systems with steady states that are diagonal in the energy eigenstate basis 
(i.e.~$[\hat{H}, \hat{\rho}_{\rm ss} ] = 0$)\footnote{The Hamiltonian in a Lindblad master equation is in general not unique.  Our condition holds for the case where jump operators are chosen to be traceless.}.  
Such states are in some sense trivial, as they can be found by solving a classical (Pauli) master equation, obtained by setting all energy-eigenstate coherences in Eq.~(\ref{eq:qme}) to zero. The resulting steady state can thus be interpreted classically:  the Hamiltonian plays no role, and steady-state probabilities are determined by balancing incoherent transition rates between different eigenstates. A simple proof of this result is presented in App.~\ref{app:SQDB_is_diagonal}.

The upshot is that CQDB is found in an extremely limited class of systems, and is not a useful tool for finding non-trivial quantum steady states; in fact, the presence of CQDB makes it impossible to have such a state.
In App.~\ref{app:QubitCQDB}, we show explicitly that an extremely simple model of a Rabi-driven dissipative qubit fails to have CQDB (though it will possess the hidden time-reversal symmetry we introduce in the next section).

\section{Hidden time-reversal symmetry and generalized detailed balance}
\label{subsec:HiddenTRS}

\subsection{Basic formulation}
\label{subsec:HiddenTRSBasics}

As we have discussed, the simple CQDB condition of Eq.~(\ref{eq:qdb_wT}) corresponds directly to microscopic time-reversal symmetry \cite{carmichael_detailed_1976}.  We now consider something more general, the notion of {\it hidden} time-reversal symmetry, which we will formulate in a doubled version of our original system (in rough analogy to the classical construction in Sec. \ref{sec:classical-doubled-system}).  This unusual symmetry will connect directly to an abstract variant of quantum detailed balance (so called ``SQDB-$\theta$'') studied in the mathematical physics literature \cite{fagnola_generators_2007,fagnola_generators_2010}, and which was recently linked to entanglement  \cite{duvenhage_balance_2018}.  Our work complements these studies by providing a direct physical motivation for this definition and connects it explicitly to symmetry.  More importantly, we show that this formulation has great practical utility: it allows us to solve for non-trivial steady states.  This connection was not previously known.  

Our starting point is again the Lindblad master equation of Eq.~(\ref{eq:qme}) and particular steady state $\hat{\rho}_{\rm ss}$, which we write in diagonal form as 
\begin{align}
    \hat{\rho}_{\rm ss} & = \sum_n p_n |n\rangle\langle n|.
    \label{eq:rhoss_diag}
\end{align}
Throughout this work, we will assume that $\hat{\rho}_{\rm ss}$ is full rank, and further, that the $p_n$ have no degeneracies.  

We next construct a purification of this state:  an entangled pure state $\ket{\psi_{T}}$ of a doubled version of our system that yields $\hat{\rho}_{\rm ss}$ if we trace out the auxiliary system $B$.  We take system $B$ to have the same Hilbert space as the original system $A$.
To construct $\ket{\psi_{T}}$, we first chose an anti-unitary operator $\hat{T}$ which will define our hidden time-reversal symmetry.  We then use this choice to construct $\ket{\psi_{T}}$ in a manner that mimics the classical doubled-system state in Eq.~(\ref{eq:delta_correlated_state}):  we pair each pointer state $\ket{n}$ in the original system with its time-reversed partner in the auxiliary system.  We thus have   
\begin{align}
    |\psi_{T}\rangle \equiv 
        \sum_n 
        \sqrt{p_n} \ket{n}_A \left( \hat{T} \ket{n}_B\right)
        \equiv 
                \sum_n 
        \sqrt{p_n} \ket{n}_A \ket{\tilde{n}}_B
    \label{eq:tfd_wT}
\end{align}
With this definition, $\ket{\psi_{T}}$ is invariant under a gauge change of the $\hat{\rho}_{\rm ss}$ eigenkets $\ket{n} \rightarrow e^{i \alpha_n} \ket{n}$.
The state $\ket{\psi_{T}}$ has the form of a so-called ``thermofield double state''; such states have been studied in many different contexts \cite{Takahashi1996}, though usually without including a time-reversal operation in its definition.  We stress that the form of $ |\psi_{T}\rangle$ is contingent on the choice of $\hat{T}$.

We next specify the dynamics of the doubled system:  as we did in the classical case, we take subsystem $A$ to evolve as per the master equation in Eq.~(\ref{eq:qme}), but take the auxiliary subsystem $B$ to have no dynamics at all.  The
dynamics of the joint system thus follows the Lindblad master equation:
\begin{align}
    \frac{d}{dt} \hat{\rho}_{AB}(t) = \left[ \mathcal{L} \otimes \mathds{1}_B \right] \hat{\rho}_{AB}(t)
    \label{eq:DoubledMEQ}
\end{align}
where $\mathcal{L}$ is our original single-system Liouvillian, and $\mathds{1}_B$ denotes the unit superoperator acting on subsystem $B$.  Starting the doubled system at $t=0$ in the pure state $\hat{\rho}_{AB}(0) = \ket{\psi_{T}} \bra{\psi_{T}}$, this dynamics leaves the reduced density matrices of each subsystem invariant.  It does not however leave the full joint state $\hat{\rho}_{AB}$ time invariant. The result is that single-system expectation values are time-independent, but correlations between them can evolve.

We will now use these evolving inter-system correlations to define the notion of a hidden time reversal symmetry.  We define single-subsystem operators in the natural way, i.e. $\hat{X}_A \equiv \hat{X} \otimes \hat{1}$, $\hat{X}_B \equiv \hat{1} \otimes \hat{X}$. Starting in the state $\ket{\psi_{T}}$ and evolving as per Eq.~(\ref{eq:DoubledMEQ}), hidden TRS holds if all inter-system correlations obey the following symmetry:
\begin{equation}
    \mathrm{Tr } \left[ \hat{X}_A \hat{Y}_{B} \, \hat{\rho}_{AB}(t) \right] = 
     \mathrm{Tr } \left[ \hat{Y}_A \hat{X}_B \, \hat{\rho}_{AB}(t) \right]
     \label{eq:HiddenTRSSchrod}
\end{equation}
Stated explicitly, the symmetry requires that inter-system correlations at any time $t\geq0$ are invariant if we exchange which subsystem each quantity is measured in.  Note that as system $B$ has no dynamics, we can equivalently write this condition in the Heisenberg picture.  Defining
\begin{equation}
    C_{XY}^\text{TFD}(t)\equiv 
    \begin{cases}
        \langle \psi_{T}| \hat{X}_A(t)\hat{Y}_B|\psi_{T}\rangle & t \geq 0 \\
        \langle \psi_{T}| \hat{Y}_A(-t)\hat{X}_B|\psi_{T}\rangle & t < 0
    \end{cases}
    \label{eq:TFD-corr-func-def-C_XY}
\end{equation}
the condition for having hidden TRS condition then becomes
\begin{align}
    C_{XY}^{\rm TFD}(t) & = C_{XY}^{\rm TFD}(-t)
    \label{eq:doubledqdb_wT}
\end{align}
Note that for a general system that does not have hidden TRS, the definition in Eq.~(\ref{eq:TFD-corr-func-def-C_XY}) implies that not only can $C_{XY}^{\rm TFD}(t)$ be asymmetric, it can even fail to be continuous at $t=0$.  We also stress again that the hidden-TRS condition above is contingent on the choice of anti-unitary $\hat{T}$.  As we will see, there exist physical systems where hidden-TRS is in some sense degenerate:  there are a whole family of distinct operators $\hat{T}$ for which Eq.~(\ref{eq:doubledqdb_wT}) holds.  

Despite mirroring the classical doubled-system construction, in the quantum case Eq.~(\ref{eq:doubledqdb_wT}) gives us something truly new: there are systems that fail to satisfy the Onsager symmetry condition of Eq.~(\ref{eq:qdb_wT}), but nonetheless satisfy the generalized condition in Eq.~(\ref{eq:doubledqdb_wT}).  In these cases, we describe the particular anti-unitary operator $\hat{T}$ used to define $\ket{\psi_{T}}$ as a {\it hidden time-reversal symmetry}. Note that {\it if} a system has hidden TRS, the steady state is invariant under the corresponding time-reversal operation:
\begin{equation}
    \hat{T} \hat\rho_{\rm ss} \hat{T}^{-1} = \hat{\rho}_{\rm ss}
    \label{eq:HiddenTRSRhoSS}
\end{equation}
This follows directly by assuming hidden TRS (i.e.~Eq.~(\ref{eq:HiddenTRSSchrod})), and taking the choice $\hat{Y}=1$.  The only way the remaining condition can hold for all $\hat{X}$ is if Eq.~(\ref{eq:HiddenTRSRhoSS}) holds.

To gain intuition for the role of entanglement in our formulation of hidden TRS, it is useful to express the doubled correlation function in terms of the eigenstates of $\hat{\rho}_{\rm ss}$, c.f.~Eq.~(\ref{eq:rhoss_diag}).  
Using Eq.~(\ref{eq:TFD-corr-func-def-C_XY}), we have
\begin{align}
 C_{XY}^\text{TFD}(t>0)&= \sum_{nm} \sqrt{p_np_m}\langle n|\hat{X}(t) |m\rangle \langle \tilde{n}| \hat{Y}|\tilde{m}\rangle \nonumber\\
    &= C_{XY}^\text{cl}(t) +  C_{XY}^\text{en}(t), \label{eq:TFD-pointer-state-expansion}
\end{align}
where
\begin{align}
    C_{XY}^\text{cl}(t)&\equiv \sum_n p_n \langle n|\hat{X}(t) |n\rangle \langle \tilde{n}| \hat{Y}|\tilde{n}\rangle \label{eq:TFD-classical-correlation}\\   
    C_{XY}^\text{en}(t)&\equiv \sum_{n \neq m} \sqrt{p_n p_m} \langle n|\hat{X}(t) |m\rangle \langle \tilde{n}| \hat{Y}|\tilde{m}\rangle.\label{eq:TFD-entanglement-correction}
\end{align}
For times $t<0$, Eqs.~(\ref{eq:TFD-pointer-state-expansion})-(\ref{eq:TFD-entanglement-correction}) are defined analogously to $C_{XY}^{\rm TFD}(t)$ itself. 

The above separation has a direct physical significance, as it is the second contribution in Eq.~(\ref{eq:TFD-entanglement-correction}) that encodes the contribution from quantum entanglement.  To see this explicitly, 
suppose that we initially had prepared our doubled system in a state that only had classical correlations (but not entanglement) between systems $A$ and $B$, i.e.
\begin{align}
    \hat{\rho}_{AB}(0)  = \sum_n p_n|n,\tilde{n}\rangle \langle n,\tilde{n}|.
\end{align}
In this case, our {\it entire} inter-system correlation functions would be given by Eq.~(\ref{eq:TFD-classical-correlation}):
\begin{align}
    \mathrm{Tr } \left[ \hat{X}_A(t) \hat{Y}_{B} \, \hat{\rho}_{AB} \right] &=\sum_n p_n\langle n|\hat{X}(t) |n\rangle \langle \tilde{n}| \hat{Y}|\tilde{n}\rangle\nonumber\\
    &=C_{XY}^\text{cl}(t).\label{eq:HiddenTRS_diagonal}
\end{align}
We can thus view $C_{XY}^\text{en}(t)$ in Eq.~(\ref{eq:TFD-entanglement-correction}) as the extra contribution to the TFD 
correlator that stems from having quantum entanglement (as opposed to purely classical correlations).

\subsection{CQDB as a special case of hidden TRS}

At this point, it is natural to ask whether there is any simple relation between our notion of hidden TRS and conventional CQDB (which we stress is directly connected to microscopic time-reversal of the complete system-plus-environment \cite{carmichael_detailed_1976}).  The first key result here is that {\it CQDB is a special case of hidden-TRS}: any system satisfying CQDB automatically has a hidden-TRS, but {\it the converse is not true}. Recall from  
Sec.~\ref{subsec:CQDBtrivial} that systems satisfying CQDB necessarily have somewhat trivial steady states (in that they are diagonal in the energy eigenstate basis).  In contrast, there are many systems with have hidden TRS but not CQDB, and have steady states with non-zero coherences between energy eigenstates.  This is a crucial result:  hidden TRS does not preclude having a non-trivial steady state.  

To understand the origin of the above result, it is useful to introduce an effective dimensionless Hermitian Hamiltonian defined by the steady-state density matrix $\hat{\rho}_{\rm ss}$, the so-called modular Hamiltonian:
\begin{align}
    \hat{H}_\rho \equiv -\log \hat{\rho}_{\rm ss}.
\end{align}
Any system possessing CQDB has an extremely tight constraint on its dynamics \cite{bratteli_unbounded_1978}:  dynamical evolution generated by the full Liouvillian $\mathcal{L}$ must commute with evolution generated by $\hat{H}_{\rho}$ (i.e. by the unitary $e^{-i \hat{H}_\rho t}$). This symmetry allows one to demonstrate that CQDB is a special case of  hidden TRS, using the same methods introduced to study a class of generalized QDB conditions in  Refs.~\cite{fagnola_generators_2007,fagnola_generators_2010}.  In Appendix \ref{app:SQDBtoGeneral}, we give a non-technical proof of this result:  for any system with modular symmetry, CQDB and hidden TRS are equivalent \cite{fagnola_generators_2007}, and thus CQDB implies hidden TRS.

The second key result connecting hidden TRS and CQDB involves the limit of vanishing incoherent dynamics.  Consider a system that possesses hidden-TRS even as the strength of the incoherent dynamics is tuned to zero (i.e.~replace the jump operators $\hat{c}_l \rightarrow \lambda \hat{c}_l$ in the master equation, and let $\lambda \rightarrow 0$).  In such systems, hidden TRS reduces to CQDB in the limit of vanishing dissipation.  Hence, while in many cases finite dissipation destroys CQDB, hidden TRS can continue to be a symmetry.  

This result can also be understood using the modular Hamiltonian.  In the limit of vanishing dissipation, the full dynamics is completely generated by the Hermitian Hamiltonian $\hat{H}$, i.e.
\begin{align}
    \mathcal L[\hat{\rho}] \sim -i[\hat{H},\hat\rho].
\end{align}
Further, in this limit the steady state $\hat{\rho}_{\rm ss}$ must commute with $\hat{H}$ in order to be stationary.  These two facts then necessarily imply that in this asymptotic zero dissipation limit, the full Liouvillian commutes with the modular Hamiltonian.  As discussed above, this then implies that our system has CQDB in the zero-dissipation limit.  This proves the desired result:  systems with hidden TRS always recover CQDB in the weak dissipation limit.  At a more physical level, systems with hidden TRS do not necessarily have (single system) correlation functions that are time-symmetric.  However, the lack of time-symmetry vanishes in the zero-dissipation limit.


\subsection{Hidden TRS has observable consequences for a single system}
\label{subsec:SingleSystemExperiment}

Hidden TRS might initially seem to be experimentally irrelevant in most settings, as it is defined in terms of TFD correlators (c.f.~Eq.~(\ref{eq:TFD-corr-func-def-C_XY})) that require one to prepare two copies of the system of interest in an initial entangled state.  This is usually extremely challenging (though see the recent trapped ion experiment in Ref.~\cite{Monroe2020}).  Formally, one could define this correlator as a single-system quantity involving a {\it state-dependent} observable.
We first introduce a superoperator $\mathcal J$ which acts on single-system operators:
\begin{align}
    \mathcal J[\hat{X}]_A|\psi_T\rangle&\equiv \hat{X}_B|\psi_T\rangle
    \label{eq:JDefinition}
\end{align}
$\mathcal J$ is well-defined and unique when $\hat\rho_{ss}$ is full-rank, and is given by the explicit formula (see App.~\ref{app:FormOfJ})
\begin{align}
    \mathcal J[\hat{X}]&=\hat{\rho}_{\rm ss}^{1/2}\tilde{X}^\dag\hat{\rho}_{\rm ss}^{-1/2}.
    \label{eq:JConstruction}
\end{align}
The TFD correlator in Eq.~(\ref{eq:TFD-corr-func-def-C_XY}) can then be written as a single system correlator, e.g. for $t>0$:
\begin{align}
    C^\text{TFD}_{XY}(t)= \langle \hat{X}(t) \mathcal J[\hat{Y}]\rangle 
        = \textrm{tr} \left( \hat{X}(t) \mathcal J[\hat{Y}] \hat{\rho}_{\rm ss}  \right)
\end{align}
Using this correspondence, the defining symmetry condition of hidden TRS in Eq.~(\ref{eq:doubledqdb_wT}) can be written in a manner that only involves a single system:
\begin{align}
    \textrm{tr} \left( \hat{X}(t) \mathcal J[\hat{Y}] \hat{\rho}_{\rm ss}  \right) = 
    \textrm{tr} \left( \hat{Y}(t) \mathcal J[\hat{X}] \hat{\rho}_{\rm ss}  \right)
    \label{eq:SingleSystemCorrSymmetry}
\end{align}
The above condition is formally equivalent to one of the many generalized quantum detailed balance conditions discussed in Ref.~\cite{fagnola_generators_2007,fagnola_generators_2010} (so-called SQDB-$\theta$).  It is also referred to as ``KMS detailed balance'' \cite{Carlen2017}.

In general, the symmetry condition in Eq.~(\ref{eq:SingleSystemCorrSymmetry}) is not helpful as an experimental tool, as it involves an operator whose very form depends on the system state $\hat{\rho}_{\rm ss}$, i.e. it is nonlinear in $\hat{\rho}_{\rm ss}$.  This is reminiscent of the problem of measuring Renyi entropies \cite{Ekert2002,Islam2015,Kaufman2016}.  Remarkably, all hope is not lost.  As we will show in Sec.~\ref{sec:ExactSols}, systems with hidden TRS are guaranteed to have their effective Hamiltonian $\hat{H}_{\rm eff}$ and jump operators $\hat{c}_k$ transform very simply under the action of $\mathcal{J}$.  For correlation functions involving these operators, Eq.~(\ref{eq:SingleSystemCorrSymmetry}) becomes a standard (and often simple) single-system correlator.

We thus have a key result that makes it possible to directly and simply test for hidden TRS in experiment:  hidden TRS implies that only a certain restricted class of correlation functions (directly identifiable from the master equation) are guaranteed to have Onsager time-symmetry.

\subsection{Example: Hidden TRS in dissipative Rabi-driven qubit}
\label{subsec:QubitHiddenTRS}

We show in App.~\ref{app:QubitCQDB} that a simple Rabi-driven qubit with loss fails to respect CQDB:  its correlation functions do not exhibit Onsager time-symmetry, regardless of how one tries to define time-reversal.  Here we show that this system nonetheless possesses a hidden TRS. Working in the rotating frame set by the drive, and making a rotating wave approximation, the master equation has the form of Eq.~(\ref{eq:qme}) with $M=1$ and
\begin{equation}
  \hat{H}= \Delta\hat{\sigma}_z + \frac{\Omega}{2}\hat{\sigma}_x, 
  \,\,\,
  \hat{c}_1 = \sqrt{\kappa} \hat{\sigma}_{-}.
  \label{eq:qubit-H-and-c_1}
\end{equation}
For simplicity, we consider the resonant-driving case $\Delta = 0$ in what follows.

This system has a unique steady state, and corresponding thermofield doubled states can be  constructed according to Eq.~(\ref{eq:tfd_wT}).  This construction depends on the choice of the hidden TRS operator $\hat{T}$; as discussed in App.~\ref{app:QubitCQDB}, the requirement that the steady state be TRS invariant (c.f.~Eq.~(\ref{eq:HiddenTRSRhoSS})) constrains $\hat{T}$ to the form:
\begin{equation}
    \hat{T} = \left[\frac{\sin(\psi/2)}{\sqrt{4b^2+1}} \left(\hat{1} - 2ib \hat{\sigma}_x \right) + i\cos(\psi/2) \hat{\sigma}_z \right] \hat{K}_z,
    \label{eq:qubit-general-TRS}
\end{equation}
where $b = \Omega / \kappa$, $\hat{K}_z$ is the complex conjugation operator acting in the $\hat{\sigma}_z$ basis, and $\psi$ is at this stage an arbitrary real parameter.  

To determine whether our system has hidden-TRS, we must find a $\hat{T}$ such that Eq.~(\ref{eq:doubledqdb_wT}) holds (i.e.~intra-system TFD correlators have a time symmetry).  We thus compute TFD correlators between different pairs of Pauli operators.  Here we will consider only the correlation function $C_{yz}^{\rm TFD}(t)$, and we leave the other two correlation functions to App.~\ref{app:qubit-h-trs}. 

Using Eq.~(\ref{eq:TFD-pointer-state-expansion}) we can decompose this into the classical correlation $C_{yz}^{\rm cl}(t)$ and the entanglement correction $C_{yz}^{\rm en}(t)$ as: $C_{yz}^{\rm TFD}(t) = C_{yz}^{\rm cl}(t) + C_{yz}^{\rm en}(t)$. The classical correlation is independent of $\psi$, and its time asymmetry is nonzero irrespective of how $\hat{T}$ is chosen:
\begin{equation}
    C_{yz}^{\rm cl}(t) - C_{yz}^{\rm cl}(-t) = -\frac{32b^5\sin\left(\frac{\alpha}{4}\kappa t\right)e^{-\frac{3}{4}\kappa t}}{\alpha(2b^2+1)(4b^2+1)}.
    \label{eq:qubit-tfd-cl-time-asymmetry}
\end{equation}
Here, $\alpha = \sqrt{16 b^2 -1}$.
We thus see that were we to neglect the entanglement correction, the system could never have hidden TRS. 

\begin{figure}[t]
     \centering
    \includegraphics[width=0.99\columnwidth]{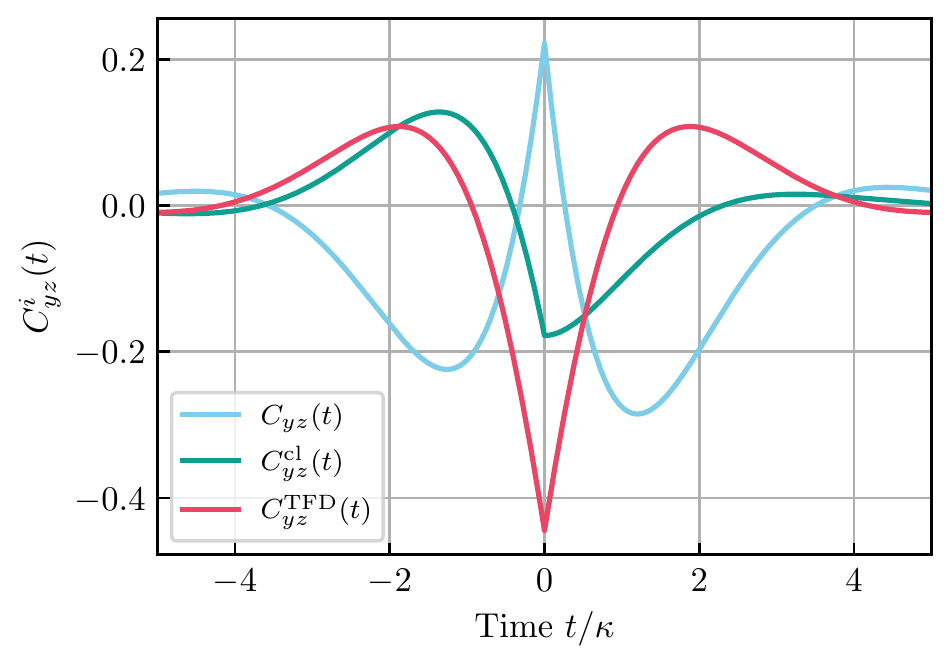}
     \caption{{\bf Correlation functions and hidden-TRS in a driven qubit.} 
    Stationary, connected $\langle \sigma_y(t) \sigma_z(0) \rangle$ correlation functions for the dissipative Rabi-driven qubit system in Eq.~(\ref{eq:qubit-H-and-c_1}), for a drive $\Omega$ equal to the decay rate $\kappa$.    Blue:  the standard single-system correlation function $C_{yz}(t)$ is asymmetric as a function of time, reflecting the fact that this system {\it does not} satisfy conventional quantum detailed balance.  Red:  Two-qubit correlator for a system prepared in a TFD state corresponding to the hidden-TRS operator $\hat{T}$ defined in Eq.~(\ref{eq:qubit-hidden-trs}). All TFD correlators symmetric in time, reflecting the presence of hidden TRS.  Green:  ``classical'' part of the TFD correlator
    (c.f.~Eq.~(\ref{eq:TFD-pointer-state-expansion})), which has no time symmetry.  The lack of symmetry shows that the importance of entanglement in the definition of hidden-TRS.    }
\label{fig:qubit-hidden-trs}
 \end{figure}

Now, we look at the time asymmetry of the entanglement correction, which is dependent on $\psi$:
\begin{equation}
    C_{yz}^{\rm en}(t) - C_{yz}^{\rm en}(-t) = -\frac{32b^5\sin\left(\frac{\alpha}{4}\kappa t\right)e^{-\frac{3}{4}\kappa t}}{\alpha(2b^2+1)(4b^2+1)}\cos\psi.
    \label{eq:qubit-tfd-en-time-asymmetry}
\end{equation}
Comparing Eqs.~(\ref{eq:qubit-tfd-cl-time-asymmetry}) and (\ref{eq:qubit-tfd-en-time-asymmetry}), we see that for the TRS with $\psi=\pi$, the entanglement correction to $C_{yz}^{\rm TFD}(t)$ modifies the classical correlation in just the right way to cancel the net time asymmetry. We see just how stark the effect is in Fig. \ref{fig:qubit-hidden-trs} which compares the full TFD correlation function with the classical correlation terms for $b=1$ at the TRS $\psi=\pi$. For reference, the single qubit correlation function $C_{yz}(t)$ for $b=1$ at $\psi=\pi$ is also included. This result shows the importance of the entanglement correction to restoring a notion of detailed balance to the Rabi-driven qubit and highlights the fact that the notion of hidden TRS has a distinctly quantum nature.

From the above, we conclude that our model does have a unique hidden TRS, described by the anti-unitary operator $\hat{T}_h$
\begin{equation}
    \hat{T}_{\rm h} = \frac{1}{\sqrt{4b^2+1}}(\hat{1} - 2ib\hat{\sigma}_x)\hat{K}_z.
    \label{eq:qubit-hidden-trs}
\end{equation}
In App. \ref{app:qubit-h-trs} we confirm that the remaining two correlation functions are time symmetric for this TRS. 

It is interesting to consider the form of $\hat{T}_h$ in various limits.  For weak Rabi driving (i.e.~$b \to 0$), Eq.~(\ref{eq:qubit-hidden-trs}) reduces to $\hat{T}_h=\hat{K}_z$.  In this limit the qubit system in fact satisfies SDQB with $\hat{T}=\hat{K}_z$ (i.e.~all correlation functions have standard Onsager time-symmetry). In the strong drive limit $b \gg 1$, $\hat{T}_h \rightarrow  -i\hat{\sigma}_x \hat{K}_z$.  Up to a phase, this just complex conjugation in the $\hat{\sigma}_y$ basis. To make sense of this, consider the steady state Eq.~(\ref{eq:qubit-driven-ss}) in this limit. To first order in small $b^{-1} \ll 1$, the steady state reduces to $\hat{\rho}_{\rm ss} = \frac{\hat{1}}{2}+\frac{b^{-1}}{2}\hat{\sigma}_y$.  The form of the hidden-TRS operator in this limit thus directly reflects the eigenvectors of $\hat{\rho}_{\rm ss}$. Furthermore, one can show that for any $b$, the hidden TRS $\hat{T}_{\rm h}$ corresponds to complex conjugation in the steady state eigenbasis.

\section{Hidden time reversal symmetry and dynamical constraints}
\label{sec:DynamicalEquiv}

We have now introduced our notion of hidden TRS (c.f.~Eqs.~(\ref{eq:HiddenTRSSchrod}),  (\ref{eq:doubledqdb_wT})), and demonstrated that this symmetry can hold even when the more standard CQDB symmetry is broken.  It still however may seem that hidden-TRS is nothing more than a formal curiosity.  We show here that this is not the case:  hidden-TRS is a symmetry that has direct operational utility in helping us understand complex phenomena, as it enables the exact solution of steady-states of non-trivial systems.  In particular, it is the symmetry condition that enables the surprising but powerful coherent quantum absorber method introduced in Ref.~\cite{stannigel_driven-dissipative_2012} and extended in Ref.~\cite{roberts_driven_2019}.  

\subsection{Equivalent subsystem dynamics and hidden TRS as a self-dual condition}

We start by demonstrating that  the hidden TRS condition can also expressed as a kind of {\it dynamical} equivalence between the two subsystems in our TFD state.  Consider a general system and a TFD state which does not necessarily satisfy the hidden TRS condition of Eq.~(\ref{eq:doubledqdb_wT}).  We stress that the TFD state is defined completely by the steady state of interest $\hat{\rho}_{\rm ss}$ and choice of anti-unitary $\hat{T}$.  We will take $\hat{\rho}_{\rm ss}$ to be full rank in what follows, and will consider intra-system correlations in this TFD state as defining a bilinear form:
\begin{align}
    \langle \langle \hat{X},\hat{Y} \rangle \rangle_{\mathcal T} \equiv \langle \psi_{T}|\hat{X}_A\hat{Y}_B|\psi_{T}\rangle. 
    \label{eq:bilinear_form}
\end{align}
where $\hat{X}$, $\hat{Y}$ denote arbitrary single-system operators.  This bilinear form can then be used to define the dual $\mathcal{E}^*$ of any given
single-system superoperator $\mathcal{E}$ via
\begin{align}
    \langle \langle \mathcal{E}[\hat{X}], \hat{Y} \rangle \rangle_{\mathcal T} \equiv
    \langle \langle \hat{X}, \mathcal{E}^*[\hat{Y}] \rangle 
    \rangle_{\mathcal T}.
    \label{eq:DualDefn}
\end{align}

Of particular interest is the case where $\mathcal{E}$ is the adjoint evolution operator $\mathcal{E}_t = \exp[\bar{\mathcal{L}}t]$ defined in Eq.~(\ref{eq:CorrFuncDefn}) in terms of the adjoint Liouvillian $\bar{\mathcal{L}}$ (c.f.~Eq.~(\ref{eq:LAdjointEqn})).  The LHS of Eq.~(\ref{eq:DualDefn}) then describes the correlation of a subsystem-$A$ operator at time $t$ and a subsystem $B$ operator at time zero.  In this case, the dual $\left( \mathcal{E}_t \right)^*$ has a direct physical interpretation:  it represents an alternate and equivalent time evolution of subsystem $B$ that would result in the same inter-system correlation.  This dual time evolution can be written as 
$\left( \mathcal{E}_t \right)^* = \exp \left[ \bar{\mathcal{L}}^* t \right]$.  
Thus, for a given subsystem-$A$ dynamics $\bar{\mathcal{L}}$, we have a corresponding ``mirrored'' dynamics $\bar{\mathcal{L}}^*$ for subsystem-$B$, defined by the constraint that it yield identical inter-system correlations, i.e.
\begin{align}
    \left \langle \left( \exp\left[ \bar{\mathcal{L}}t \right] \hat{X} \right)_A \hat{Y}_B \right \rangle_T =
        \left \langle \hat{X}_A \left( \exp\left[ \bar{\mathcal{L}}^* t \right] \hat{Y} \right)_B \right \rangle_T 
    \label{eq:MirrorCorrelations}
\end{align}

These notions now give an extremely transparent way to rephrase the hidden-time reversal condition of Eq.~\eqref{eq:doubledqdb_wT}:  
{\it the original system-$A$ dynamics and its mirrored version must be identical}, that is $\bar{\mathcal{L}}$ is self-dual, 
\begin{equation}
    \bar{\mathcal{L}}^* = \bar{\mathcal{L}}.
    \label{eq:LSelfDual}
\end{equation}
To see this, note first that if a system satisfies the hidden TRS condition of  Eq.~\eqref{eq:doubledqdb_wT}, then the bilnear form in Eq.~(\ref{eq:bilinear_form}) must be symmetric, i.e.    
$ \langle \langle \hat{X},\hat{Y} \rangle \rangle_{\mathcal T} = \langle \langle \hat{Y},\hat{X} \rangle \rangle_{\mathcal T}$; this follows from the $t=0$ limit of Eq.~\eqref{eq:doubledqdb_wT}. This in turn implies that the steady state $\hat{\rho}_{\rm ss}$ of the original master equation must be invariant under our hidden time-reversal operator $\hat{T}$ (i.e.~consider the case $\hat{Y} = \hat{1}$).  Combining these two conditions lets us express the hidden-TRS condition of Eq.~(\ref{eq:doubledqdb_wT}) as:
\begin{align}
    \langle \psi_T| \hat{X}_A(t)\hat{Y}_B(0)|\psi_T\rangle =  \langle \psi_T| \hat{X}_A(0)\hat{Y}_B(t)|\psi_T\rangle.\label{eq:pairing}
\end{align}
where for either subsystem, $\hat{O}(t) = \mathcal{E}_t[\hat{O}]$.  This now looks more like a standard Onsager-type correlation function symmetry, except that the two operators are measured on different subsystems.    Finally, comparing this equation against Eq.~(\ref{eq:DualDefn}) directly yields the self-duality condition in Eq.~(\ref{eq:LSelfDual}) \footnote{Note that we are implicitly using the fact that our bilinear form is non-degenerate (as $\hat{\rho}_{\rm ss}$ is full rank), something which guarantees the uniqueness of the dual.}.

\subsection{Hidden TRS as a symmetry of the Liouvillian}

We now show that the hidden TRS condition can be viewed as a dynamical symmetry that directly constrains the system's adjoint Liouvillian $\bar{\mathcal{L}}$.  To do this, we step back and consider a general system and $\hat{T}$, such that hidden TRS is not necessarily satisfied.  We then explicitly construct the dual Liouvillian $\bar{\mathcal{L}}^*$ that generates the mirrored-system dynamics, by considering each term in Eq.~(\ref{eq:LAdjointEqn}).  Our construction will explicitly make use of the exchange superoperator $\mathcal{J}$ introduced in Eqs.~(\ref{eq:JDefinition}) and (\ref{eq:JConstruction}); recall that this superoperator lets us convert subsystem-$A$ into corresponding subsystem-$B$ operators (and vice-versa) such that TFD expectation values are preserved.

The exchange superoperator allows us to efficiently express the desired dual of the adjoint Liouvillian $\bar{\mathcal{L}}$.  This is done using the following relations, that follow directly from the definition of $\mathcal{J}$ in Eq.~\eqref{eq:JDefinition}:
\begin{align}
    \langle \langle \hat{X}\hat{O}, \hat{Y} \rangle \rangle_{\mathcal T} &= 
        \langle\langle \hat{X}, \hat{Y}\mathcal J[\hat{O}] \rangle 
        \rangle_{\mathcal T},~\\
    \langle\langle \hat{O}^\dag\hat{X}, \hat{Y} \rangle \rangle_{\mathcal T} &= 
        \langle \langle \hat{X}, \mathcal J[\hat{O}]^\dag\hat{Y}
        \rangle \rangle_{\mathcal T}.
\end{align}
We can thus obtain an explicit expression for the desired dual $\bar{\mathcal{L}}^*$ as
\begin{align}
    \bar{\mathcal L}^{*}[\hat{A}] &\equiv 
        i(\mathcal J[\hat{H}_\text{eff}]^\dag \hat{A} - 
            \hat{A} \mathcal J[ \hat{H} _\text{eff}])\nonumber\\
    &~~~~~~~~~~~~~~~~~~~~~~~~~+\sum_{l}\mathcal J[\hat{c}_l]^\dag 
    \hat{A} \mathcal J[\hat{c}_l].\label{eq:dual_lindbladian}
\end{align}
Recall $\hat{H}_{\rm eff}$ is the effective non-Hermitian Hamiltonian, and $\hat{c}_l$ the jump operators in our original Lindblad master equation Eq.~(\ref{eq:qme}).  We thus see that the properties of the system-$B$ ``mirrored dynamics'' are encoded in the exchange superoperator $\mathcal{J}$.

We now ask what constraints ensue when we insist that the hidden TRS condition holds, and hence $\bar{\mathcal{L}}^* = \bar{\mathcal{L}}$. For two Liouvillians (each defined with traceless jump operators) to be equivalent, the effective Hamiltonians for each must be identical (up to an additive real constant), and the jump operators related by a unitary mixing matrix (see e.g. Ref.~\cite{parthasarathy_introduction_1992}). 

Hence, insisting that our system has hidden TRS leads to the following constraint equations:
\begin{align}
    \mathcal J[\hat{H}_\text{eff}] = \hat{H}_\text{eff}+E,~~~\mathcal J[\hat{c}_l] = 
        \sum_{k=1}^M  U_{lk} \hat{c}_k,
    \,\,\,\,\,
    U^2 = 1. \label{eq:qdb_jumps}
\end{align}
$E$ is a real number,  and $U_{lk}$ is a $M \times M$ unitary matrix.  The last constraint on $U$ (i.e.~that it is involutory) follows from the fact that if hidden TRS holds, then the steady state is itself invariant under $\hat{T}$ (c.f.~Eq.~(\ref{eq:HiddenTRSRhoSS})). This resulting additional symmetry of the TFD state then implies (via Eq.~(\ref{eq:JDefinition})) that two exchanges yield the identify superoperator:  $\mathcal J^2 =\mathds{1}$.  This immediately constrains the unitary matrix $U_{lk}$ to have purely real eigenvalues
\footnote{Note that the condition in Eq.~(\ref{eq:qdb_jumps}) on the jump operators implies the exchange superoperator acts as a unitary on the subspace spanned by the Lindblad operators $c_l$. Since $\mathcal J$ squares to one, the unitary operation in Eq. \eqref{eq:qdb_jumps}, which is the restriction of $\mathcal J$ to the subspace spanned by the Lindblad operators $c_l$, {\it also} squares to one. It follows that its eigenvalues can only be $1$ or $-1$. }.  

Eqs.~(\ref{eq:qdb_jumps}) represent necessary and sufficient conditions for our system to have a hidden TRS.  They are however somewhat unwieldy, as they directly involve the exchange superoperator, which is itself a function of $\hat{T}$ and the steady state $\hat{\rho}_{\rm ss}$.  We can eliminate the explicit appearance of $\mathcal{J}$ by using the fact that since $\hat{\rho}_{\rm ss}$ is full rank, the TFD state is {\it separating} (see e.g.~\cite{bratteli_operator_1997}): if two subsystem-$A$ operators have the same action on the TFD state, then they {\it must} be identical operators.  Stated explicitly:  
\begin{align}
    \left( \hat{X}_A|\psi\rangle = \hat{Y}_A|\psi\rangle \right) 
        \Leftrightarrow
            \left( \hat{X}=\hat{Y} \right).
    \label{eq:SeparatingCondition}
\end{align}

Using this, we can eliminate $\mathcal{J}$ from each equation in Eqs.~(\ref{eq:qdb_jumps}) by taking each side of each equation to be a system-$A$ operator, and applying it to the TFD state $\ket{\psi_T}$.  Using the definition of the exchange operator on the resulting state, this gives us an equivalent but more useful set of constraint equations:
\begin{align}
    \hat{H}_{{\rm eff},B} \ket{\psi_T} & = 
        (\hat{H}_{{\rm eff},A} + E) \ket{\psi_T} 
            \label{eq:StateConstraint1} \\
    \hat{c}_{l,B} \ket{\psi_T} & = 
        \left(\sum_{k=1}^M U_{lk} \hat{c}_{k,A} \right) \ket{\psi_T} 
        \label{eq:StateConstraint2}
        \\
    U^2 & = U U^\dagger = 1
        \label{eq:StateConstraint3}
\end{align}
Eqs.~(\ref{eq:StateConstraint1})-(\ref{eq:StateConstraint3}) are the main result of this subsection: {\it they express the existence of a hidden TRS symmetry directly as a constraint on the Hamiltonian and jump operators that define our open system dynamics}. 
Heuristically, these conditions imply that the action of $\hat{H}_{\rm eff}$ and the jump operators are ``almost'' the same whether they act on subsystem $A$ or $B$.  
Hidden TRS requires that these equations must hold for some pure state $\ket{\psi_T}$ of the doubled system, some constant $E$ and some involutory $M \times M$ unitary matrix $U$.  One can view this as a generalization of the classical detailed balance condition in Eq.~(\ref{eq:DBRates}).  While the classical condition only involves transition rates, our quantum conditions above constrain both the incoherent dynamics generated by the $\hat{c}_l$ operators, as well as the coherent system Hamiltonian $\hat{H}$.  

We stress that the above equations are equivalent to those derived in Ref.~\cite{fagnola_generators_2007} when considering the abstract ``SQDB-$\theta$'' version of quantum detailed balance.  By phrasing these conditions directly in terms of the thermofield double state, we will be able to directly exploit them as a means for efficiently {\it finding} unknown steady states (something that was not considered previously).

Finally, we note that Eq.~(\ref{eq:StateConstraint1}) - Eq.~(\ref{eq:StateConstraint2}) tells us that if hidden-TRS holds, than the action of the exchange superoperator is extremely simple when acting on $\hat{H}_{\rm eff}$ or the jump operators $\hat{c}_l$.  This means that for doubled-system TFD correlation functions involving these operators can be directly converted to single-system correlation functions.  As discussed extensively in Sec.~\ref{subsec:SingleSystemExperiment}, this gives a direct means for experimentally testing for hidden-TRS in a single system:  hidden-TRS ensures that a certain reduced class of correlation functions will obey Onsager-like time symmetry (c.f.~Eq.~(\ref{eq:SingleSystemCorrSymmetry})).


\section{Hidden TRS as a route to exact solutions}
\label{sec:ExactSols}

\subsection{Basic idea}

Classical detailed balance has a profound operational utility:  it provides an extremely efficient method for finding the steady state of a given dynamical model (i.e.~so called potential solutions of Fokker-Planck equations \cite{gardiner_stochastic_2009}). It is thus natural to ask whether something similar is possible using our notion of hidden TRS.  
If a system satisfies this symmetry, does this directly provide a method for solving for the steady state?  As we now show, the answer is a resounding yes.  The existence of hidden TRS places a strong constraint on the form of our dynamics via Eqs.~(\ref{eq:StateConstraint1}) - (\ref{eq:StateConstraint3}).  {\it These equations also provide an efficient method for finding an unknown steady state.}  To see this, we change perspective, and view $\ket{\psi_T}$ in these equations as an unknown pure state of a doubled version of the original system.  The goal is then to {\it find} a pure state $\ket{\psi_T}$, constant $E$, and unitary matrix $U$ such that Eqs.~(\ref{eq:StateConstraint1}) - (\ref{eq:StateConstraint3}) are satisfied.  If we are able to do this, then as we will show, our system has hidden TRS, and the desired system-$A$ steady state $\hat{\rho}_{\rm ss}$ is obtained by tracing out system-$B$ from $\ket{\psi_T}$.  Conversely, if we cannot do this, then our system does not have hidden TRS, and there is no generic simple route to finding the steady state.

We stress that the above procedure for finding the steady state is simpler than a direct brute-force approach.  Suppose our original system has a Hilbert space dimension $d$.  Without assuming hidden TRS, solving for the steady state of Eq.~(\ref{eq:qme}) reduces to the problem of solving for the null space of a matrix with dimensions $d^2 \times d^2$.
Without additional assumptions, this matrix does not have any obvious sparseness properties.  
In contrast, with the assumption of hidden TRS, we need to solve  Eqs.~(\ref{eq:StateConstraint1}) and  (\ref{eq:StateConstraint2}).  Each of these $M+1$ equations also involves a $d^2 \times d^2$ matrix.  However, each of these matrices has a simplified structure as there are no terms corresponding to an interaction between the two subsystems.  As a result, there can be at most $\mathcal{O}[d^3]$ non-zero matrix elements.  In addition, our constraint equations decouple the effective Hamiltonian physics (Eq.~(\ref{eq:StateConstraint1})) from the incoherent ``jump'' physics (Eq.~(\ref{eq:StateConstraint2})).
This effective non-interacting property provides a considerable simplification, as we will exploit more fully in the next section.

\subsection{Connection to perfect quantum absorbers}
\label{sec:absorber}

As we show below, the presence of hidden TRS guarantees that we can construct a simple mirrored system that {\it perfectly absorbs} everything emitted by the main system into its environment.  Such absorbing systems have been studied previously as a method for deriving exact solutions of certain Lindblad master equations \cite{stannigel_driven-dissipative_2012,roberts_driven_2019}.  Our discussion here will provide a generalization of this ``coherent quantum absorber'' method to systems with multiple jump operators, and also show that its success is indeed intimately connected to hidden time-reversal symmetry.  \\

To establish this connection, we again consider a general system described by the master equation Eq.~(\ref{eq:qme}) with a steady state $\hat{\rho}_{\rm ss}$.  We also construct a doubled system as in Sec.~\ref{subsec:HiddenTRSBasics} with a TFD state given by Eq.~(\ref{eq:tfd_wT}).  To start, we do not assume that the system has hidden TRS.  As discussed in Sec.~\ref{sec:DynamicalEquiv}, for a given subsystem $A$ dynamics (generated by $\bar{\mathcal{L}}$), we can always construct a corresponding ``mirrored'' dynamics on subsystem $B$ (generated by $\bar{\mathcal{L}}^*$), such that 
{\it either} evolution generates the same time-dependent inter-system correlations, c.f.~Eq.~(\ref{eq:MirrorCorrelations}).

 \begin{figure}
     \centering
    \includegraphics[width=0.90\columnwidth]{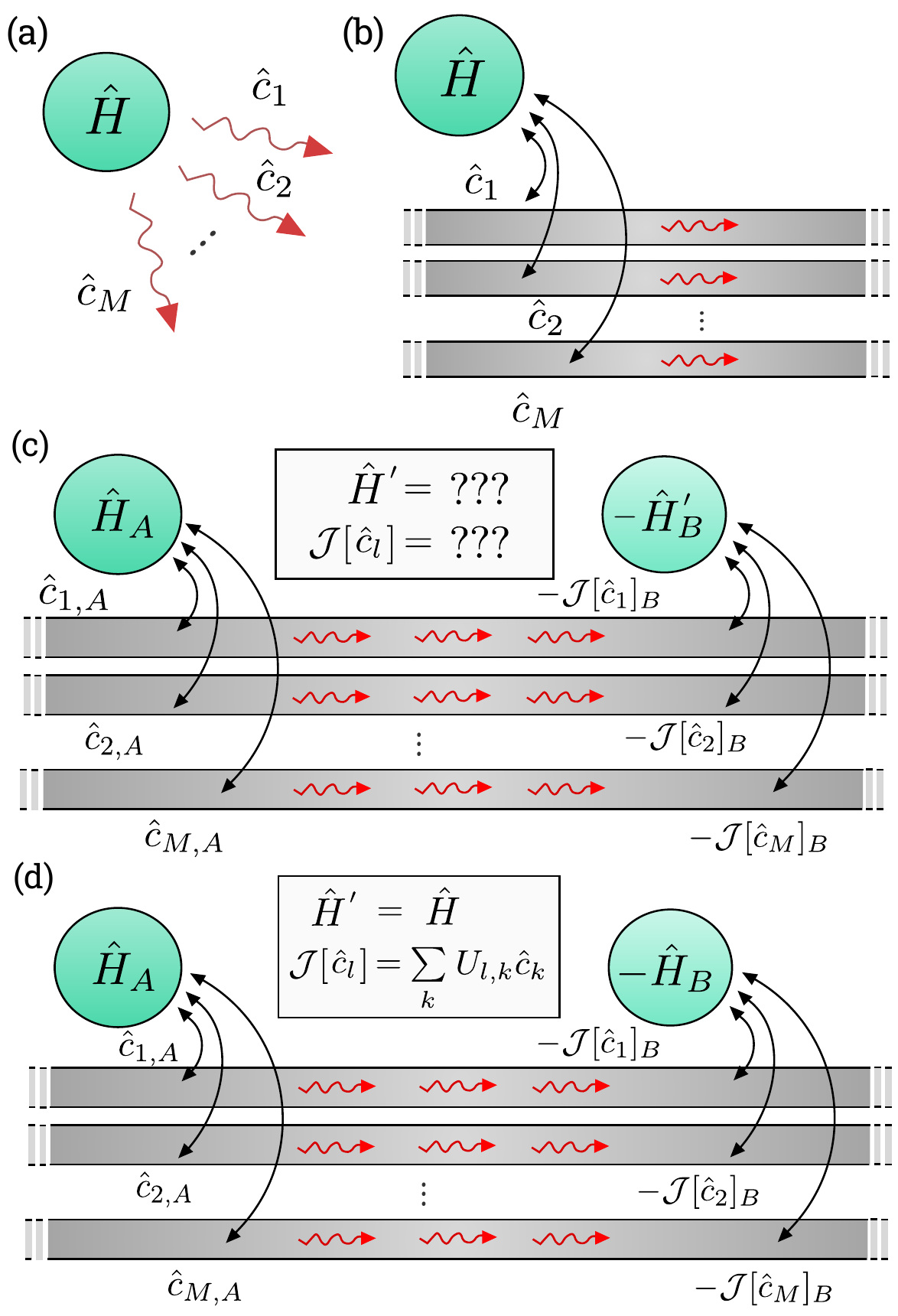}
     \caption{ {\bf Hidden TRS and perfect quantum absorbers}. (a) A Markovian quantum system evolves according to a master equation in Lindblad form, with Hamiltonian $\hat{H}$ and jump operators $\hat{c}_l$. 
     (b) A particular realization of the environment as a collection of unidirectional waveguides. 
     (c) The dual Lindbladian $\overline{\mathcal L}^*$ always formally solves the "perfect absorber" problem for the Lindbladian $\overline{\mathcal L}$ depicted in panel (a): when a system described by $\overline{\mathcal L}^*$ placed downstream, it absorbs {\it all} of the output radiation (red squiggly arrows) emitted by the original $A$ systems.  As a result, the two quantum systems $A$ and $B$ relax into a pure entangled state (which has the general form of a thermofield double state). In general, the Hamiltonian $\hat{H}'$ and jump operators $\mathcal{J}[\hat{c}_l]$ of the $B$ system are extremely complex and difficult to find.  
     (d) If the master equation in panel (a) has hidden TRS, then it is extremely easy to construct the Hamiltonian and jump operators of the absorber $B$ system. }
\label{fig:reservoir_engineering_picture}
 \end{figure} 

Somewhat remarkably, this mirrored dynamics is also {\it exactly} what is needed make subsystem-$B$ a ``perfect absorber'' of energy and information emitted by subsystem-$A$ into its environment (Fig.~\ref{fig:reservoir_engineering_picture}).  This can be established by using the exchange superoperator $\mathcal{J}$ introduced in Eq.~(\ref{eq:JDefinition}), which converts the action of a subsystem-$A$ operator acting on the TFD state to a subsystem-$B$ operator (and vice-versa).  From the definition of $\mathcal{J}$ we have:
\begin{align}
    \hat{H}_{{\rm eff},A} \ket{\psi_T} &= 
    \mathcal{J} [ \hat{H}_{{\rm eff}} ]_B \ket{\psi_T} \\
    \hat{c}_{l,A} \ket{\psi_T} &= 
    \mathcal{J} [ \hat{c}_{l} ]_B \ket{\psi_T}
\end{align}
where $\hat{H}_{\rm eff}$ is the effective Hamiltonian in our master equation, and $\hat{c}_l$ are the jump operators.  

As shown in App.~\ref{app:CascadedMapping}, these equations can be re-written as:
\begin{align}
    \hat{H}_{AB} \ket{\psi_T} = 0, 
    \,\,\,\,\,\,\,\,
    (\mathcal J[\hat{c}_l]_B - \hat{c}_{l,A})|\psi_T\rangle&=0. 
    \label{eq:GeneralDarkState}
\end{align}
Here,  the Hermitian Hamiltonian $\hat{H}_{AB}$ describes an {\it interaction} between the two subsystems in our doubled system:
\begin{align}
    \hat{H}_{AB} = 
        \hat{H}_A - \hat{H}'_B-\frac{i}{2} \sum_{l=1}^M(\hat{c}_{l,A}^\dag \mathcal J[\hat{c}_l]_B - h.c.) 
\end{align}
with
\begin{align}
    \hat{H}'  \equiv  \textrm{Re} \left[\mathcal J[\hat{H}_\text{eff}]\right],
    \,\,\,\,\,
    \hat{H}  \equiv  \textrm{Re} \left[\hat{H}_\text{eff} \right]
\end{align}
We denote the Hermitian part of an operator $\hat{A}$ as $\textrm{Re} [\hat{A} ]$.  Note that 
$\hat{H}$ is nothing but the Hermitian Hamiltonian in our original master equation.

Eq.~(\ref{eq:GeneralDarkState}) has an extremely suggestive form:  it tells us that $\ket{\psi_T}$ is necessarily a zero-energy eigenstate of a Hermitian Hamiltonian describing a doubled system with an inter-system coupling, and that it is also annihilated by particular combinations of jump operators. Together, these conditions imply that $\ket{\psi_T}$ is a zero energy pure-state, steady-state of the 
{\it cascaded} doubled system sketched in Fig.~\ref{fig:reservoir_engineering_picture}c.  In this cascaded system \cite{gardiner_driving_1993,carmichael_quantum_1993}, there is an independent chiral (directional) waveguide associated with each jump operator $\hat{c}_l$ in our original master equation. These channels mediate a directional coupling between systems $A$ and $B$, with $B$ downstream from $A$.  Using the standard theory of cascaded quantum systems \cite{gardiner_driving_1993,gardiner_quantum_2000}, the full master equation for this system is:
\begin{align}
    \partial_t\hat{\rho}_{AB}
        &=
            -i[\hat{H}_{AB},\hat{\rho}_{AB}] 
    +
        \sum_{l=1}^M \mathcal{D}\Big[\hat{c}_{l,A}-\mathcal J[\hat{c}_l]_B \Big]\hat{\rho}_{AB}.
        \label{eq:CascadedMEq}
\end{align}
Here $\hat{\rho}_{AB}$ is the density matrix of the doubled system, and $\mathcal{D}[\hat{z}] \hat{\rho} = \hat{z}\hat{\rho}\hat{z}^\dag -\{\hat{z}^\dag \hat{z}, \hat{\rho}\}/2$ is the standard Lindblad dissipation operator.
One can easily verify that if $\ket{\psi_T}$ satisfies Eq.~(\ref{eq:GeneralDarkState}), then it is a steady state of Eq.~(\ref{eq:CascadedMEq}).  

We thus have established the desired connection:  the same formal construction that gives us a correlation-conserving mirrored dynamics on subsystem-$B$ also tells us the precise dynamics that is needed for subsystem-$B$ to be a perfect absorber for subsystem-$A$.  We stress that for each possible choice of candidate time-reversal operator $\hat{T}$, we have a different TFD state, a different mirrored-dynamics (i.e.~Hamiltonian $\hat{H}^\prime$, jump operators $\mathcal J[\hat{c}_l]$), and hence a different possible coherent quantum absorber.

\subsection{Hidden TRS and simple absorbing dynamics}

The cascaded master equation in Eq.~(\ref{eq:CascadedMEq}) in principle provides a route for finding the steady state of the physical system $A$.  If one could find the steady state of this master equation, then tracing out system $B$  necessarily yields a steady state of the original single-system master equation.  One could simplify this procedure by trying to find a pure state solution to Eq.~(\ref{eq:CascadedMEq}).
Of course, there is an obvious problem to this approach:  the construction of Eq.~(\ref{eq:CascadedMEq}) is contingent on {\it already} knowing the steady state $\hat{\rho}_{\rm ss}$, as this is needed to construct 
the exchange superoperator $\mathcal{J}$.  

Things simplify considerably though in the case where our system possesses a hidden TRS.  In this case, we can use Eqs.~(\ref{eq:StateConstraint1})-(\ref{eq:StateConstraint3}) to dramatically simplify the cascaded master equation for our system.    The system-$B$ jump operators and Hamiltonian are then given by 
\begin{equation}
    \mathcal J[\hat{c}_l] \rightarrow \left( \hat{d}_l = \sum_{m=1}^M U_{lm} \hat{c}_m \right),~
    \textrm{Re} \left[\mathcal J[\hat{H}_\text{eff}] \right]\to \hat{H} +E
    \label{eq:CQAJumpOp}
\end{equation}
for some involutory unitary $M\times M$ matrix $U$ and real constant $E$, and the Hamiltonian of the coupled system becomes
\begin{align}
    \hat{H}_{AB} = 
        \hat{H}_{A} - \hat{H}_B - 
            \frac{i}{2}
                \sum_{l=1}^M
                \left( 
                    \hat{c}_{l,A}^\dag  \hat{d}_{l,B} -h.c.
                    \right),
    \label{eq:CQAHamiltonian}
\end{align}
where $E$ is now implicitly absorbed into an energy-shift of the dark state. 

We can now view this as a {\it method} for finding an unknown steady state of our original system-$A$ master equation in Eq.~(\ref{eq:qme}).  If we assume the existence of hidden TRS, finding this steady state is equivalent to finding a involutory $M \times M$ unitary matrix $U$ and energy $E$, such that the cascaded master equation in Eq.~(\ref{eq:CascadedMEq}) (with the simplifications of Eqs.~(\ref{eq:CQAJumpOp}) and (\ref{eq:CQAHamiltonian})) yields a pure-state, steady state.  This pure state then gives us the desired system-$A$ steady state by just tracing over system $B$.

The technique detailed above is a generalized version of the CQA exact solution method introduced in Ref.~\cite{stannigel_driven-dissipative_2012} for solving master equations with a single jump operator.  Our extension to systems with multiple jump operators involves a new object, the involutory unitary matrix $U$.  We have shown that this solution technique is thus intimately connected to the notion of a hidden TRS, and thus to the generalized SQDB-$\theta$ quantum detailed balance conditions introduced earlier on purely formal grounds \cite{fagnola_generators_2007,fagnola_generators_2010}.  As far as we know, this is the first example of this notion of quantum detailed balance having an operational utility.


\section{Hidden-TRS in nonlinear driven-dissipative quantum cavities}
\label{sec:CavityHiddenTRS}

 \begin{figure}
     \centering
    \includegraphics[width=0.999\columnwidth]{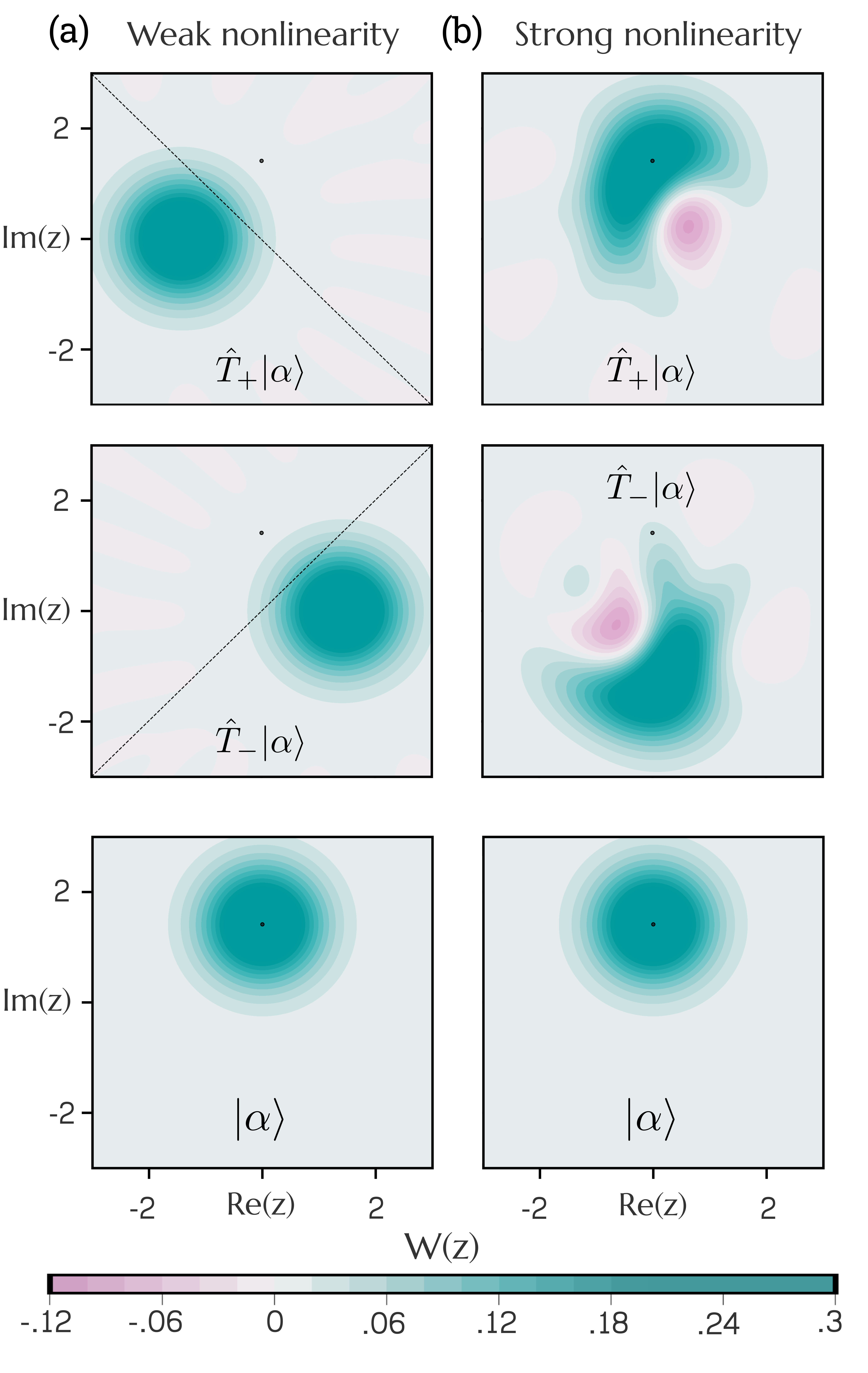}
     \caption{ {\bf Hidden time-reversal symmetry operations in a parametrically-driven Kerr cavity}. 
     For vanishing single-photon drive $\Lambda_1$, the driven-dissipative cavity model in 
     Eqs.~(\ref{eq:KerrHamiltonian})-(\ref{eq:KerrJumpOps}) has two distinct hidden time-reversal symmetries, corresponding to anti-unitary operators $\hat{T}_\pm$. Here, we plot the Wigner functions of the states $\hat{T}_{\sigma} \ket{\alpha}$, where $\ket{\alpha}$ is a coherent state (amplitude $\alpha = \sqrt{2}i$, black dots). 
     (a) For weak nonlinearity, $K = 5\times 10^{-4} \kappa_1$, and two-photon drive $\Lambda_2 = 6.25 \times 10^{-5} \kappa_1$, $\hat{T}_{\pm}$ are simple phase-space reflections about the axes $\theta = \pm \arg (\Lambda_2/i\kappa)$ (indicated by dashed black lines). (b) For strong nonlinearity, $K = \kappa_1, \Lambda_2 = \kappa/8$, hidden-TRS operations become highly non-Gaussian, as indicated by the presence of significant Wigner negativity in the final states. }
\label{fig:hidden_TRS}
 \end{figure} 

At this stage, we have established the basic notion of hidden TRS.  This symmetry can hold even if the more conventional CQDB condition (Sec.~\ref{subsec:CQDB}) is broken. Moreover, it directly enables a simple but powerful method for finding non-trivial steady states (Sec.~\ref{sec:ExactSols}).  We have illustrated these ideas by explicitly considering hidden TRS in a model of a dissipative Rabi-driven qubit (see App.~\ref{app:QubitCQDB} and \ref{subsec:QubitHiddenTRS}).  In this section, we turn to a more complex class of models.  These describe a bosonic mode (canonical annihilation operator $\hat{a}$) with a Kerr or Hubbard-$U$ type interaction, subject to both one and two particle coherent driving, as well as one and two-particle losses.  The system is described by a Lindblad master equation Eq.~\eqref{eq:qme} with a coherent Hamiltonian:  
\begin{align}
    \hat{H} &= \frac{K}{2}\hat{a}^\dagger \hat{a}^\dag \hat{a}\hat{a} +\Delta\hat{a}^\dagger\hat{a}
        \label{eq:KerrHamiltonian}\\
    &~~~~~~~~~~~~~~+ \bigg(\Lambda_1\hat{a}^\dag + \frac{\Lambda_2}{2}\hat{a}^\dag\hat{a}^\dag+ h.c.\bigg) \nonumber
\end{align}
and with jump operators
\begin{align}
    \hat{c}_1  = \sqrt{\kappa_1} \hat{a}, \,\,\,\,\, \hat{c}_2  = \sqrt{\kappa_2} \hat{a}^2
    \label{eq:KerrJumpOps}
\end{align}
This model describes a dissipative cavity mode driven both with linear and parametric drives that have commensurate frequencies (working within rotating-wave approximations, and in a rotating frame that eliminates the time dependence of the drives).  It is a ubiquitous system, having both been studied extensively in quantum optics, and more recently in the field of superconducting quantum circuits as a route to error-corrected quantum memories \cite{mirrahimi_dynamically_2014,leghtas_confining_2015,wang_schrodinger_2016,puri_engineering_2017,Grimm2020}.

It is well known that the steady state of this class of systems can be found analytically using quantum-optical phase space methods \cite{drummond_generalised_1980,wolinsky_quantum_1988,bartolo_exact_2016}; more recent work has shown that these exact solutions can be derived more directly (and even extended) using the
coherent quantum absorber method (CQA) \cite{stannigel_driven-dissipative_2012,roberts_driven_2019}.  An underlying explanation however for {\it why} these models are solvable has been lacking.  We now have such an explanation:  this class of models possess hidden-TRS, which directly leads to their solvability.

In what follows, we discuss the nature of hidden-TRS in these systems, showing that hidden-TRS is present even though (generically) CQDB does not hold.  Crucially, we show that the presence of hidden-TRS has observable consequences even in experiments on a single system:  while generic correlation functions do not exhibit a time-symmetry, there is a special class of correlators that do. In Sec.~\ref{sec:complex_P}, we will show how hidden-TRS directly enables the required symmetry exploited in the complex-$P$ function phase space methods that were first used to solve these systems \cite{drummond_generalised_1980,wolinsky_quantum_1988}.

\subsection{Multiple non-trivial hidden-TRS symmetries}

We start by noting that our general driven-dissipative resonator problem does not satisfy CQDB, and thus its correlation functions do not all exhibit a simple time-symmetry.  An example of such a lack of correlation function symmetry is shown in Fig.~\ref{fig:J_invariant_corr}.  More generally, as discussed in Sec.~\ref{subsec:CQDBtrivial}, CQDB can only hold if the system's steady state commutes with $\hat{H}$.  This condition is violated except in the vanishing dissipation limit $\kappa_1, \kappa_2 \rightarrow 0^+$.  

Despite the lack of CQDB, these systems {\it always} possess hidden-TRS, which explains their solvability.  The specific nature however of the symmetry operator (or operators) depends on the particular version of the model.  Consider first the most common case, where there is no two-photon loss, $\kappa_2 = 0$.  To determine whether our system has hidden-TRS, we consider a doubled two-cavity system and a two-cavity state $\ket{\psi_T}$. The question is whether this state could represent a TFD state constructed using an anti-unitary operator $\hat{T}$ which describes a hidden-TRS (c.f. Eq.~(\ref{eq:tfd_wT})).  From Eqs.~(\ref{eq:StateConstraint1})-(\ref{eq:StateConstraint3}), such a state must satisfy:
\begin{align}
    \hat{a}|\psi_T\rangle&=u\hat{b}|\psi_T\rangle\label{eq:the_u_eqn}\\
     \hat{H}_{\text{eff},A}|\psi_T\rangle&=(\hat{H}_{\text{eff},B}+E)|\psi_T\rangle
     \label{eq:KerrHeffEqn}
\end{align}
for some real energy $E$ and constant $u = \pm 1$.  Here (as always) $\hat{H}_{\rm eff} = \hat{H} - i \kappa_1 \hat{a}^\dagger \hat{a}/2$ is the effective non-Hermitian Hamiltonian in our master equation.
If we can find a two-cavity state $\ket{\psi_T}$ satisfying the above equations, then we are {\it guaranteed} both to have hidden TRS, and to be able to solve for our system using the CQA method of Sec.~\ref{sec:ExactSols}.

If we have a non-zero single-photon drive $\Lambda_1$, one can only solve Eqs.~(\ref{eq:the_u_eqn})-(\ref{eq:KerrHeffEqn}) if $u=1$ and $E = 0$.  With these choices, there is a unique solution 
for the two-cavity state $\ket{\psi_T}$.  This was explicitly found and expressed as a confluent hypergeometric function in Ref.~\cite{roberts_driven_2019}, which demonstrated that this model can be solved using CQA.  Hence, the system has a unique hidden-TRS operator $\hat{T}$ in this case.  As we will see in the next section, this gives us more than just a way to understand the solvability of the model:  it also directly lets us predict a surprising correlation function symmetry.  

It is also interesting to consider the special case where there is no single-photon drive, $\Lambda_1\to 0$.  Because of the single photon loss, the system still has a unique steady state.  However, there are now {\it two distinct}
hidden-TRS symmetries $\hat{T}_\pm$, each corresponding to distinct TFD states $|\psi_T^\pm \rangle$:
\begin{align}
    |\psi_T^\pm\rangle = \sum_n \sqrt{p_n} |n\rangle_A\hat{T}_\pm |n\rangle_B.
\end{align}
We stress that both these states each yield the same $\hat{\rho}_{\rm ss}$ when the auxiliary second cavity is traced out.  Formally, the two TFD states (and corresponding $\hat{T}_\pm$) are found by solving Eqs.~(\ref{eq:the_u_eqn})-(\ref{eq:KerrHeffEqn}) for $E=0,u=1$ and $E=0,u=-1$.  The explicit states can 
be found analytically in terms of Bessel functions \cite{roberts_driven_2019}.

We thus have our first example of a physical system with multiple, distinct hidden TRS symmetries; other examples are listed in Table \ref{tab:thermal-trs}.  Using the explicit forms of the TFD states, we can explicitly compute the action of the hidden TRS symmetry operators $\hat{T}_{+}$ and $\hat{T}_{-}$.  In general, their action is highly non-trivial (as can be seen in Fig.~\ref{fig:hidden_TRS}, where we show their action in phase space on an initial coherent state).  

In the limit of vanishing nonlinearity $K \rightarrow 0$, the hidden-TRS operators $\hat{T}_{\sigma}$ take a simple form.  In this case, the two TFD states limit to simple two-mode squeezed states:
\begin{align}
    |\psi_T^\pm\rangle \underset{K\to 0}{\sim} e^{\frac{\Lambda_2}{2\Delta_\text{eff}}(\hat{a}^\dagger\pm \hat{b}^\dagger)^2}|0,0\rangle,
\end{align}
where $\Delta_\text{eff}\equiv \Delta +i\kappa_1/2$ and $\ket{0,0}$ is the two-cavity vacuum state. Expanding out the exponential allows us to pick out the corresponding hidden time-reversal operators, which correspond to simple phase-space reflections about the axes $\theta = \pm\arg (\Lambda_2/\Delta_\text{eff})$ in phase space:
\begin{align}
    \hat{T}_\pm \underset{K\to 0}{\sim}e^{\arg (\pm \Lambda_2/\Delta_\text{eff}) \hat{a}^\dagger \hat{a}}\hat{K},
\end{align}
where here, $\hat{K}$ denotes complex-conjugation in the Fock basis. For non-zero Kerr, the corresponding time-reversal operations become highly nontrivial and non-Gaussian, and must be extracted via a numerical Schmidt decomposition. In Fig. \ref{fig:hidden_TRS}, we show the action of $\hat{T}_{\pm}$ for both $K = 5\times 10^{-4}\kappa_1$ (weak nonlinearity) and $K=\kappa_1$ (strong nonlinearity).

\subsection{Experimental consequences of hidden-TRS}

Our finding that driven-dissipative nonlinear cavities possess a hidden-TRS does more than simply explain why these systems are exactly solvable:  it also lets us predict observable phenomena that are accessible in a standard single-system experiment.  Recall our discussion in Sec.~\ref{subsec:SingleSystemExperiment}:  while hidden-TRS (by definition) guarantees a symmetry of doubled-system TFD correlation functions, for certain operators, this directly implies time-symmetry of standard, single-system correlators.  In particular, these special operators are ones that transform simply under the exchange operator $\mathcal{J}$.  By virtue of Eqs.~(\ref{eq:StateConstraint1})-(\ref{eq:StateConstraint3}), the effective Hamiltonian $\hat{H}_{\rm eff}$ and jump operators $\hat{c}_k$ are guaranteed to be such special operators.

\begin{figure}
     \centering
    \includegraphics[width=0.999\columnwidth]{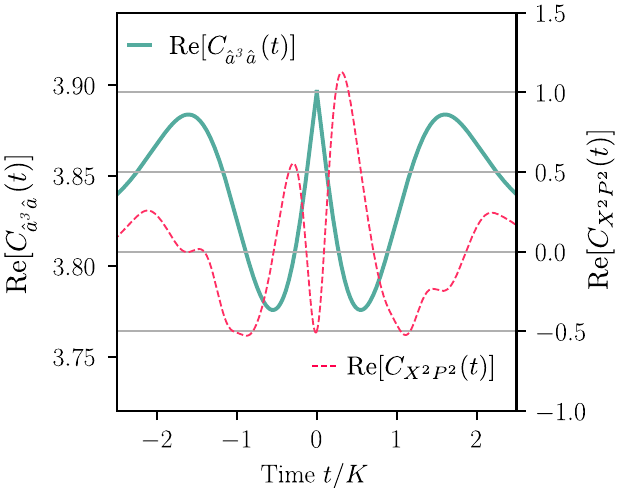}
     \caption{ {\bf Time symmetry of special correlation functions in a driven Kerr resonator}. 
     Real part of the connected, steady-state correlation function $C_{a^3,a}(t)\equiv \langle \hat{a}^3(t)\hat{a}\rangle$ (c.f.~Eq.~(\ref{eq:probe})) 
     for a parametrically driven nonlinear cavity 
     with $\Lambda_2 = K$, $\kappa_1 = 0.4K$ and $\kappa_2 = \Lambda_1 = \Delta = 0$. 
     This correlation function is symmetric in $t$, something that is guaranteed by the existence of hidden TRS.  We also plot another quartic correlation function $C_{X^2P^2}(t)$ (where $\hat{X}$, $\hat{P}$ are canonical quadrature operators).  This correlator is clearly asymmetric as a function of time.  Hidden-TRS only ensures that a certain restricted class of correlators are time symmetric (in contrast to the more commonly studied CQDB which guarantees all correlators exhibit a form of time-symmetry).  }
\label{fig:J_invariant_corr}
 \end{figure}

As a specific example, consider the following steady-state, single-system correlation function:
\begin{align}
    C_{a^3,a}(t)& \equiv \begin{cases}\langle \hat{a}^3(t) \hat{a}(0)\rangle&t\geq 0\\
    \langle \hat{a}(-t) \hat{a}^3(0)\rangle&t<0.
    \end{cases}
    \label{eq:probe}
\end{align}
If we set $\kappa_2= 0 $, Eqs.~(\ref{eq:StateConstraint1})-(\ref{eq:StateConstraint3}) ensure that $\mathcal{J}[\hat{a}] = \pm \hat{a}$.  From the definition of the exchange operator, it follows that $\mathcal{J}[\hat{a}^m] = (-1)^m \hat{a}^m$.  As a result, hidden-TRS guarantees 
(via Eq.~(\ref{eq:SingleSystemCorrSymmetry})) the above correlator has an Onsager-like time symmetry:  
$C_{a^3,a}(t) = C_{a^3,a}(-t)$.  We stress that this correlation function symmetry is special:  unlike the case with CQDB, most correlation functions {\it will not} exhibit any time-symmetry.  This behavior is shown explicitly in  Fig.~\ref{fig:J_invariant_corr}, where we contrast the correlator $C_{a^3,a}(t)$ (time-symmetric) with a more generic correlator involving quadrature operators $\hat{X} = (\hat{a} + \hat{a}^\dagger ) / \sqrt{2}$, $\hat{P} = -i(\hat{a} - \hat{a}^\dagger ) / \sqrt{2}$.  Hidden-TRS does not enforce any special symmetry of this latter correlator; hence, as expected, it is manifestly not symmetric in time.  We stress that even though $\mathcal{J}[\hat{a}^2]$ is simple, this does not imply that $\mathcal{J}[\hat{a}^\dag \hat{a}]$ is simple.  

We thus have a clear experimental test for confirming the existence of hidden-TRS in this class of systems.

\section{Breaking of hidden-TRS by thermal fluctuations and interactions}
\label{sec:ThermalHiddenTRSBreaking}

We have now demonstrated that hidden-TRS holds in two very different zero-temperature dissipative models:  a Rabi-driven qubit with loss (Sec.~\ref{subsec:QubitHiddenTRS}), and a driven nonlinear cavity with one and possibly two photon loss processes (Sec.~\ref{sec:CavityHiddenTRS}).  Within the setting of our Lindblad master equations, zero temperature corresponds to only having dissipators that remove (and not add) excitations. 

The natural next question is to ask what happens to hidden-TRS if we introduce a non-zero temperature to the above systems. This corresponds to adding dissipative processes that can add excitations.  We show that in a generic setting where there is both coherent (Hamiltonian) driving as well as nonlinearity, introducing such thermal dissipators can break hidden-TRS.  The only exceptions to this are the case of no driving (where the system is effectively classical), or the case of no nonlinearity (where the steady state is Gaussian).   Our results here suggest that for a generic nonlinear driven-dissipative system, hidden-TRS is a symmetry associated with vacuum fluctuations, and hence only emerges as one approaches the zero-temperature limit.    

Our work here is inspired by and complements seminal studies from Dykman and co-workers of related phenomena in driven nonlinear oscillators \cite{DykmanKrivoglaz1979,Dykman1988,marthaler_switching_2006,dykman_periodically_2012,Guo2013,LingzhenThesis,Yaxing2019}.  These works studied the basic nonlinear resonator model of Eqs.~(\ref{eq:KerrHamiltonian})-(\ref{eq:KerrJumpOps}) in the limit of weak dissipation, where the quantum master equation can be reduced to a simpler Pauli master equation (i.e.~one can drop off-diagonal elements of the density matrix in the energy eigenstate basis).  The resulting classical master equation was found to satisfy the classical detailed balance condition of Eq.~(\ref{eq:DBRates}) at zero-temperature; in a semiclassical limit, this could be shown analytically.  Further, it was shown that this classical detailed balance failed to hold at non-zero temperatures, and that in the semiclassical limit, the corresponding transition temperature became exponentially small.  Our work extends these results:  by formulating detailed balance in completely quantum manner using hidden TRS, we are not limited to weak-damping or semiclassical regimes.  We also discuss how the breaking of hidden-TRS by thermal fluctuations is contingent on having driving and nonlinearity; without both these ingredients, there is no symmetry breaking.  Finally, we discuss how this symmetry breaking could be directly probed in experiment by measuring the time-symmetry of correlation functions.

\subsection{Rabi-driven qubit subject to thermal dissipation}

A driven-dissipative qubit is a simple example to illustrate the breaking of hidden TRS due to thermal fluctuations. While this model has hidden-TRS at zero temperature (c.f.~Sec.~\ref{subsec:QubitHiddenTRS}), this symmetry is broken in the presence of {\it both} coherent driving and thermal fluctuations.  The master equation is of the form Eq.~(\ref{eq:qme}) but now with $M=2$ and
\begin{align}
    \hat{H} &= \Delta\hat{\sigma}_z +\frac{\Omega}{2}\hat{\sigma}_x, \nonumber \\
    \hat{c}_1 &= \sqrt{\kappa(1+\bar{n}_{\rm th})}\hat{\sigma}_{-}, \,\,\, \hat{c}_2 = \sqrt{\kappa\bar{n}_{\rm th}}\hat{\sigma}_{+}.
\end{align}
where $\bar{n}_{\rm th}$ 
represents the bath thermal occupancy at the qubit frequency.  

\subsubsection{Thermal dissipation with no drive}

In the absence of a Rabi drive (i.e.~$\Omega = 0$), the unique steady state of our master equation has the thermal equilibrium form:
\begin{equation}
    \hat{\rho} = \frac{1+\bar{n}_{\rm th}}{1+2\bar{n}_{\rm th}}
    \ket{g}\bra{g} + 
    \frac{\bar{n}_{\rm th}}{1+2\bar{n}_{\rm th}}
    \ket{e}\bra{e}.
    \label{eq:qubit-undriven-thermal-ss}
\end{equation}
where $\ket{g}, \ket{e}$ denote $\hat{\sigma}_z$ eigenstates.  This steady state
commutes with $\hat{H}$, and it is easy to confirm that the system has CQDB.  Due to the lack of coherences, the problem is analogous to a classical two-state system; hence, the presence of detailed balance is not surprising.

Formally, the system still possesses a set of hidden-TRS symmetries; this symmetry is however not unique.  There is a one parameter family of hidden TRS operators
\begin{equation}
    \hat{T} = (e^{i \psi}|e \rangle\langle e| + |g \rangle\langle g|)\hat{K}_z.
\end{equation}
For each $\psi$ there is a corresponding $U$ matrix (c.f. Eq.~(\ref{eq:qdb_jumps})) 
\begin{equation}
    U_\psi = \begin{pmatrix} 0 & e^{-i\psi} \\ e^{i \psi} & 0 \end{pmatrix}
\end{equation}
for which the dynamical constraints Eqs.~(\ref{eq:StateConstraint1})-(\ref{eq:StateConstraint3}) are satisfied.

The presence of hidden TRS in the finite-temperature, undriven qubit system implies
that it can be solved using the coherent absorber method of Sec.~\ref{sec:absorber}.  The qubit-plus-absorber system has the cascaded Hamiltonian Eq.~(\ref{eq:CQAHamiltonian}) where $\hat{H}_A$ is the qubit Hamiltonian $\hat{H}=\frac{1}{2}\omega_0 \hat{\sigma}_z$ acting on the physical qubit $A$ and $\hat{H}_B$ is the  Hamiltonian acting on the auxiliary qubit $B$ (the absorber). The cascaded system also has the collective jump operators
\begin{align}
    \hat{C}_1 &= \hat{c}_{1,A} - e^{-i\psi}\hat{c}_{2,B}, \\
    \hat{C}_2 &= \hat{c}_{2,A} - e^{i\psi}\hat{c}_{1,B}. 
\end{align}

The pure state which is simultaneously dark with respect to $\hat{H}_{AB}$, $\hat{C}_1$, and $\hat{C}_2$ is
\begin{equation}
    |\psi_0\rangle = \frac{1}{\sqrt{1+2\bar{n}_{\rm th}}}\left(\sqrt{1+\bar{n}_{\rm th}}|gg\rangle + e^{i\psi}\sqrt{\bar{n}_{\rm th}}|ee\rangle\right).
\end{equation}
After tracing out the absorber system, the single site steady state density matrix is precisely Eq. \eqref{eq:qubit-undriven-thermal-ss}.

\subsubsection{Thermal dissipation with a non-zero drive}

We now ask what happens to our thermal qubit when a non-zero drive is added ($\Omega \neq 0$).  For simplicity we take $\Delta = 0$ (resonant driving), and define the dimensionless driving $b' \equiv \Omega/\kappa(1+2\bar{n}_{\rm th})$.  With this definition, the steady state $\hat{\rho}_{ss,T}$ of the driven qubit with thermal dissipation is given by the zero temperature result in Eq.~(\ref{eq:qubit-driven-ss}) with the simple substitution $b \rightarrow b'$.

Furthermore the eigensystem of the Liouvillian at finite temperature is obtained from the zero temperature results in Eqs.~(\ref{eq:qubit-mode-r0})-(\ref{eq:qubit-mode-r3}) by replacements $\hat{\rho}_{\rm ss} \to \hat{\rho}_{ss,T}$, $b \to b'$, and $\kappa \to \kappa(1+2\bar{n}_{\rm th})$. Finally, the permissible TRS of the finite temperature system are given by Eq.~(\ref{eq:qubit-general-TRS}) with $b \to b'$.

We consider the TFD correlation function $C_{yz}^{\rm TFD}(t)$ defined in Eq.~(\ref{eq:TFD-corr-func-def-C_XY}) for Pauli operators $\hat{\sigma}_y,\hat{\sigma}_z$. As in the zero-temperature case, we decompose this into classical correlation and the entanglement correction using Eq.~(\ref{eq:TFD-pointer-state-expansion}), and we look at the time asymmetry of each. At finite temperature, the classical correlation asymmetry picks up new temperature dependent terms:
\begin{align}
    &C_{yz}^{\rm cl}(t) - C_{yz}^{\rm cl}(-t) =  \nonumber \\
    &-\frac{8b'\sin\left(\frac{\alpha}{4}\kappa t\right)e^{-\frac{3}{4}\kappa t}}{\alpha(2b^{\prime 2}+1)(4b^{\prime 2}+1)} \left[ 4b^{\prime 4} +(4b^{\prime 2} +1)\eta_{\rm th} \right]
    \label{eq:qubit-thermal-cl-time-asymmetry}
\end{align}
where we have defined 
\begin{equation}
    \eta_{\rm th} = 1 - \frac{1}{(1+2\bar{n}_{\rm th})^2}.
\end{equation}

In contrast, the $\psi$-dependent entanglement correction is
\begin{align}
    &C_{yz}^{\rm en}(t) - C_{yz}^{\rm en}(-t) =  \nonumber \\
    &-\frac{8b'\sin\left(\frac{\alpha}{4}\kappa t\right)e^{-\frac{3}{4}\kappa t}}{\alpha(2b^{\prime 2}+1)(4b^{\prime 2}+1)} \left[2b^{\prime 2}\sqrt{4b^{\prime 4} +(4b^{\prime 2} +1)\eta_{\rm th}}\right]\cos\psi,
    \label{eq:qubit-thermal-en-time-asymmetry}
\end{align}
which also gains temperature-dependent terms. 

\begin{figure}[t]
     \centering
    \includegraphics[width=0.99\columnwidth]{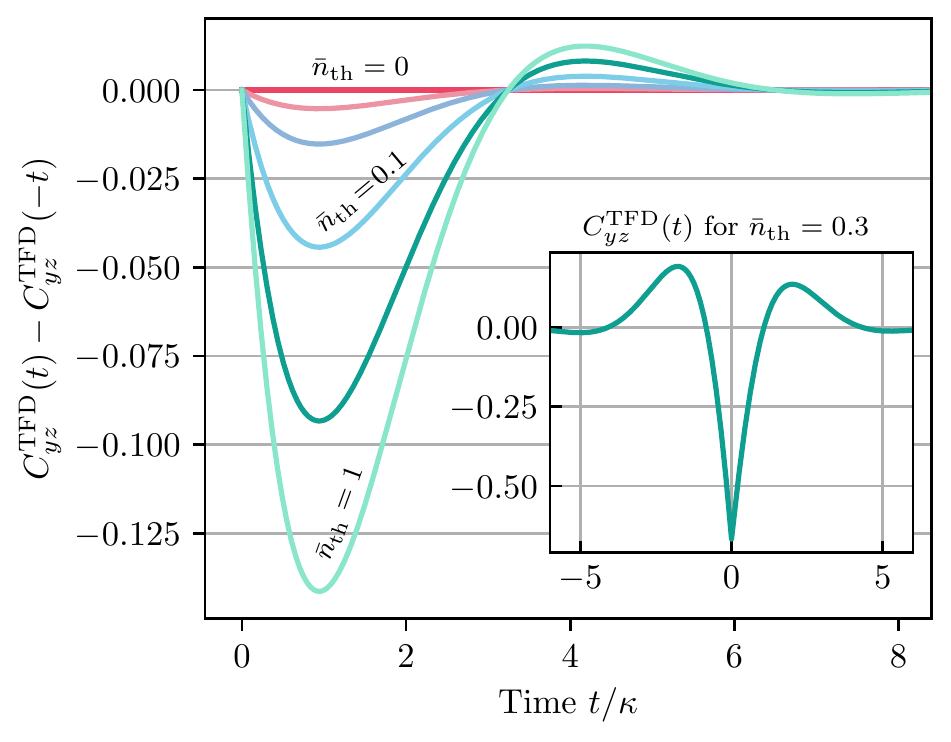}
     \caption{{\bf Hidden-TRS breaking in a driven qubit.} 
     {\bf Inset:} The doubled-system TFD $\sigma_y$-$\sigma_z$ connected correlation function
     $C_{yz}^{\rm TFD}(t)$ (c.f.~Eq.~(\ref{eq:TFD-corr-func-def-C_XY})) as a function of time, for a Rabi-driven dissipative qubit for a non-zero temperature corresponding to $\bar{n}_{\rm th} = 1$.  The TFD state is defined by the hidden-TRS operator $\hat{T}$ (c.f.~Eq.~(\ref{eq:qubit-hidden-trs})). 
     {\bf Main plot:} The time asymmetry of the TFD correlation function $C_{yz}^{\rm th}(t) - C_{yz}^{\rm th}(-t)$ versus time $t$ for various temperatures. The values of $\bar{n}_{\rm th}$ are in order from top to bottom: 0, 0.01, 0.03, 0.1, 0.3, and 1. The $\bar{n}_{\rm th} = 0$ trace is identically zero which reflects the presence  hidden TRS at zero temperature.
     The onset of asymmetry heralds the breaking of hidden-TRS with the introduction of thermal fluctuations.  
     All functions are computed for a resonant Rabi drive with amplitude 
     $\Omega = \kappa (1 + 2 \bar{n}_{\rm th})$, where $\kappa$ is the loss rate.}
\label{fig:qubit-thermal-breaking}
 \end{figure}

At zero temperature the above expressions reduce to Eqs.~(\ref{eq:qubit-tfd-cl-time-asymmetry})-(\ref{eq:qubit-tfd-en-time-asymmetry}) so for $\psi=\pi$, $C_{yz}^{\rm TFD}(t)$ has time symmetry, and our system has a hidden-TRS.  However, as soon as $\bar{n}_{\rm th}$ is non-zero, hidden-TRS is broken.  For non-zero temperature, there is choice of $\psi$ for which the time-asymmetry of the correlation function.  To see this explicitly, suppose we set the time asymmetry of $C_{yz}^{\rm TFD}(t)$ to zero and attempt to solve for $\cos\psi$.  We obtain
\begin{equation}
    \cos\psi \overset{?}{=} -\sqrt{1 + \eta\frac{4b^{\prime 2}+1}{4b^{\prime 4}}}
\end{equation}
For any finite temperature $\eta>0$ so the right-hand side has a magnitude greater than 1, thus there is no solution for $\psi$.  This shows explicitly that for finite temperature and non-zero driving, the driven-dissipative qubit problem has no hidden-TRS.  This breaking of correlation function time symmetry is shown explicitly in Fig.~\ref{fig:qubit-thermal-breaking}.  At a heuristic level, this symmetry breaking is a result of the classical contribution growing faster with $\bar{n}_{\rm th}$ than the entanglement contribution, breaking the cancellation that occurs at $\bar{n}_{\rm th}=0$.

\subsection{Parametrically-driven nonlinear cavity at finite temperature}

 \begin{figure}[t]
     \centering
    \includegraphics[width=0.99\columnwidth]{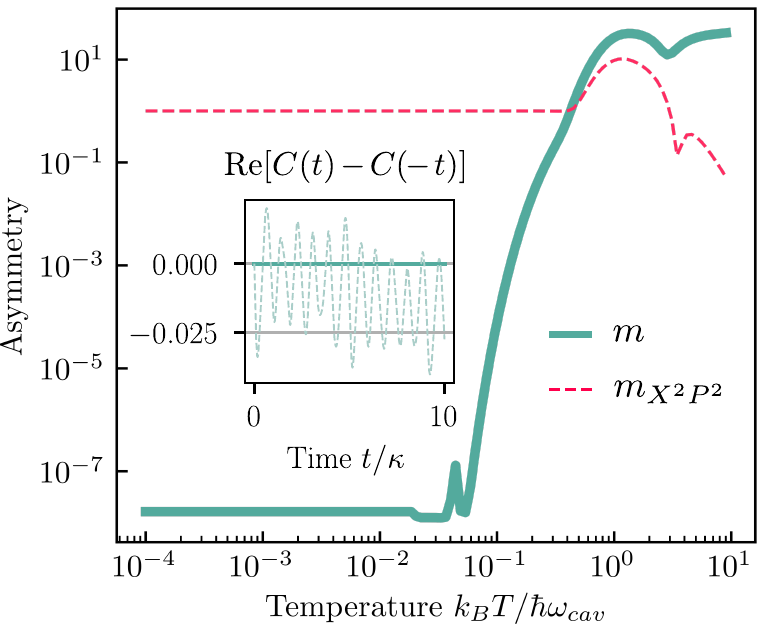}
     \caption{{\bf Hidden-TRS breaking in a driven nonlinear cavity.} {\bf Main plot}: total correlation function time-asymmetry 
     $m(T)$ vs. temperature for a parametrically-driven Kerr 
     resonator, c.f.~Eq.~(\ref{eq:mDefinition}).
     Solid green:  integrated asymmetry for the special correlation function $C_{a^2,H_\text{eff}}(t)$ (c.f.~Eq.~(\ref{eq:a2HeffCorrelator})), 
     which is guaranteed to be symmetric if hidden-TRS holds.  There is a sudden onset of asymmetry above a threshold temperature, indicating a sharp temperature at which hidden-TRS is broken.  In contrast, we also plot the total time asymmetry  of a correlation function whose behavior is not constrained by hidden-TRS, function $C_{X^2,P^2}(t)$ (dashed red curve); here $\hat{X}$ and $\hat{P}$ are standard quadrature operators.  
     This correlator is asymmetric already at zero temperature, and shows no strong temperature dependence.
     {\bf Inset}:  real-part of the correlation function asymmetry of 
     $C(t)\equiv C_{a^2,H_\text{eff}}(t)$ for $\bar{n}_\text{th} = 0$ (solid green), and $\bar{n}_\text{th} = 0.2$ (dashed green).  For all plots we take $\Lambda_2 = 3K$, $\kappa_1 = 0.01 K$, $\Delta = \Lambda_1 = \kappa_2 = 0$.}
\label{fig:asymmetry}
 \end{figure} 

The qubit example above corresponds to a system where the strength of the nonlinearity is essentially infinite.  We now consider a driven-dissipative system where the strength of the nonlinearity is tuneable:  the parametrically-driven damped nonlinear cavity of Sec.~\ref{sec:CavityHiddenTRS}, but now with thermal dissipation.
For simplicity, we consider where there is only a parametric (two-photon) drive, and there are only single-photon dissipation processes.  We then have a Lindblad master equation with a Hamiltonian given by Eq.~(\ref{eq:KerrHamiltonian}) (with $\Lambda_1 = \Delta = 0$), and dissipators
\begin{align}
    \hat{c}_1 &= \sqrt{\kappa(1+\bar{n}_{\rm th})}\hat{a}, \,\,\, \hat{c}_2 = \sqrt{\kappa\bar{n}_{\rm th}}\hat{a}^\dag.
\end{align}
As is standard in the derivation of quantum optics master equations, the thermal occupancy $\bar{n}_{\rm th}$ corresponds to a Bose-Einstein factor evaluated at the bath temperature $T$ and cavity resonance frequency $\omega_{\rm cav}$: 
$\bar{n}_{\rm th} = 1/\left(\exp[ \hbar \omega_{\rm cav} / k_B T] - 1 \right)$.
For simplicity, we ignore two-photon dissipative processes (i.e.~$\kappa_2 = 0$).

At non-zero temperature, hidden TRS is broken
unless one or both of the nonlinearity $K$ or drive $\Lambda_2$ are zero.  To see this, note that for hidden-TRS to hold,  Eq.~(\ref{eq:StateConstraint2}) requires that for some two-mode state $\ket{\psi_T}$ and $2 \times 2$ involutory unitary matrix $U$ the jump operators must satisfy
\begin{align}
    \bigg[\hat{c}_{1,A} - (U_{11}\hat{c}_{1,B} + U_{12}\hat{c}_{2,B})\bigg]|\psi_T\rangle&=0,\\
    \bigg[\hat{c}_{2,A} - (U_{21}\hat{c}_{1,B} + U_{22}\hat{c}_{2,B})\bigg]|\psi_T\rangle&=0.
\end{align}
The equations imply that $\ket{\psi_T}$ is annihilated by two independent canonical annihilation operators.  As such, this state must be Gaussian, which in turn implies that $\hat{\rho}_{\rm ss}$ must be Gaussian.  However, this steady state is Gaussian {\it only} if one or both of $\Lambda_2$ and $K$ are zero.  We thus have an important conclusion:  {\it the combination of thermal fluctuations, driving and nonlinearity together can break hidden-TRS.}  Note that for a more explicit proof that hidden-TRS does not hold, one can explicitly try to solve both Eqs.~(\ref{eq:StateConstraint1}) and (\ref{eq:StateConstraint2}); for both $\Lambda_2$ and $K$ non-zero, one can confirm that it is impossible to solve these equations.

It is interesting to consider the simple case of an undriven, linear thermal cavity (i.e.~$K=0$, $\Lambda_2 = 0$, $\bar{n}_{\rm th} > 0$).  In this case, the steady state is essentially classical (no Fock-state coherences), and it is well known that CQDB holds \cite{agarwal_open_1973}.  Formally, our system also has hidden-TRS, implying that this system can be solved using the CQA method.  This can be explicitly shown by solving Eqs.~(\ref{eq:StateConstraint1})-(\ref{eq:StateConstraint3}).  We find that solutions are possible when $E=0$ and when $U$ is chosen to have the form 
\begin{align}
    U = \begin{pmatrix}0&e^{i\theta}\\ e^{-i\theta}&0\end{pmatrix}.
\end{align}
Here, $\theta$ is a real parameter.  We thus have a continuous family of distinct hidden-TRS operators $\hat{T}_\theta$ (see Tab. \ref{tab:thermal-trs}), in contrast to the pair of hidden time-reversal operators $\hat{T}_\sigma$ seen for nonzero parametric driving and zero temperature (i.e. $K=0, \Lambda_1 = 0, \bar{n}_\text{th}=0$).

For each possible hidden-TRS operator $\hat{T}_{\theta}$, we have a corresponding thermofield double state.  These always have the form of Gaussian, two-mode squeezed states:
\begin{align}
    |\psi_T(\theta)\rangle \equiv \sum_{n=0}^\infty e^{-n(\beta /2 + i\theta)}|n,n\rangle.
\end{align}

Returning now to the more interesting case where we add both parametric driving and nonlinearity, we can study how thermal fluctuations break the hidden-TRS that is present at zero temperature.  We focus on an experimentally-accessible quantity that shows this symmetry breaking:  the time-asymmetry of the steady-state correlation function
\begin{align}
    C_{a^2,H_\text{eff}}(t)&\equiv \begin{cases}\langle \hat{a}^2(t) \hat{H}_\text{eff}(0)\rangle&t\geq 0 \\ \langle \hat{H}_\text{eff}(-t) \hat{a}^2(0)\rangle&t<0\end{cases},
    \label{eq:a2HeffCorrelator}
\end{align}
where as always, $\hat{H}_{\rm eff}$ is the non-Hermitian effective Hamiltonian associated with our master equation.  As discussed above, at zero temperature hidden-TRS guarantees that this special correlator has time-symmetry.  This time-symmetry is lost as $\bar{n}_{\rm th}$ is increased.  

To see this explicitly, we plot the total time-{\it asymmetry} vs. temperature, which we define as:
\begin{align}
    m(T) &\equiv \int_0^\infty  |C_{a^2,H_\text{eff}}(t) - C_{a^2,H_\text{eff}}(-t)| dt.
    \label{eq:mDefinition}
\end{align}
As shown in Fig.~\ref{fig:asymmetry}, the total asymmetry remains zero as long as the temperature $T$ is small, but then suddenly jumps at a critical ``transition'' temperature, consistent with a sudden breaking of detailed balance. 

This temperature at this transition can be understood heuristically as corresponding to having the thermal excitation rate $\kappa \bar{n}_{\rm th}$ be comparable to the dissipative gap of the zero-temperature system.  This dissipative gap $\Gamma$ (i.e.~slow relaxation rate) corresponds to switching between two coherent states $\ket{\pm \alpha}$ with $|\alpha|=\sqrt{(\Lambda_2^2-\kappa^2/4)/K^2}$ \cite{puri_engineering_2017}.  One finds that $\Gamma$ is exponentially small due to the small overlap of these coherent states \cite{mirrahimi_dynamically_2014}:   
\begin{equation}
    \Gamma = \kappa |\alpha|^2 e^{-2|\alpha|^2}
\end{equation}
Setting this rate equal to $\kappa \bar{n}_{\rm th}$ using the parameters in Fig.~\ref{fig:asymmetry} yields a temperature $k_B T / \hbar \omega_c \approx 0.2$; this is consistent with the temperature scale for hidden-TRS breaking.  

We stress that even at zero temperature, most system correlation functions do not exhibit any time-symmetry.  Such correlation functions do not show any dramatic behavior as a function of temperature.  As an example, we plot the asymmetry of the correlation function $C_{X^2,P^2}(t)$, defined as
\begin{align}
    m_{X^2P^2}(T) &\equiv \int_0^\infty  |C_{X^2,P^2}(t) - C_{X^2,P^2}(-t)| dt,\label{eq:m_XP_definition}
\end{align}
in Fig. \ref{fig:asymmetry}; here, $\hat{X}$, $\hat{P}$ are standard quadrature operators.

\section{Hidden TRS and phase-space methods: a quantum-classical correspondence}
\label{sec:complex_P}

\begin{table}[]
    \centering
    \begin{tabular}{|p{43mm}|p{43mm}|}\hline
         {\bf Hidden TRS}  & {\bf Classical TRS} \\ \hline
         $|\tilde{\psi}\rangle\equiv \hat{T}|\psi\rangle$ & $(\tilde{\alpha},\tilde{\beta}) \equiv (U\alpha,U\beta)$ \\ \hline
         $C^\text{TFD}_{X,Y}(t)\equiv\langle \hat{X}_A(t) \hat{Y}_B(0)\rangle$  & $C^{\mathcal P}_{X,Y}(t)\equiv\overline{X(t) \tilde{Y}(0)}$\\\hline
         $\mathcal J[\hat{a}]\equiv U\hat{a},$\newline $\mathcal J[\hat{H}_\text{eff}] \equiv \hat{H}_\text{eff}+E.$& Potential conditions \newline ($U$ must be unity)\\  \hline
    \end{tabular}
    \caption{{\bf Dictionary connecting hidden TRS and an effective classical notion of TRS in the effective phase space used in the complex-$P$ function method.}  We list objects/conditions commonly appearing in the solution of quantum master equations via quantum detailed balance (i.e.~hidden TRS), and their counterparts in the language of classical detailed balance in the complex-$P$ representation. This correspondence only exists for multimode bosonic systems coupled to local zero-temperature dissipation.}
    \label{tab:dictionary}
\end{table}

In this final section, we turn to driven-dissipative systems comprised of one or more bosonic modes, and connect our notion of hidden TRS to phase space methods that are well known in quantum optics, and have been used to solve non-trivial problems using an effective Fokker-Planck equation in an expanded phase space.  The focus is on Lindblad master equations of the form 
\begin{align}
    \partial_t\hat{\rho} = -i[\hat{H}, \hat{\rho}] + \sum_j \kappa_j \mathcal D[\hat{a}_j]\label{eq:local_qme}
\end{align}
Here, $[\hat{a}_j,\hat{a}_k^\dag] =\delta_{jk}$ is a standard set of independent bosonic modes, $\kappa_j$ represents a decay rate for each mode, and $\hat{H}$ is an arbitrary bosonic quantum many-body Hamiltonian. We will establish that for this restricted class of models, the fully quantum notion of hidden-TRS (described by an anti-unitary operator $\hat{T}$) coincides with a classical notion of time-reversal in an expanded phase space, i.e. an involution of the form $(x,p) \to (\tilde{x},\tilde{p})$ (where $x,p$ are classical phase space coordinates).  Hence, the effective detailed balance properties (and potential conditions) of a complex-$P$ Fokker-Planck equation is directly tied to hidden-TRS.  This allows us to directly understand the success of the complex-$P$ method in solving several non-trivial driven cavity problems \cite{drummond_generalised_1980}.  Interestingly, we show that for extended models, this correspondence no longer necessarily holds.  For example, by simply adding two-photon loss processes, there exist hidden-TRS operators that have no correspondence to a simple operation in an extended phase space.

The context of our discussion will be the complex-$P$ phase-space representation of the general bosonic master equation in Eq.~(\ref{eq:local_qme}). This is a particular non-diagonal expansion of the system's density matrix in terms of coherent states that can be used to convert the master equation into a well-behaved, Fokker-Planck-like equation (see \cite{MilburnWalls} for a pedagogical introduction).  Consider the single-mode case first for simplicity.  We consider a doubled phase space described by complex coordinates $(\alpha, \beta)$, and chose appropriate integration contours $\mathcal C,\mathcal C'$ for each of these variables.  This lets us express the density matrix as
\begin{align}
    \hat{\rho} = \int_{\mathcal C}d\alpha \int_{\mathcal C'}d\beta \frac{|\alpha\rangle \langle \beta^*|}{\langle \alpha|\beta^* \rangle}P(\alpha,\beta).\label{eq:def-of-p}
\end{align}
where $P(\alpha,\beta)$ is the complex-$P$ quasi-distribution function.  
Using standard techniques \cite{drummond_generalised_1980}, one can often convert the Lindblad master equation for $\hat{\rho}$ into a Fokker-Planck equation for this function, which is required to be nonsingular on the integration surface $\mathcal C\times \mathcal C'$ defined by the contours.  The resulting equation has the standard form:
\begin{align}
    \partial_t P(\alpha,\beta)&= \partial_\mu\bigg[ C^\mu(\alpha,\beta)P(\alpha,\beta)\bigg]\nonumber\\
    &~~~~~~~~~~~~~+ \partial_\mu\partial_\nu\bigg[ D^{\mu\nu}(\alpha,\beta)P(\alpha,\beta)\bigg]\label{eq:holomorphic_fpe}
\end{align}
Here, $C^\mu(\alpha,\beta)$ represents a generalized drift vector, and $D^{\mu\nu}(\alpha,\beta)$ a generalized diffusion tensor. The above derivatives are holomorphic derivatives \cite{drummond_generalised_1980}, and Einstein summation notation is implied.

We can now state our quantum-classical correspondence principle: {\it if the quantum master equation Eq. \eqref{eq:local_qme} has a hidden quantum time-reversal symmetry $\hat{T}$ and corresponds to a well-defined Fokker-Planck equation in the complex-$P$ representation, then this associated Fokker-Planck equation has a well-defined classical TRS corresponding to $\hat{T}$.} In the case where this classical TRS operation is trivial (i.e.~the identity operation), this symmetry then correspond to a standard detailed balance condition, meaning that the drift and diffusion matrices satisfy potential conditions \cite{gardiner_stochastic_2009}.  This directly enables an efficient solution for the steady-state distribution function $P(\alpha, \beta)$, and hence the steady-state density matrix.

The fact that the complex-$P$ Fokker-Planck equations satisfy potential conditions is precisely the property that enabled exact solutions of a variety of nonlinear driven cavity models \cite{drummond_quantum_1980}.  Our result shows that this surprising property is directly tied to a more general, and fully quantum symmetry:  hidden-TRS.  In what follows, we will describe precisely how to construct the classical time-reversal operator corresponding to a hidden TRS $\hat{T}$, and then show how this correspondence can be broken by considering higher-order Markovian loss channels.

\subsection{Detailed balance in generalized $P$-representations}

We start by defining a notion of time-reversal symmetry that is meaningful for the complex-$P$ distribution function $P(\alpha,\beta)$.  We stress the well-known fact that this distribution function is in general complex valued, and thus does not represent a meaningful probability distribution.  Nonetheless, we can formally use it to define quantities analogous to expectation values and correlation functions.

Note first that the expectation value of a holomorphic function $X(\alpha, \beta)$ defined on our complex phase space is defined as:
\begin{align}
    \overline{X}\equiv \int_{\mathcal C} d\alpha \int_{\mathcal C'} d\beta P_{ss}(\alpha,\beta) X(\alpha,\beta).
\end{align}
We can also define a time-evolved function $X(\alpha,\beta; t)$ defined by the solution of the dual Fokker-Planck equation
\begin{align}
    \partial_tX(\alpha,\beta) &= -\partial_\mu\bigg[ C^\mu(\alpha,\beta)X(\alpha,\beta)\bigg]\nonumber\\
    &~~~~~~~~~~~~~+ \partial_\mu\partial_\nu\bigg[ D^{\mu\nu}(\alpha,\beta)X(\alpha,\beta)\bigg].
\end{align}

With these definitions in hand, we define time-reversal symmetry in our doubled, complex classical phase space as the existence of a phase-space involution
\begin{align}
    (\alpha,\beta)\to \widetilde{(\alpha,\beta)},
\end{align}
such that all two-time correlation functions (calculated as defined above) are time-symmetric:
\begin{align}
    C^{\mathcal P}_{X,Y}(t)\equiv \begin{cases}\overline{X(t)Y(0)}&t\geq 0\\\overline{\tilde{Y}(-t)\tilde{X}(0)}&t<0 \end{cases},\label{eq:complex_p_cdb}
\end{align}
Here $X,Y$ are any two holomorphic functions, and the time-reversed functions are given as
\begin{align}
    \tilde{X}(\alpha,\beta)\equiv X(\widetilde{\alpha,\beta})
    \label{eq:clasical_trs_complexp}
\end{align}
where $\widetilde{(\alpha,\beta)}$ is the time-reversed counterpart to $(\alpha,\beta)$, i.e. another point in the integration surface on which the complex Fokker-Planck evolution is taking place. All we require is that this time-reversal operation squares to the identity, namely, time-reversing a point twice recovers the original point on $\mathcal C\times \mathcal C'$. 

For complex-$P$ distributions, one can establish a generalization of a standard result in classical probability theory, which we rigorously establish in Appendix \ref{app:pot_app}: 
in the limit where the time-reversal operation is just the identity,  the classical detailed balance condition Eq. \eqref{eq:complex_p_cdb} is equivalent to the potential conditions on the Fokker-Planck equation \cite{risken_fokker_1996,gardiner_stochastic_2009}.  Recall that these conditions correspond to the having the (pseudo)probability current vanish in the steady state at every point in phase space, where the {\it pseudoprobability current} $J^\mu (\alpha,\beta,t)$ is defined by rewriting the Fokker-Planck equation as a continuity equation:
\begin{align}
    \partial_t P(\alpha,\beta;t) \equiv \partial_\mu J^\mu(\alpha,\beta;t).
\end{align}
This constraint allows a direct method for solving for the steady state in terms of a potential function.  

Note that in the case where the time-reversal operation $ (\alpha,\beta)\to  \widetilde{(\alpha,\beta)}$ is {\it not} the identity, there is no simple potential-condition method for solving Fokker-Planck equations, unless the time-reversal symmetry is known beforehand (see e.g.~Ref.~\cite{risken_fokker_1996}.)

\subsection{Constructing the classical TRS corresponding to a hidden TRS $\hat{T}$}
We briefly outline the correspondence here in the single-mode case, and leave the discussion of the multimode case to App.~\ref{app:the_correspondence}. For these systems, hidden TRS implies (among other constraints) the constraint 
\begin{align}
    \mathcal J[\hat{a}]=u\hat{a}, \label{eq:the_key}
\end{align}
where $\mathcal{J}$ is the exchange superoperator as always, and $u=\pm 1$. In App.~\ref{app:the_correspondence} we show that, under the assumption that Eq.~\eqref{eq:the_key} holds, hidden TRS is equivalent to classical detailed balance in the complex-$P$ representation with respect to the following classical time-reversal operation:
\begin{align}
\widetilde{(\alpha,\beta)}= (u\alpha,u\beta)\label{eq:the_classical_trs}
\end{align}
Note that the fact that $u$ squares to one, an intrinsic property of the exchange superoperator, ensures that this is a valid classical time-reversal operation in the effective phase space used for the complex-$P$ representation.

This surprising correspondence is the result of TFD correlation functions of normally-ordered operators coinciding with complex-$P$ correlations.  More precisely, if  $\mathcal J[\hat{a}]=u  \hat{a}$ with $u=\pm 1$, then we have (see App.~\ref{app:the_correspondence}):
\begin{align}
    C_{X,\tilde{Y}}^{\mathcal P}(t)= C^\text{TFD}_{\hat{X},\hat{Y}}(t)\label{eq:the_goods}
\end{align}
where $X,Y$ are the classical representatives of $\hat{X},\hat{Y}$ in the complex-$P$ representation. Explicitly, without loss of generality, $\hat{X},\hat{Y}$ can be written as
\begin{align}
    \hat{X} &= \sum_{n,m}\chi_{n,m}(\hat{a}^\dag)^n \hat{a}^m,\\
    \hat{Y} &= \sum_{n,m}\lambda_{n,m}(\hat{a}^\dag)^n \hat{a}^m.
\end{align}
In terms of the normally-ordered expressions above, the classical representatives $X,Y$ have the following form:
\begin{align}
    X(\alpha,\beta) \equiv \sum_{n,m}\chi_{n,m}\beta^n \alpha^m,\\ 
    Y(\alpha,\beta) \equiv \sum_{n,m}\lambda_{n,m}\beta^n \alpha^m.
\end{align}
Finally, the classical time-reversal operation used to define the reversibility of the Fokker-Planck equation is given in Eq. \eqref{eq:the_classical_trs}. Therefore, in this situation, hidden (quantum) TRS is equivalent to classical TRS in the complex-$P$ representation.

\subsection{Breakdown of the correspondence principle: going beyond phase-space methods}

The simplest situations in which the above correspondence principle breaks down is in systems with higher-order loss dissipators, e.g. a system of the form
\begin{align}
    \frac{d}{dt}\hat{\rho} = -i[\hat{H},\hat{\rho}]+\kappa \mathcal D[\hat{a}^2]\hat{\rho}.\label{eq:two_ph_qme}
\end{align}
An example of this is the driven cavity problem considered in Sec. \ref{sec:CavityHiddenTRS}, in the regime where all single-photon terms are set to zero:  $\Lambda_1 = \kappa_1 = 0$.  We are left with a model with an interaction, detuning, two-photon drive and two-photon loss.  The full master equation in this case conserved photon number parity, and thus does not have a unique steady state. 

For this model, we still have hidden-TRS for each of the steady states.  The full set of hidden-TRS compatible TFD states, i.e.~obtained by solving Eqs. (\ref{eq:StateConstraint1})-(\ref{eq:StateConstraint3}), has the form:
\begin{align}
    |\psi_T(\gamma,\nu)\rangle \equiv \gamma |\psi_T^+\rangle + \nu |\psi_T^-\rangle.\label{eq:superposition}
\end{align}
where the individual terms $\ket{\psi_T^\pm}$ in the superposition correspond to the two simple thermofield doubled states encountered in Sec. \ref{sec:CavityHiddenTRS}, and which, upon tracing-out the auxiliary cavity $B$, correspond to a {\it single} steady state of the master equation Eq. \eqref{eq:two_ph_qme}:
\begin{align}
    \hat{\rho}_{\rm ss}^{(0)} &= \text{Tr}_B[|\psi^+_T\rangle \langle \psi_T^+|] = \text{Tr}_B[|\psi^-_T\rangle \langle \psi_T^-|].
\end{align}
According to the quantum-classical correspondence outlined in this section, this stationary state corresponds to a stationary complex-$P$ distribution with both a trivial TRS (corresponding to the potential conditions) and an inversion TRS (corresponding to something more complicated):
\begin{align}
    \hat{\rho}_{\rm ss}^{(0)}  &= \int_{\mathcal C} d\alpha \int_{\mathcal C'} d\beta \frac{|\alpha\rangle \langle \beta^*|}{\langle \alpha|\beta^*\rangle} P_{ss}(\alpha,\beta)\\
    &= \int_{\mathcal C} d\alpha \int_{\mathcal C'} d\beta \frac{|\alpha\rangle \langle \beta^*|}{\langle \alpha|\beta^*\rangle} P_{ss}(-\alpha,-\beta)
\end{align}
In light of the above observation, one might interpret the more exotic thermofield doubled state $|\psi(\gamma,\nu)\rangle$ as describing a quantum TRS which corresponds to a "superposition" of both the trivial and inversion TRS, and thus this hidden time-reversal symmetry no longer has a classical analogue.

Indeed, it is well-known that the stationary state obtained via solution of the potential conditions is not the {\it only} stationary state of the quantum master equation Eq. \eqref{eq:two_ph_qme} \cite{bartolo_exact_2016,wang_schrodinger_2016}. Tracing out the auxiliary cavity $B$ for arbitrary parameters $\gamma,\nu$ reveals a family of quantum steady states:
\begin{align}
    \hat{\rho}_{\rm ss}(\gamma,\nu) &\equiv  \text{Tr}_B[|\psi_T(\gamma,\nu)\rangle \langle \psi_T(\gamma,\nu)|].
\end{align}
While any $\hat{T}$ and TFD must yield an exchange superoperator  $\mathcal{J}$ that acts simply on $\hat{a}^2$ (via Eq. \eqref{eq:StateConstraint2}), the action on $\hat{a}$ need not be simple.  In fact, we only get a simple action when $\nu = 0$ ($\gamma=0$), in which case $\mathcal{J}[\hat{a}] = \hat{a}$ (  $\mathcal{J}[\hat{a}] = -\hat{a}$).  For the more general case, the identity Eq. \eqref{eq:the_key} is broken. The more complicated nature of the hidden TRS operator and the corresponding $\mathcal{J}$ implies that the steady states corresponding to $|\psi_T(\gamma,\nu)\rangle$ cannot be easily found using the complex-$P$ phase space solution method.

\section{Summary \& Outlook}
\label{sec:Conclusions}

In this work, we have introduced a new symmetry that can exist in driven-dissipative systems described by a Lindblad master equation: hidden time-reversal symmetry.  We have shown explicitly how this goes beyond the conventional definition of quantum detailed balance (CQDB) introduced by Agarwal \cite{agarwal_open_1973}; crucially, hidden-TRS can exist in systems whose steady states have non-zero energy-eigenstate coherences, something that makes it impossible to have CQDB.  
While hidden-TRS is most naturally formulated in terms of a doubled system prepared in a thermofield double state, we demonstrated that it has a direct observable consequence:  a certain class of single-system correlation functions are guaranteed to be time-symmetric.  This is in contrast to CQDB, which requires all correlation functions to obey a time-symmetry.  To illustrate our ideas, we have analyzed how several ubiquitous driven quantum systems (qubit and nonlinear cavity models) can have hidden-TRS despite not having CQDB.    

Perhaps most importantly, we established how hidden TRS provides a powerful means to derive analytic solutions for non-trivial steady states of quantum master equations.  In particular, hidden-TRS underlies both the coherent quantum absorber exact solution method \cite{stannigel_driven-dissipative_2012,roberts_driven_2019}, as well as phase space methods based on the complex-$P$ function \cite{drummond_generalised_1980}.  

We hope that our results will lay the groundwork for many further fruitful studies exploiting hidden-TRS as a means to understand even more complex systems.  This symmetry could provide an interesting means for finding non-trivial, exactly solvable many-body driven dissipative systems, both of bosons, qubits and possibly of fermions.  It could also lead to novel perturbative techniques for studying systems that weakly break hidden-TRS.    Finally, it would also be extremely interesting to rephrase this symmetry fully in terms of a dissipative field theory describing the system of interest (i.e.~in terms of a Keldysh action \cite{KamenevBook,Diehl2015,Diehl2016}).  This could yield further insights, and also perhaps enable an extension of these ideas into non-Markovian regimes.

\section*{Acknowledgements}

This work is supported by the Air Force Office of Scientific Research MURI program under Grant No.~FA9550-19-1-0399, and by the Army Research Office under Grant No.~W911NF-19-1-0380.  AC also acknowledges support from the Simons Foundation via a Simons Investigator Award.


\appendix

\section{Doubled-system classical detailed balance}
\label{app:DoubledClassicalDB}

In this section, we prove that classical detailed balance may be equivalently stated as the following symmetry condition:
\begin{align}
    \overline{X_A(t)Y_B(0)} = \overline{Y_A(t)X_B(0)},~~~~\forall X,Y.\label{eq:doubled}
\end{align}
The reason why the above condition is equivalent to the standard definition of detailed balance is that the doubled-system correlation function $\overline{X_A(t)Y_B(0)}$ is actually a single-system correlation function in disguise:
\begin{align}
    \overline{X_A(t)Y_B(0)}&= \sum_{n,m} \bar{p}(n)\delta_{m,\tilde{n}} X(t,n) Y(0,m)\\
   &=\sum_{n} \bar{p}(n) X(t,n) Y(0,\tilde{n})=\overline{X(t)\tilde{Y}(0)},
\end{align}
Therefore, the doubled-system correlation function $\overline{X_A(t)Y_B(0)}$ is time-symmetric for all random variables $X,Y$ if and only if
\begin{align}
    \overline{X(t)\tilde{Y}(0)}=\overline{Y(t) \tilde{X}(0)},~~~~\forall X,Y.\label{eq:cdb_twisted}
\end{align}
Now, making the replacement $Y\to \tilde{Y}$ in this single-site symmetry condition yields the definition of classical detailed balance used in the main text. Therefore, Eq. \eqref{eq:cdb_twisted} is equivalent to classical detailed balance:
\begin{align}
    \overline{X(t)Y(0)}=\overline{\tilde{Y}(t) \tilde{X}(0)},~~~~\forall X,Y.\label{eq:std_def}
\end{align}
Note that we have implicitly used the fact that $Y\to \tilde{Y}$ is a bijection of the algebra of random variables. In summary, we have shown that the doubled definition Eq. \eqref{eq:doubled} of classical detailed balance is completely equivalent to the standard definition Eq. \eqref{eq:std_def}.

\section{CQDB rules out stationary coherences between energy eigenstates}
\label{app:SQDB_is_diagonal}

We now demonstrate that systems with CQDB have steady state density matrices that are always guaranteed to be diagonal in the energy eigenbasis. There are many references that show this explicitly \cite{kossakowski_quantum_1977, bratteli_unbounded_1978, alicki_detailed_1976}. However, here we will assume an intermediate result, namely that CQDB implies {\it modular symmetry} \cite{bratteli_unbounded_1978}, that is, a symmetry of the driven-dissipative dynamics with respect to the unitary dynamics generated by the modular Hamiltonian
\begin{align}
    \hat{H}_{\rho}\equiv -\log \hat{\rho}_{\rm ss}.
\end{align}
The reason for taking this symmetry-based perspective is that it informs most of the central results in the theory of quantum detailed balance \cite{fagnola_generators_2007, fagnola_generators_2010}. 

Indeed, once the above symmetry is established, the proof that steady states with CQDB are diagonal in the energy eigenbasis is very easy. We provide here a simple argument that works in the finite-dimensional case. We begin with Lindblad's original expression for the effective Hamiltonian as a classical average \cite{lindblad1976}:
\begin{align}
    i\hat{H}_\text{eff} = \overline{\bar{\mathcal L}[U^\dag]\cdot U}
\end{align}
where here, $U$ is a Haar-random unitary, and $\bar{\mathcal L}$ is the Heisenberg-picture Lindbladian, which generates time-evolution of observables. We now observe how the effective Hamiltonian evolves under the modular (dynamical) group $\hat{O}(t)\equiv \hat{\rho}_{\rm ss}^{it} \hat{O}\hat{\rho}_{\rm ss}^{-it}$:
\begin{align}
    i\hat{H}_\text{eff}(t)= \overline{\hat{\rho}_{\rm ss}^{it}\bar{\mathcal L}[\hat{U}^\dag]\cdot \hat{U} \hat{\rho}_{\rm ss}^{-it}}.
\end{align}
The above identity holds for any Lindbladian, and thus contains no hidden assumptions about the system in question. Now suppose, however, that $\bar{\mathcal L}$ satisfies CQDB, and thus has modular symmetry. Then, in particular, we have
\begin{align}
    \hat{\rho}_{\rm ss}^{it}\bar{\mathcal L}[\hat{U}^\dag]=\bar{\mathcal L}[\hat{\rho}_{\rm ss}^{it}\hat{U}^\dag\hat{\rho}_{\rm ss}^{-it}]\hat{\rho}_{\rm ss}^{it}.
\end{align}
Substituting the above identity into the Haar-random average defining the effective non-hermitian Hamiltonian yields a simple result - $\hat{H}_\text{eff}$ is time-independent with respect to the unitary dynamics generated by the modular Hamiltonian:
\begin{align}
    i\hat{H}_\text{eff}(t) = \overline{\bar{\mathcal L}[\hat{U}^\dag(t)]\cdot \hat{U}(t)} = \overline{\bar{\mathcal L}[\hat{U}^\dag]\cdot \hat{U}} = i\hat{H}_\text{eff}(0).
\end{align}
Therefore, the effective Hamiltonian commutes with the modular Hamiltonian, and thus the effective Hamiltonian commutes with the steady state itself: $[\hat{H}_\text{eff},\hat{\rho}_{\rm ss}]=0$. By taking the anti-hermitian part of the commutator, we immediately have that $\hat{H}$ commutes with $\hat{\rho}_{\rm ss}$.

\section{Example of broken CQDB: dissipative Rabi-driven qubit}
\label{app:QubitCQDB}

 \begin{figure}[t]
     \centering
    \includegraphics[width=0.99\columnwidth]{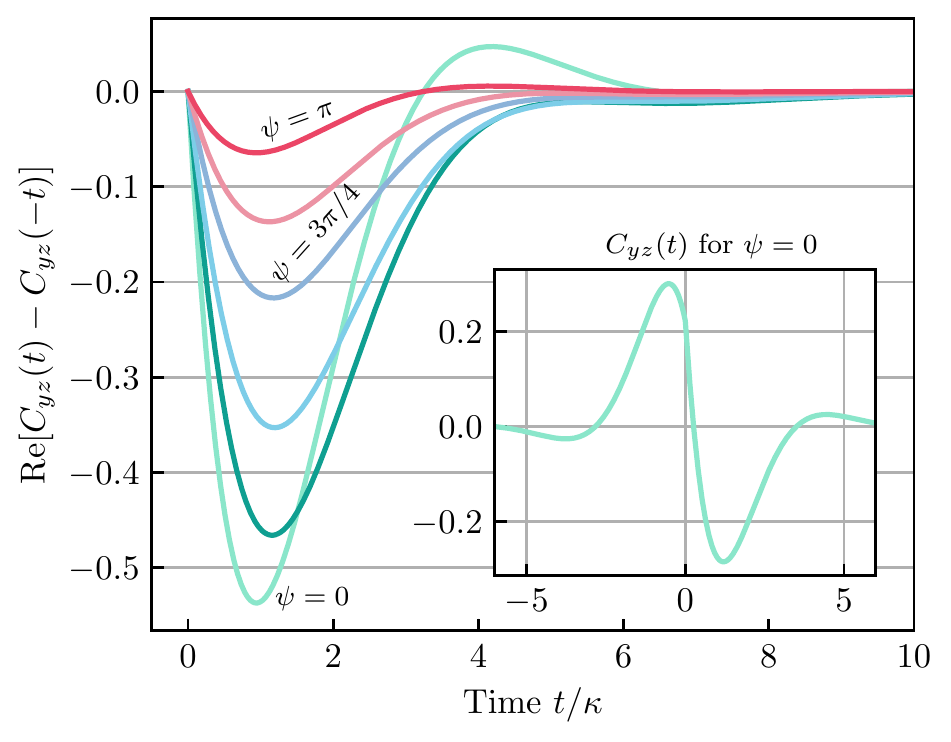}
     \caption{\textbf{The Rabi-driven qubit violates conventional quantum detailed balance.} \textbf{Inset:} The correlation function $C_{yz}(t)$ is shown as a function of time for TRS $\psi = 0$. Although time symmetry is not explicitly ruled out for $\psi=0$, the lack of time symmetry is clear. We show only the connected part which decays to zero for $|t|\gg \kappa$. \textbf{Main plot:} The time asymmetry of the symmetrized correlation function ${\rm Re}(C_{yz}(t) - C_{yz}(-t))$ as a function of time $t/\kappa$ for different permissible TRS. The correlation function is complex in general for $\psi \neq 0,\pi$; however, the time asymmetry is manifest even in the real part alone. The values of $\psi$ in order from top to bottom are: $\pi$, $5\pi/6$, $3\pi/4$, $5\pi/8$, $\pi/2$, and $0$. All functions are computed for Rabi drive strength $b=1$.} 
\label{fig:qubit-sqdb}
 \end{figure} 

To make the ideas of sections \ref{subsec:CQDB} and \ref{subsec:CQDBtrivial} more concrete, we consider simple but ubiquitous system which does not satisfy CQDB, does have hidden TRS.
\subsection{Violation of CQDB via correlation function asymmetry}
Our example is a qubit (with Pauli operators $\hat{\sigma}_{x,y,z}$) subject to a coherent Rabi drive in the presence of loss. Working in the rotating frame set by the drive, and making a rotating wave approximation, the master equation has the form of Eq.~(\ref{eq:qme}) with $M=1$ and
\begin{equation}
  \hat{H}= \Delta\hat{\sigma}_z + \frac{\Omega}{2}\hat{\sigma}_x, 
  \,\,\,
  \hat{c}_1 = \sqrt{\kappa} \hat{\sigma}_{-}.
  \label{eq:qubit-H-and-c_1_APPENDIX}
\end{equation}
Here $\Delta$ is the detuning of the drive from the qubit splitting frequency, $\Omega$ is the Rabi frequency, and $\kappa$ is the decay rate of the qubit excited state. The steady state for this system is easy to find and given in many textbooks, see e.g. Ref. \cite{gardiner_quantum_2000}: 
\begin{align}
    \hat{\rho}_{\rm ss}=\frac{\hat{1}}{2} &- \frac{4 \Delta \Omega}{16\Delta^2 + 2 \Omega^2 + \kappa^2}\hat{\sigma}_x + \frac{\Omega \kappa}{16\Delta^2 + 2 \Omega^2 + \kappa^2}\hat{\sigma}_y \nonumber\\
    &- \frac{16 \Delta^2 +\kappa^2}{2(16\Delta^2 + 2\Omega^2 + \kappa^2)}\hat{\sigma}_z
    \label{eq:qubit-driven-ss}
\end{align}
Given the external driving, this steady state does not correspond to thermal equilibrium, and hence a priori there is no reason to expect that it will satisfy CQDB.  While this may seem obvious, we will now show that CQDB is broken explicitly, by directly uncovering correlation function time-asymmetry in this system.  We stress that the CQDB condition in Eq.~(\ref{eq:qdb_wT}) is contingent on the choice of time-reversal operator $\hat{T}$.
We will take a general approach here (and throughout this paper):  we do not pre-select the definition of $\hat{T}$ using on additional knowledge of our system, but rather ask where there is {\it any} possible anti-unitary $\hat{T}$ which would give rise to a symmetry.  
Hence to truly rule out CQDB, one must check Eq.~(\ref{eq:qdb_wT}) for all permissible choices of $\hat{T}$. We will thus show that CQDB does not hold no matter what choice is made for $\hat{T}$.   

In what follows, for simplicity we assume a resonant drive (i.e.~$\Delta = 0$), and introduce the dimensionless Rabi frequency $b\equiv \Omega/\kappa$; CQDB is violated even for non-resonant drives, see Sec.~\ref{app-subsec:qubit-any-detuning} below. The first step is to parameterize all possible TRS operators.  Since CQDB only holds if $\hat{\rho}_{\rm ss}$ is itself invariant under TRS, this constrains the form of TRS.  The only permissible TRS operators are then parameterized by a single phase $\psi$ and have the form:
\begin{equation}
    \hat{T} = \left[\frac{\sin(\psi/2)}{\sqrt{4b^2+1}} \left(\hat{1} - 2ib \hat{\sigma}_x \right) + i\cos(\psi/2) \hat{\sigma}_z \right] \hat{K}_z
    \label{eq:qubit-general-TRS_APPENDIX}
\end{equation}
where $\hat{K}_z$ is the complex conjugation operator acting in the $\hat{\sigma}_z$ basis. 

To show that CQDB cannot hold in this system, it is sufficient to show that at least one correlation function violates Eq.~(\ref{eq:qdb_wT}) for each TRS angle $\psi$. As our main object of study, we introduce the correlation function defined for positive and negative times:
\begin{equation}
    C_{yz}(t) = 
    \begin{cases}
    \langle \hat{\sigma}_y(t) \hat{\sigma}_z(0) \rangle & t\geq 0 \\ 
    \langle \tilde{\sigma}_z(-t) \tilde{\sigma}_y(0) \rangle & t < 0
    \end{cases}
    \label{eq:StandardCorr}
\end{equation}
where $\tilde{\sigma}_j = \hat{T} \hat{\sigma}_j \hat{T}^{-1}$. CQDB holds, then Eq.~(\ref{eq:qdb_wT}) implies time symmetry: $C_{yz}(t) = C_{yz}(-t)$. In what follows we show that the properties of the TRS and the eigenmodes of the Liouvillian do not allow for the time symmetry of $C_{yz}(t)$ and other correlation functions.

The Liouvillian of the driven qubit system is readily diagonalized.  Letting $\lambda_j$ denote its eigenvalues and $\hat{r}_j$ the corresponding right-eigenvectors, we find:
\begin{align}
    \lambda_0 = 0 &\qquad \hat{r}_0 = \hat{\rho}_{\rm ss} \label{eq:qubit-mode-r0} \\
    \lambda_1 = -\frac{\kappa}{2} &\qquad \hat{r}_1 = \hat{\sigma}_x \label{eq:qubit-mode-r1} \\
    \lambda_2 = -\frac{\kappa}{4} \left( 3 + i \alpha \right)  &\qquad \hat{r}_2 = \hat{\sigma}_{y} + \frac{(1+i\alpha)}{4b}\hat{\sigma}_{z} \label{eq:qubit-mode-r2} \\
    \lambda_3 = -\frac{\kappa}{4} \left( 3 - i \alpha \right) &\qquad \hat{r}_3 = \frac{(1+i\alpha)}{4b}\hat{\sigma}_{y} + \hat{\sigma}_{z} \label{eq:qubit-mode-r3}
\end{align}
where $b\equiv \Omega/\kappa$, $\alpha = \sqrt{16b^2 - 1}$ is the dimensionless damped Rabi frequency. For $b < 1/4$ the qubit is overdamped and the frequency is imaginary: $\alpha \to i\sqrt{1 - 16b^2}$. The essential feature of the eigensystem is that the $\hat{\sigma}_x$ coherence behaves differently from the $\hat{\sigma}_y$ coherence or the classical population ($\hat{\sigma}_z$). The $\hat{\sigma}_x$ coherence decays exponentially with rate $(-\lambda_1) = \kappa/2$. The $\hat{\sigma}_y$ coherence and the classical populations decay with $(-{\rm Re}\,\lambda_{2,3})>\kappa/2$ for any finite $b$. This implies that $\langle  \hat{\sigma}_x (t) \hat{\sigma}_k (0)\rangle$ has a time dependence that is always different from $\langle \hat{\sigma}_y (t) \hat{\sigma}_k (0)\rangle$ or $\langle \hat{\sigma}_z (t) \hat{\sigma}_k (0)\rangle$ for any $\hat{\sigma}_k$.

With this in mind, we turn to the time-reversed operator $\tilde{\sigma}_z$ which, in general is a linear combination of all three Pauli operators. In particular, the $\hat{\sigma}_x$ component is
\begin{equation}
    \tilde{\sigma}_z = -(2b/\sqrt{4b^2+1})\sin\psi \hat{\sigma}_x +\cdots
\end{equation}
We can see that for any $\psi \neq 0,\pi$, for which $\sin\psi \neq 0$, the expression $C_{yz}(t<0)$ will have terms measuring the decay of $\hat{\sigma}_x$ coherence. Thus the time dependence at $t<0$ \emph{must be} qualitatively different from that at $t>0$. Even at $\psi=0,\pi$ the correlation function is generically \emph{not} time symmetric. As an example, we show a generic plot of the time asymmetry of $C_{yz}(t)$ in Fig.~\ref{fig:qubit-sqdb} for $b=1$ computed for various $\psi$. Although the argument breaks down at $\psi=0,\pi$, we can show definitively that CQDB cannot hold by considering $C_{xz}(t)$, which is defined analogously with $C_{yz}(t)$. Since at $\psi=0,\pi$ there are no $\hat{\sigma}_x$ terms in $\tilde{\sigma}_z$, the time dependence of $C_{xz}(t<0)$ must be qualitatively different from the time dependence of $C_{xz}(t>0)$. Therefore we conclude that the Rabi-driven qubit does not satisfy CQDB.

\subsection{Violation of CQDB for any detuning}
\label{app-subsec:qubit-any-detuning}

In the preceding section we restricted our analysis to the resonantly driven qubit for which $\Delta=0$ primarily because the diagonalization of the Liouvillian becomes unwieldy for $\Delta\neq0$. Here we show by an alternate route that for any detuning, the Rabi driven system violates CQDB.

Recall from Appendix \ref{app:SQDB_is_diagonal} that a system which satisfies CQDB must have a steady state that is diagonal in the energy eigenbasis. The commutator $[\hat{H},\hat{\rho}_{\rm ss}]$ is equivalent to taking the cross product between the Hamiltonian ``vector'' $(\Omega/2,0,\Delta)$ and the traceless part of $\hat{\rho}_{\rm ss}$. Imposing the constraint that the commutator is zero requires the form of $\hat{\rho}_{\rm ss}$ to be
\begin{equation}
\hat{\rho}_{\rm ss}=\frac{1}{2}\left(\hat{1}+\alpha\hat{H}\right)
\end{equation}
where $\alpha$ is a real constant of proportionality. Crucially, the above expression is linear in $\Delta$ and $\Omega$, whereas the true steady state is a quadradic rational function of these parameters \emph{and} the decay rate $\kappa$, c.f. Eq.~(\ref{eq:qubit-driven-ss}). Therefore we conclude that for any detuning, the Rabi-driven qubit does not satisfy CQDB. Furthermore, one can numerically diagonalize the Liouvillian at any drive detuning and verify that there does not exist any TRS for which the steady state is invariant and for which all correlation functions are time symmetric.

\subsection{Permissible TRS}
\label{app-subsec:qubit-permissible-TRS}

Here we show how the permissible TRS of Eq.~\eqref{eq:qubit-general-TRS_APPENDIX} are determined from the steady state Eq.~\eqref{eq:qubit-driven-ss}. We require only that $\hat{T}$ leaves the steady state invariant as a necessary condition of CQDB. All possible TRS take the form $\hat{T}=\hat{V}\hat{K}_{\rho}$ for unitary $\hat{V}$ and complex conjugation $\hat{K}_{\rho}$ in the steady state eigenbasis such that $\hat{T}\hat{\rho}_{\rm ss}\hat{T}^{-1}=\hat{\rho}_{\rm ss}$. This condition implies that the action of $\hat{T}$ on the eigenstates of $\hat{\rho}_{\rm ss}$ (i.e. pointer states) is restricted to be
\begin{align}
\hat{T}|1\rangle= & |1\rangle\\
\hat{T}|2\rangle= & e^{i\psi}|2\rangle
\end{align}
up to a global phase. Thus in the steady state eigenbasis the permissible TRS take the form $\hat{T}=(|1\rangle\langle1|+e^{i\psi}|2\rangle\langle2|)\hat{K}_{\rho}$ for any $\psi$, where $\hat{K}_{\rho}$ is complex conjugation in the steady state eigenbasis and the pointer states are assumed time-reversal invariant: $\hat{K}_{\rho}|n\rangle=|n\rangle$.

As the final step, we represent $\hat{T}$ in the familiar $\hat{\sigma}_{z}$ basis. Given the change of basis unitary $\hat{U}$ that diagonalizes the steady state, $\hat{U}\hat{\rho}_{\rm ss}\hat{U}^{\dagger}=\mathrm{diag}(p_{1},p_{2})$, the permissible TRS are given in the $\hat{\sigma}_{z}$ basis as
\begin{equation}
\hat{T}=\hat{U}^{\dagger}\hat{V}\hat{K}_{\rho}\hat{U}=\hat{U}^{\dagger}\hat{V}\hat{U}^{*}\hat{K}_{z}.
\end{equation}
For completeness $\hat{U}$ is given in terms of $b\equiv\Omega/\kappa$
and $s\equiv\sqrt{4b^{2}+1}$ as
\begin{equation}
\hat{U}=\frac{1}{\sqrt{2s}}\left(\begin{array}{cc}
i\sqrt{s-1} & \sqrt{s+1}\\
i\sqrt{s+1} & -\sqrt{s-1}
\end{array}\right).
\end{equation}
Substituting this into the expression for $\hat{T}$ above recovers Eq.~\eqref{eq:qubit-general-TRS_APPENDIX} (which is also Eq.~(\ref{eq:qubit-general-TRS}) in the main text).

\section{Explicit construction of exchange superoperator $\mathcal J$}
\label{app:FormOfJ}

By definition, the exchange superoperator is supposed to act on a single site observable $\hat{O}$ and produce a new single-site observable, $\mathcal J[\hat{O}]$, which, upon acting on site $B$ of the thermofield double, produces the same state as one would obtain by acting on site $A$ with the observable $\hat{O}$:
\begin{align}
    \mathcal J[\hat{O}]_B|\psi_T\rangle \equiv \hat{O}_A|\psi_T\rangle\label{eq:def-of-J}
\end{align}
One may interpret the above equation as the vectorization of an operator equation, according to the rules \begin{align}
    \hat{O}_A|\psi_K\rangle &\to \hat{O}\hat{\rho}^{1/2}_{ss} ,\\   \hat{O}_B|\psi_K\rangle &\to \hat{\rho}^{1/2}_{ss} \tilde{O}^T,
\end{align}
where $|\psi_K\rangle$ is the thermofield doubled state where the time-reversal operation $\hat{T}\equiv \hat{K}$ is complex-conjugation in the eigenbasis of the steady state. Explicitly:
\begin{align}
    |\psi_K\rangle = \sum_{n}\sqrt{p_n}|n\rangle|n\rangle.
\end{align}
Furthermore, $\hat{O}\to \hat{O}^T$ denotes matrix transposition in the eigenbasis of the steady state.

An arbitrary time-reversal operation can be decomposed as $\hat{T}\equiv \hat{V}\hat{K}$, where $\hat{V}$ commutes with $\hat{\rho}_{\rm ss}$. Explicitly, an arbitrary thermofield doubled state corresponding to a hidden TRS must always have the expression
\begin{align}
    |\psi_T\rangle &=\sum_{n}\sqrt{p_n}|n\rangle  e^{i\theta_n}|n\rangle,\\ &\equiv \sum_{n}\sqrt{p_n}|n\rangle \hat{V}\hat{K}|n\rangle.
\end{align}
Under the above rules, the thermofield doubled state with an arbitrary time-reversal operation  for a unitary $\hat{V}$ satisfies the following identities:
\begin{align}
    \hat{O}_A|\psi_T\rangle& = \hat{O}_A\hat{V}_B|\psi_K\rangle \to \hat{O}\hat{\rho}_{\rm ss}^{1/2}\hat{V}^T\label{eq:op_identity1}\\
    \hat{O}_B|\psi_T\rangle& = (\hat{O}\hat{V})_B|\psi_K\rangle \to \hat{\rho}_{\rm ss}^{1/2}(\hat{O}\hat{V})^T\label{eq:op_identity2}
\end{align}
We now verify the formula $\mathcal J[\hat{O}]= \hat{\rho}_{\rm ss}^{1/2} \tilde{O}^\dag\hat{\rho}_{\rm ss}^{-1/2}$ used in the main text. First of all, any hidden TRS must leave the steady state invariant, and we can rewrite this formula as
\begin{align}
\mathcal J[\hat{O}]\equiv \hat{\rho}_{\rm ss}^{1/2} \hat{O}^T\hat{\rho}_{\rm ss}^{-1/2}
\end{align}
We can then proceed to verify Eq. \eqref{eq:def-of-J} directly:
\begin{align}
    (\hat{\rho}_{\rm ss}^{1/2} \hat{O}\hat{\rho}_{\rm ss}^{-1/2})_B|\psi_T\rangle&\to  \hat{\rho}^{1/2}_{ss}\hat{V}^T(\hat{\rho}_{\rm ss}^{1/2} \hat{O}^T\hat{\rho}_{\rm ss}^{-1/2})^T.
\end{align}
Taking advantage of the fact that $\hat{V}$ commutes with $\hat{\rho}_{\rm ss}$, we have
\begin{align}
    \hat{\rho}^{1/2}_{ss}\hat{V}^T(\hat{\rho}_{\rm ss}^{1/2} \hat{O}^T\hat{\rho}_{\rm ss}^{-1/2})^T = \hat{O}\hat{V}^T\hat{\rho}_{\rm ss}^{1/2}= \hat{O}\hat{\rho}_{\rm ss}^{1/2}\hat{V}^T,
\end{align}
which is just the operator representation of $\hat{O}_A|\psi_T\rangle$ according to the rules Eqs. (\ref{eq:op_identity1}-\ref{eq:op_identity2}).

\section{From CQDB to hidden TRS}
\label{app:SQDBtoGeneral}

In this appendix, we will show that, for systems with modular symmetry, CQDB and hidden TRS are equivalent. Since CQDB already implies modular symmetry by itself \cite{kossakowski_quantum_1977, bratteli_unbounded_1978}, this will demonstrate that CQDB necessarily implies hidden TRS. Therefore, CQDB is a strict subphenomenon of hidden TRS.

Indeed, consider an arbitrary driven-dissipative system described by a Lindblad master equation. The definition of the exchange superoperator tells us that
\begin{align}
    C_{X,\mathcal J[Y]}(t)&=   C_{X,Y}^\text{TFD}(t),~~~~\forall t\geq 0.
\end{align}
The above correlation function, for negative times, is harder to rephrase as a two-site quantity:
\begin{align}
    C_{X,\mathcal J[Y]}(-t)&=\langle(\hat{\rho}^{-1/2}_{ss}\hat{Y}\hat{\rho}^{1/2}_{ss})(t)\tilde{X}^\dag \rangle. \label{eq:trying}
\end{align}
However, the above expression simplifies considerably in systems with modular symmetry. Indeed, let us now assume that the driven-dissipative system in question is symmetric with respect to the modular Hamiltonian $\hat{H}_\rho$, as is always the case with systems that have CQDB \cite{kossakowski_quantum_1977}. Then, we can write $(\hat{\rho}^{-1/2}_{ss}\hat{Y}\hat{\rho}^{1/2}_{ss})(t) = \hat{\rho}^{-1/2}_{ss}\hat{Y}(t)\hat{\rho}^{1/2}_{ss}$. Substituting this identity into Eq. \eqref{eq:trying}, we get
\begin{align}
    C_{X,\mathcal J[Y]}(-t)&=C_{X,Y}^\text{TFD}(-t),~~~~\forall t\geq 0.
\end{align}
Therefore, for any driven-dissipative system with modular symmetry, 
\begin{align}
    C_{X,\mathcal J[Y]}(t)&=   C_{X,Y}^\text{TFD}(t)~~~~\forall t\in \mathbb{R}.
\end{align}
Therefore, since $\mathcal J$ is a bijection of the observable algebra, CQDB and hidden TRS are equivalent for this class of systems, as time-symmetry of one set of correlation functions implies time-symmetry of the other.

\section{Complex-$P$ and hidden TRS correspondence theorem}
\label{app:the_correspondence}
We now consider a general many-body bosonic Lindblad master equation, with, say, $n$ bosonic modes. Recall that classical detailed balance for this class of master equations can be formulated in the complex-$P$ representation as the time-symmetry of the correlation function
\begin{align}
    C^{\mathcal P}_{X,Y}(t)\equiv \begin{cases}\overline{X(t)Y(0)}&t\geq 0\\ \overline{\tilde{Y}(-t) \tilde{X}(0)}&t< 0\end{cases}\label{eq:complex_p_cdb_app}
\end{align}
where, here, $X(\vec{\alpha},\vec{\beta}),Y(\vec{\alpha},\vec{\beta})$ are arbitrary multivariate holomorphic functions, and averages are understood to be taken with respect to the stationary complex-$P$ distribution, e.g.
\begin{align}
    \overline{X}\equiv \int_{\Sigma} d^n\alpha d^n\beta\, P_{ss}(\vec{\alpha},\vec{\beta}) X(\vec{\alpha},\vec{\beta}),
\end{align}
where $\Sigma$ is a $2n$-dimensional closed integration surface (the many-body analogue of the pair of integration contours $\mathcal C,\mathcal C'$ mentioned in the main text). Furthermore, $X(\vec{\alpha},\vec{\beta})\to \tilde{X}(\vec{\alpha},\vec{\beta})$ is a well-defined classical time-reversal operation, that is, induced by a bijection of the phase-space squaring to one (c.f. Eq.~\eqref{eq:clasical_trs_complexp} in the main text).

Recall that hidden TRS for a quantum system is defined as the time-symmetry of the thermofield doubled correlation function, which, e.g.~for positive times, is given by a symmetric bilinear form:
\begin{align}
    C_{\hat{X},\hat{Y}}^\text{TFD}(t) \equiv \langle \!\langle \hat{X}(t),\hat{Y}\rangle\!\rangle_{\mathcal T},\label{eq:pos_T}
\end{align}
and for negative times (with a time-reversal invariant steady state), is expressible using the same bilinear form:
\begin{align}
    C_{\hat{X},\hat{Y}}^\text{TFD}(-t) \equiv \langle \!\langle \hat{X},\hat{Y}(t)\rangle\!\rangle_{\mathcal T},\label{eq:neg_T}
\end{align}
where the relevant bilinear form (which depends on a particular choice of time-reversal operator $\hat{T}$) is defined in the main text as follows:
\begin{align}
    \boxed{\langle \!\langle \hat{X},\hat{Y}\rangle\!\rangle_{\mathcal T}\equiv C^\text{TFD}_{\hat{X},\hat{Y}}(t\equiv 0)}
\end{align}
Clearly, the condition of hidden TRS is equivalent to identifying the left-hand side of Eq. \eqref{eq:pos_T} with the left-hand side of Eq. \eqref{eq:neg_T}, which, upon examining the right-hand sides of said equations, is equivalent to the symmetry of the Heisenberg-picture time-evolution superoperator  $\mathcal E_t \equiv e^{-t\bar{\mathcal L}}$ with respect to the bilinear form $\langle\!\langle \cdot,\cdot\rangle\!\rangle_{\mathcal T}$.

One can form an analogous construction to express classical correlations in the complex-$P$ distribution in terms of a bilinear form. Indeed, for any pair of normally-ordered quantum observables $\hat{X},\hat{Y}$ and a classical time-reversal operation $X\to \tilde{X}$, one can write
\begin{align}
    C_{X,\tilde{Y}}^{\mathcal P}(t)=\langle \!\langle \hat{X}(t),\hat{Y}\rangle\!\rangle_{\mathcal P},\label{eq:pos_TP}
\end{align}
and for negative times (with a stationary complex-$P$ distribution which is time-reversal invariant), we can also write:
\begin{align}
     C_{X,\tilde{Y}}^{\mathcal P}(-t)=\langle \!\langle \hat{X},\hat{Y}(t)\rangle\!\rangle_{\mathcal P},\label{eq:neg_TP}
\end{align}
Where we have introduced a new symmetric bilinear form
\begin{align}
    \boxed{\langle \!\langle \hat{X}, \hat{Y}\rangle\!\rangle_{\mathcal P}\equiv C_{X,\tilde{Y}}^{\mathcal P}(t\equiv 0)}
\end{align}
where $X,Y$ are the normally-ordered symbols of $\hat{X},\hat{Y}$, i.e. the classical representatives of the observables $\hat{X},\hat{Y}$ in the complex-$P$ representation. Explicitly, if $\hat{X}=\sum_{IJ} c_{IJ}(\hat{a}_1^\dag)^{i_1}\cdots(\hat{a}_n^\dag)^{i_n} \hat{a}_1^{j_1}\cdots \hat{a}_n^{j_n}$, then the normally-ordered symbol $X$ is
\begin{align}
    X = \sum_{IJ} c_{IJ} \beta^{i_1}\cdots \beta^{i_n}\alpha_1^{j_1}\cdots \alpha_n^{j_n},~~~~\text{etc..}
\end{align}
where $I\equiv \{i_1,\cdots ,i_n\}$ and $J\equiv \{j_1,\cdots, j_n\}$ are multi-indices. The condition of classical detailed balance in the complex-$P$ representation, i.e. the time-symmetry of Eq.~\eqref{eq:complex_p_cdb_app}, is equivalent to identifying the left-hand side of Eq.~\eqref{eq:pos_TP} with the left-hand side of Eq. \eqref{eq:neg_TP}, which is equivalent to the symmetry of the Heisenberg-picture time-evolution superoperator $\mathcal E_t \equiv e^{-t\bar{\mathcal L}}$ with respect to the bilinear form $\langle\!\langle \cdot,\cdot\rangle\!\rangle_{\mathcal P}$.

With these definitions in hand, we return to the general problem of interest:  a Markovian multi-mode bosonic system where each mode is subject to loss.  For simplicity, we start by assuming each mode has the same loss rate, implying a master equation of the form
\begin{align}
    \frac{d}{dt}\hat{\rho} = -i[\hat{H},\hat{\rho}]+ \sum_{j=1}^N \kappa\mathcal D[\hat{a}_j].\label{eq:general_manybody_qme}
\end{align}
If this system has hidden TRS with respect to a particular quantum time-reversal operation $\hat{T}$, then as discussed in the main text (c.f.~Eq.~(\ref{eq:qdb_jumps})) there must exist a $N \times N$ unitary matrix $U$ such that:
\begin{align}
    \mathcal J[\hat{a}_j] = \sum_{kj}U_{kj}\hat{a}_k,~~~~U^2\equiv 1. \label{eq:fagnola_manybody}
\end{align}
Here (as always) $\mathcal{J}$ is the exchange superoperator.

What we will show in this section is a remarkable coincidence between the two families of bilinear forms, under the above jump operator constraint. That is, Eq. \eqref{eq:fagnola_manybody} implies that we can identify both bilinear forms, i.e.
\begin{align}
     \boxed{\langle \!\langle \hat{X}, \hat{Y}\rangle\!\rangle_{\mathcal P}\equiv \langle \!\langle \hat{X}, \hat{Y}\rangle\!\rangle_{\mathcal T}~~~~\forall \hat{X},\hat{Y}}
\end{align}
where the classical time-reversal operation on the complex-$P$ side is none other than the change-of-Kraus representation given in Eq.~\eqref{eq:fagnola_manybody}: $\widetilde{(\vec{\alpha},\vec{\beta})} \equiv (U \vec{\alpha}, U\vec{\beta})$. In what follows, we first prove this result in the single mode case, then extend to the multi-mode case where each mode has an identical damping rate $\kappa$.  Finally, we extend the result to the more general case where each mode has a different damping rate $\kappa_j$.  

\subsubsection{Single-mode case}

Consider Eq.~(\ref{eq:general_manybody_qme}) in the single mode limit $N=1$.  
If this system has hidden TRS with respect to a particular quantum time-reversal operation $\hat{T}$, then the jump operator constraint reduces to $\mathcal J[\hat{a}] = u\hat{a}$, with $u=\pm 1$ a scalar quantity. We show in what follows that this in turn implies the following identity:
\begin{align}
    \langle \!\langle \hat{X}, \hat{Y}\rangle\!\rangle_{\mathcal P}\equiv \langle \!\langle \hat{X}, \hat{Y}\rangle\!\rangle_{\mathcal T}, \label{eq:what_we_wanted_to_prove}
\end{align}
where the bilinear form on the right-hand side is defined using the classical time-reversal operation $\widetilde{(\alpha,\beta)} = (u\alpha,u\beta)$. To see this, note that both the left- and right-hand sides are bilinear with respect to $\hat{X},\hat{Y}$, and so it suffices to verify the above identity in a basis of normally-ordered monomials. That is, without loss of generality, we may assume that
\begin{align}
    \hat{X} &= (\hat{a}^\dag)^k\hat{a}^l,~~~~~\hat{Y} = (\hat{a}^\dag)^p\hat{a}^q.
\end{align}
We begin by computing, e.g.
\begin{align}
    \langle \!\langle \hat{X},\hat{Y}\rangle\!\rangle_{\mathcal T}&=\langle \psi_T|(\hat{a}^\dag)^k\hat{a}^l (\hat{b}^\dag)^p \hat{b}^q|\psi_T\rangle. 
\end{align}
Substituting-in the definition of the exchange superoperator, we get
\begin{align}
    \langle \!\langle \hat{X},\hat{Y}\rangle\!\rangle_{\mathcal T}&=\langle(\hat{a}^\dag)^k\hat{a}^l \mathcal J[(\hat{a}^\dag)^p \hat{a}^q]\rangle,
\end{align}
where, here, $\langle\hat{O}\rangle\equiv \text{Tr}[\hat{\rho}_{\rm ss}\hat{O}]$ denotes a steady-state expectation value. Now, we utilize the fact that $\mathcal J[\hat{O}\hat{O}']=\mathcal J[\hat{O}']\mathcal J[\hat{O}]$ for generic $\hat{O},\hat{O}'$:
\begin{align}
    \langle \!\langle \hat{X},\hat{Y}\rangle\!\rangle_{\mathcal T}&=\langle (\hat{a}^\dag)^k\hat{a}^l (u\hat{a})^q\mathcal J[(\hat{a}^\dag)^p]\rangle.
\end{align}
Finally, by direct computation, we also have that $\mathcal J[\hat{O}^\dag] = \hat{\rho}_{\rm ss}\mathcal J[\hat{O}]^\dag \hat{\rho}_{\rm ss}^{-1}$ for generic $\hat{O}$, and so (via the cyclic nature of the trace)
\begin{align}
    \langle \!\langle \hat{X},\hat{Y}\rangle\!\rangle_{\mathcal T} &= \langle (u\hat{a}^\dag)^p (\hat{a}^\dag)^k  \hat{a}^l (u\hat{a})^q \rangle\nonumber\\
    &=\int_{\mathcal C} d\alpha \int_{\mathcal C'} d\beta\, P_{ss}(\alpha,\beta) \tilde{\beta}^p \beta^k \alpha^l \tilde{\alpha}^q
\end{align}
Immediately, we recognize here the normally-ordered symbols of $\hat{X},\hat{Y}$, which are, explicitly:
\begin{align}
    \tilde{Y}(\alpha,\beta) = \tilde{\beta}^p\tilde{\alpha}^q,~~~X(\alpha,\beta) = \beta^k\alpha^l.
\end{align}
With the above observation, we have thus proved Eq. \eqref{eq:what_we_wanted_to_prove}, which establishes the equivalence of {\it generalized} quantum detailed balance, that is, hidden TRS, with classical detailed balance in the complex-$P$ representation. Note that the potential conditions, as well as the original CQA method, as originally formulated in \cite{stannigel_driven-dissipative_2012}, both correspond in this context to the special case of a trivial TRS, i.e. $U=1$.

\subsubsection{Many-body case}

Now, consider the multi-mode master equation Eq. \eqref{eq:general_manybody_qme}. If this system has hidden TRS with respect to a particular quantum time-reversal operation $\hat{T}$, then we have the constraint Eq. \eqref{eq:fagnola_manybody} on the jump operators, which we write as
\begin{align}
    \mathcal J[\hat{a}_j] = \hat{\alpha}_j,~~~\hat{\alpha}_j \equiv \sum_{jk}U_{jk}\hat{a}_k 
\end{align}
What we will demonstrate in this section is that this jump operator constraint implies the following identity:
\begin{align}
    \langle \!\langle \hat{X}, \hat{Y}\rangle\!\rangle_{\mathcal P}\equiv \langle \!\langle \hat{X}, \hat{Y}\rangle\!\rangle_{\mathcal T}, \label{eq:what_we_wanted_to_prove_many_body}
\end{align}
where the bilinear form on the right-hand side is defined using the classical time-reversal operation $\widetilde{(\vec{\alpha},\vec{\beta})} = (U\vec{\alpha},U\vec{\beta})$. To see this, note that both the left- and right-hand sides are bilinear with respect to $\hat{X},\hat{Y}$, and so it suffices to verify the above identity in a basis of normally-ordered monomials. That is, without loss of generality, we may assume that
\begin{align}
    \hat{X} = (\hat{a}_1^\dag)^{k_1}\cdots (\hat{a}_n^\dag)^{k_n}\hat{a}_1^{l_1}\cdots \hat{a}_n^{l_n},\\
    \hat{Y} =  (\hat{a}_1^\dag)^{p_1}\cdots (\hat{a}_n^\dag)^{p_n}\hat{a}_1^{q_1}\cdots \hat{a}_n^{q_n}.
\end{align}
We begin by computing, e.g.
\begin{align}
    \langle \!\langle \hat{X},\hat{Y}\rangle\!\rangle_{\mathcal T}&=\langle \psi_T|(\hat{a}_1^\dag)^{k_1}\cdots (\hat{a}_n^\dag)^{k_n}\hat{a}_1^{l_1}\cdots \hat{a}_n^{l_n}\\
    &~~~\cdot(\hat{b}_1^\dag)^{p_1}\cdots (\hat{b}_n^\dag)^{p_n}\hat{b}_1^{q_1}\cdots \hat{b}_n^{q_n}|\psi_T\rangle. 
\end{align}
Substituting-in the definition of the exchange superoperator, we get
\begin{align}
    \langle \!\langle \hat{X},\hat{Y}\rangle\!\rangle_{\mathcal T}&=\langle (\hat{a}_1^\dag)^{k_1}\cdots (\hat{a}_n^\dag)^{k_n}\\
    &~~~\cdot \mathcal J[(\hat{a}_1^\dag)^{p_1}\cdots (\hat{a}_n^\dag)^{p_n}\hat{a}_1^{q_1}\cdots \hat{a}_n^{q_n}]\rangle,
\end{align}
where, here, $\langle\hat{O}\rangle\equiv \text{Tr}[\hat{\rho}_{\rm ss}\hat{O}]$ denotes a steady-state expectation value. Again, we now utilize the fact that $\mathcal J[\hat{O}\hat{O}']=\mathcal J[\hat{O}']\mathcal J[\hat{O}]$ for generic $\hat{O},\hat{O}'$:
\begin{align}
    \langle \!\langle \hat{X},\hat{Y}\rangle\!\rangle_{\mathcal T}&=\langle (\hat{a}_1^\dag)^{k_1}\cdots (\hat{a}_n^\dag)^{k_n}\\
    &~~~\cdot\hat{\alpha}_1^{q_1}\cdots \hat{\alpha}_n^{q_n}\mathcal J[(\hat{a}_1^\dag)^{p_1}\cdots (\hat{a}_n^\dag)^{p_n}]\rangle.
\end{align}
Finally, by direct computation, we also have that $\mathcal J[\hat{O}^\dag] = \hat{\rho}_{\rm ss} \mathcal J[\hat{O}]^\dag \hat{\rho}_{\rm ss}^{-1}$ for generic $\hat{O}$, and so
\begin{align}
    \langle \!\langle \hat{X},\hat{Y}\rangle\!\rangle_{\mathcal T} &= \langle (\hat{\alpha}_1^\dag)^{p_1}\cdots (\hat{\alpha}_n^\dag)^{p_n} \\
    &~~~\cdot(\hat{a}_1^\dag)^{k_1} \cdots (\hat{a}_n^\dag)^{k_n}\hat{a}_1^{l_1}\cdots \hat{a}_n^{l_n}\hat{\alpha}_1^{q_1}\cdots\hat{\alpha}_n^{q_n} \rangle\nonumber\\
    &=\int_\Sigma \,d^n\alpha d^n\beta \, P_{ss}(\vec{\alpha},\vec{\beta}) \tilde{\beta}_1^{p_1}\cdots \tilde{\beta}_n^{p_n}\nonumber\\
    &~~~\cdot\beta_1^{k_1}\cdots \beta_n^{k_n} \alpha_1^{l_1}\cdots \alpha_n^{l_n} \tilde{\alpha}_1^{q_1}\cdots \tilde{\alpha}_n^{q_n} 
\end{align}
Immediately, we recognize here the normally-ordered symbols of $\hat{X},\hat{Y}$, which are, explicitly:
\begin{align}
    \widetilde{Y}(\vec{\alpha},\vec{\beta}) &= \tilde{\beta}_1^{p_1}\cdots \tilde{\beta}_n^{p_n}\tilde{\alpha}_1^{q_1}\cdots \tilde{\alpha}_n^{q_n}\\
    X(\alpha,\beta) &= \beta_1^{k_1}\cdots \beta_n^{k_n}\alpha_1^{l_1}\cdots \alpha_n^{l_n}.
\end{align}
With the above observation, we have thus proved Eq. \eqref{eq:what_we_wanted_to_prove_many_body}, which establishes the equivalence of {\it generalized} quantum detailed balance, that is, hidden TRS, with classical detailed balance in the complex-$P$ representation. 

Finally, consider the general case where each mode has a different damping rate $\kappa_j$.  In this case, the corresponding classical TRS is modified to be
\begin{align}
    U_{ij}\to \sqrt{\frac{\kappa_i}{\kappa_j}}U_{ij},
\end{align}
which can be proven by trivially repeating the steps above.


\section{Doubled system correlation function symmetry for TRS $\psi=\pi$}
\label{app:qubit-h-trs}
In the main text we showed that the TFD correlation function $C_{yz}^{\mathrm{TFD}}(t)$ is time symmetric for the TRS $\psi=\pi$. We must still explicitly verify that the other two correlation functions, $C_{xy}^{\rm TFD}(t)$ and $C_{xz}^{\rm TFD}(t)$, are time symmetric for the same TRS.

The expansion of the TFD correlation functions as in Eq.~(\ref{eq:TFD-pointer-state-expansion}) is practically useful for computing correlation functions if the time-dependent operators are easily found because it reduces the problem to computing matrix elements. For the qubit system, the time-dependent operators are readily written in terms of the Liouvillian eigenmodes as
\begin{equation}
    \hat{\sigma}_k(t) \equiv \sum_n e^{\lambda_n t}{\rm Tr}(\hat{\sigma}_k \hat{r}_n ) \hat{l}_n^\dagger.
\end{equation}
This definition ensures that the operators reproduce the correct time averages and single system correlation functions under the hypotheses of the quantum regression theorem. We can then immediately write the time evolved operators as
\begin{align}
    \hat{\sigma}_x(t) &= e^{-\kappa t/2}\hat{\sigma}_x \\
    \hat{\sigma}_y(t) &= \langle \hat{\sigma}_y \rangle + \sum_{n=2}^3 e^{\lambda_n t}{\rm Tr}(\hat{\sigma}_y \hat{r}_n) \hat{l}_n^\dagger \\
    \hat{\sigma}_z(t) &= \langle \hat{\sigma}_z \rangle + \sum_{n=2}^3 e^{\lambda_n t}{\rm Tr}(\hat{\sigma}_z \hat{r}_n) \hat{l}_n^\dagger
\end{align}
 With these in hand we proceed to compute the matrix elements.

We label the pointer states of the steady state by $|\pm\rangle$ satisfying $\hat{\rho}_{\rm ss}|\pm\rangle = p_\pm |\pm\rangle$ for eigenvalues $p_{\pm} = \frac{1}{2} (1 \pm \sqrt{4b^2 + 1}/\sqrt{2b^2 + 1})$. The eigenstates lie in the $YZ$ plane of the Bloch sphere, namely
\begin{equation}
    |+\rangle = \cos\theta|g\rangle - i\sin\theta|e\rangle
\end{equation}
for some angle $\theta$ whose precise value as a function of $b$ does not concern us (except that $0 < \theta < \pi/2$ so that these do not reduce to pure $\hat{\sigma}_y$ or $\hat{\sigma}_z$ eigenstates). The matrix elements of $\hat{\sigma}_x = \hat{\sigma}_x(0)$ and $\hat{\sigma}_x(t)$ follow immediately as
\begin{align}
    \langle\pm|\hat{\sigma}_x(t)|\pm\rangle &= 0 \\
    \langle -|\hat{\sigma}_x(t)|+ \rangle &= \langle +|\hat{\sigma}_x(t)|- \rangle^* = ie^{-\kappa t/2}
\end{align}
and the matrix elements in the time-reversed pointer state basis are easily found for the TRS $\psi=\pi$ (c.f. Eq.~(\ref{eq:qubit-general-TRS}) in the main text):
\begin{align}
    \langle\tilde{\pm}|\hat{\sigma}_x|\tilde{\pm}\rangle &= 0 \\
    \langle \tilde{-}|\hat{\sigma}_x|\tilde{+} \rangle &= \langle \tilde{+}|\hat{\sigma}_x|\tilde{-} \rangle^* = -i.
\end{align}
We thus arrive at an interesting result already: the doubled system classically correlated state has identically zero correlation functions $C_{xy}^{\rm cl}(t)=0=C_{xz}^{\rm cl}(t)$. We therefore need to compute only the off-diagonal elements of $\hat{\sigma}_y(t)$ and $\hat{\sigma}_z(t)$. Given the TRS $\psi=\pi$ and the form of the pointer states, it is straightforward to compute these in the time-reversed pointer state basis. The relevant result is that they are real and thus
\begin{align}
    \langle \tilde{-}|\hat{\sigma}_y|\tilde{+}\rangle &= \langle \tilde{+}|\hat{\sigma}_y|\tilde{-}\rangle, \\
    \langle \tilde{-}|\hat{\sigma}_z|\tilde{+}\rangle &= \langle \tilde{+}|\hat{\sigma}_z|\tilde{-}\rangle,
\end{align}
as required by their hermiticity. To see why this must be true without explicitly computing the matrix elements, note that in the time-reversed pointer state basis, $\hat{\sigma}_x(t)$ is off-diagonal and imaginary and thus plays the role of an effective ``$\hat{\sigma}_y$'' in this basis. Therefore $\hat{\sigma}_y$ and $\hat{\sigma}_z$ must be linear combinations of the effective ``$\hat{\sigma}_x$'' and ``$\hat{\sigma}_z$'' and hence their off-diagonal matrix elements must be real and equal.

Furthermore, we can draw the same conclusion about the matrix elements of the time-dependent $\hat{\sigma}_y(t)$ and $\hat{\sigma}_z(t)$:
\begin{align}
    \langle -|\hat{\sigma}_y(t)|+ \rangle &= \langle +|\hat{\sigma}_y(t)|- \rangle, \\
    \langle -|\hat{\sigma}_z(t)|+ \rangle &= \langle +|\hat{\sigma}_z(t)|- \rangle.
\end{align}
Again $\hat{\sigma}_x$ is off-diagonal and imaginary and because $\hat{l}_2$ and $\hat{l}_3$ have only $\hat{\sigma}_y$ and $\hat{\sigma}_y$ components, the time-dependent operators must remain as linear combinations of the effective ``$\hat{\sigma}_x$'' and ``$\hat{\sigma}_z$'' which have real and equal off-diagonal matrix elements.

Finally putting everything together, the TFD correlation functions are
\begin{align}
    C_{xy}^{\rm TFD}(t) &= ie^{-\kappa t/2}
    \begin{cases}
        \left[ \langle \tilde{-}|\hat{\sigma}_y|\tilde{+}\rangle - \langle \tilde{+}|\hat{\sigma}_y|\tilde{-}\rangle\right] & t\geq 0 \\
        \left[ \langle +|\hat{\sigma}_y(t)|- \rangle - \langle -|\hat{\sigma}_y(t)|+ \rangle\right] & t < 0
    \end{cases} \nonumber \\
    &= 0
\end{align}
and
\begin{align}
    C_{xz}^{\rm TFD}(t) &=ie^{-\kappa t/2}
    \begin{cases}
        \left[ \langle \tilde{-}|\hat{\sigma}_z|\tilde{+}\rangle - \langle \tilde{+}|\hat{\sigma}_z|\tilde{-}\rangle\right] & t\geq 0 \\
        \left[ \langle +|\hat{\sigma}_z(t)|- \rangle - \langle -|\hat{\sigma}_z(t)|+ \rangle\right] & t < 0
    \end{cases} \nonumber \\
    &= 0
\end{align}
which are obviously time symmetric. Therefore all TFD correlation functions of the Rabi-driven qubit are indeed time symmetric for the TRS $\psi=\pi$, Eq.~(\ref{eq:qubit-general-TRS}).

\section{Mapping to a cascaded quantum system}
\label{app:CascadedMapping}

In this section, we show that the dual Liouvillian $\overline{\mathcal L}^*$, as defined and discussed in the main text, can be reinterpreted as a "perfect absorber" of the output radiation of $\overline{\mathcal L}$ \cite{stannigel_driven-dissipative_2012}. We start with the following identity in the main text (which is just a rewriting of the definition of $\mathcal J$): 
\begin{align}
    \hat{H}_{{\rm eff},A} \ket{\psi_T} &= 
    \mathcal{J} [ \hat{H}_{{\rm eff}} ]_B \ket{\psi_T} \label{eq:G1} \\
    \hat{c}_{l,A} \ket{\psi_T} &= 
    \mathcal{J} [ \hat{c}_{l} ]_B \ket{\psi_T}\label{eq:G2}
\end{align}
where $\hat{H}_{\rm eff}$ is the effective Hamiltonian in our master equation, and $\hat{c}_l$ are the jump operators. We denote the Hermitian (anti-Hermitian) parts of an operator $\hat{A}$ as $\textrm{Re} \left[\hat{A} \right]$ 
($i \textrm{Im} \left[\hat{A} \right]$). One can then tautologically rewrite Eq. \eqref{eq:G1} as
\begin{align}
    \bigg[\hat{H}_A& - \text{Re}[\mathcal J[\hat{H}_\text{eff}]]_B\label{eq:the_beginning}\\
    &-\bigg(\frac{i}{2}\sum_{l=1}^M \hat{c}_{l,A}^\dag \hat{c}_{l,A} -i\text{Im}
    [\mathcal{J} [ \hat{H}_{{\rm eff}} ]]_B\bigg)\bigg] \ket{\psi_T} = 0. \nonumber
\end{align}
A short but nontrivial calculation in 
App.~\ref{app:antiherm} then shows that
\begin{align}
    i \textrm{Im} 
    \left[ \mathcal J[H_\text{eff}] \right] 
    &=-\frac{i}{2}\sum_{l=1}^M \mathcal J[c_l]^\dagger \mathcal J[c_l],
\end{align}
so that Eq. \eqref{eq:the_beginning} simplifies to:
\begin{align}
    \bigg[H_A -& \text{Re}[\mathcal J[H_\text{eff}]]_B\label{eq:the_middle}\\
    &-\frac{i}{2} \sum_{l=1}^M(c_{l,A}^\dagger c_{l,A} - \mathcal J[c_{l}]_B^\dagger \mathcal J[c_{l}]_B)\bigg]|\psi_{T}\rangle=0.\nonumber
\end{align}
Now, separately note the following algebraic identity (which is independent of Eqs. (\ref{eq:G1}-\ref{eq:G2}):
\begin{align}
    & (c_{l,A}^\dagger c_{l,A} - \mathcal J[c_{l}]_B^\dagger \mathcal J[c_{l}]_B)
         \\
    &= 
        \left( \hat{c}^{\pp}_{l,A} + \mathcal J[\hat{c}_{l}]_B \right)^\dag \,
        (\hat{c}^{\pp}_{l,A}-\mathcal J[\hat{c}_l]_B )\nonumber\\
    &~~~~~~~~~~~~~~~~~~~~~~~~~~~~~~~~~~+(\hat{c}_{l,A}^\dag \mathcal J[\hat{c}_l]_B - h.c.).\nonumber
\end{align}
Using the above equality along with Eq. \eqref{eq:G2} to simplify Eq.~(\ref{eq:the_middle}), one obtains:
\begin{align}
    \bigg[H_A -\text{Re}[\mathcal J[H_\text{eff}]]_B-\frac{i}{2} \sum_{l=1}^M(\hat{c}_{l,A}^\dag \mathcal J[\hat{c}_l]_B - h.c.)\bigg]|\psi_{T}\rangle&=0\nonumber, \label{eq:the_end}
\end{align}
as desired.

\section{Anti-Hermitian part of the effective Hamiltonian for an electromagnetic absorber}
\label{app:antiherm}

In this appendix, we compute the anti-hermitian part of the effective Hamiltonian for the absorber system discussed in Appendix \ref{app:CascadedMapping}:
\begin{align}
i\text{Im}\,\mathcal J[\hat{H}_\text{eff}]&=\frac{1}{2}(\mathcal J[\hat{H}_\text{eff}] - \mathcal J[\hat{H}_\text{eff}]^\dagger)\nonumber\\
&=\frac{1}{2}(\mathcal J[\hat{H}_\text{eff}] - \hat{\rho}_{\rm ss}^{-1}\mathcal J[\hat{H}_\text{eff}^\dagger]\hat{\rho}_{\rm ss})
\end{align}
We now apply $\mathcal J^2$ to both sides:
\begin{align}
i\text{Im}\,\mathcal J[\hat{H}_\text{eff}] &=\frac{1}{2}\mathcal J[\hat{H}_\text{eff} - \hat{\rho}_{\rm ss}\hat{H}_\text{eff}^\dagger\hat{\rho}_{\rm ss}^{-1}]\nonumber\\
&=\frac{i}{2}\mathcal J[-i(\hat{H}_\text{eff}\hat{\rho}_{\rm ss} - \hat{\rho}_{\rm ss}\hat{H}_\text{eff}^\dagger)\hat{\rho}_{\rm ss}^{-1}]\label{eq:fragment}
\end{align}
Now, since $\hat{\rho}_{\rm ss}$ is a steady state of the Lindbladian $\mathcal L$, we have
\begin{align}
-i(\hat{H}_\text{eff}\rho_{ss} - \hat{\rho}_{\rm ss}H_\text{eff}^\dagger) + \sum_l\hat{c}_l \hat{\rho}_{\rm ss}\hat{c}_l^\dagger \equiv 0.
\end{align}
Therefore, plugging the above identity into Eq. \eqref{eq:fragment}, we get
\begin{align}
i\text{Im}\,\mathcal J[\hat{H}_\text{eff}]&=\frac{i}{2}\mathcal J[-\sum_l\hat{c}_l \rho_{ss}\hat{c}_l^\dagger\hat{\rho}_{\rm ss}^{-1}].\nonumber\\
&=-\frac{i}{2}\sum_l\hat{\rho}_{\rm ss}^{-1}\mathcal J[\hat{c}_l^\dagger]\hat{\rho}_{\rm ss}\mathcal J[\hat{c}_l]\nonumber\\
&= -\frac{i}{2}\sum_l\mathcal J[\hat{c}_l]^\dagger\mathcal J[\hat{c}_l],
\end{align}
which is the expression utilized in Appendix \ref{app:CascadedMapping}.


\section{The potential conditions: manifestation of trivial TRS in the complex-$P$ representation}
\label{app:pot_app}

The main goal of this appendix is to reproduce a well-known result in classical probability theory which is known for Fokker-Planck equations. Specifically, these results apply to Fokker-Planck equations of the form 
\begin{align}
    \partial_t P(\vec{x},t) &\equiv\partial_\mu [C^\mu(\vec{x}) P(\vec{x},t)] +\partial_\mu\partial_\nu [D^{\mu\nu}(\vec{x}) P(\vec{x},t)],  \label{eq:FPE}
\end{align}
where $P(\vec{x},t)$ is the probability distribution of a real-valued random process on $\mathbb{R}^d$, with stationary expectation values of observables $X(\vec{x})$ described in the standard way, e.g.
\begin{align}
    \overline{X}\equiv \int d^dx  P_{ss}(\vec{x}) X(\vec{x}).
\end{align}
The statement about these Fokker-Planck equations that we wish to generalize to the complex-$P$ representation is as follows \cite{pavliotis_stochastic_2014}: provided that $D^{\mu\nu}(\vec{x})$ is positive definite for all $\vec{x}$, then the potential conditions are equivalent to detailed balance with respect to a trivial time-reversal operation $\tilde{x}\equiv \vec{x}$.  By the potential conditions, we mean the statement that the stationary probability current vanishes.

In this appendix, we generalize the above result for the pseudo-Fokker-Planck equations encountered in the complex-$P$ representation:
\begin{align}
    \partial_t P(\vec{\alpha},\vec{\beta},t) &\equiv\partial_\mu [C^\mu(\vec{\alpha},\vec{\beta}) P(\vec{\alpha},\vec{\beta})]\nonumber\\
    &+\partial_\mu\partial_\nu [D^{\mu\nu}(\vec{\alpha},\vec{\beta}) P(\vec{\alpha},\vec{\beta},t)]. \label{eq:holomorphic_fokker_planck_app}
\end{align}
A complex {\it pseudo}-probability current may then be defined for the complex-$P$ distribution in the way described in Sec. \ref{sec:complex_P}:
\begin{align}
    \partial_tP(\vec{\alpha},\vec{\beta},t) \equiv \partial_\mu J^\mu(\vec{\alpha},\vec{\beta},t)\label{eq:pseudocurrent}
\end{align}
However, there are significant departures from standard probability theory: the complex-$P$ distribution is complex-valued, and thus violates the axioms of classical probability. In particular, stationary moments are given by 
\begin{align}
    \overline{X}\equiv \int_{\Sigma} d^n\alpha d^n\beta  P_{ss}(\vec{\alpha},\vec{\beta},t) X(\vec{\alpha},\vec{\beta}).
\end{align}
Here, $\Sigma$ is a 2$n$-dimensional closed integration surface (the many-body analogue of the pair of integration contours $\mathcal C,\mathcal C'$ mentioned in the main text). Furthermore, as the diffusion tensor $D^{\mu\nu}(\vec{\alpha},\vec{\beta})$ is generically complex, it may fail to be positive-definite.

Despite these differences, in this appendix we will nonetheless prove that the standard classical result in \cite{pavliotis_stochastic_2014} still applies (however, here there will be no assumption on the diffusion tensor $D^{\mu\nu}(\vec{\alpha},\vec{\beta})$): the potential conditions are equivalent to detailed balance with respect to a trivial time-reversal operation:
\begin{align}
    \widetilde{(\vec{\alpha},\vec{\beta})} \equiv (\vec{\alpha},\vec{\beta}).
\end{align}
This result is critical for bosonic many-body systems with local onsite loss, as, with a nontrivial $U$ parameter, hidden TRS corresponds to a nontrivial classical TRS in the complex-$P$ representation, and thus potentially corresponds to a richer symmetry than that encapsulated by the potential conditions.

We first establish the forward implication: if the stationary pseudoprobability current Eq. \eqref{eq:pseudocurrent} vanishes, then a straightforward calculation shows that
\begin{align}
    \mathcal L(XP_\text{ss}) =P_\text{ss}\mathcal L^*(X),\label{eq:lemma1}
\end{align}
where $\mathcal L$ is the Liouvillian for the pseudo Fokker-Planck equation Eq. \eqref{eq:holomorphic_fokker_planck_app}, and $\mathcal L^*$ is the adjoint Liouvillian, obtained via integration by parts \cite{drummond_generalised_1980}. The proof almost exactly follows \cite{pavliotis_stochastic_2014}, except while carrying out the calculation we notice that positive-definiteness of the diffusion tensor $D^{\mu\nu}(\vec{\alpha},\vec{\beta})$ is not needed in the proof, and thus the assumption of positive-definiteness may be relaxed.

From Eq. \eqref{eq:lemma1}, detailed balance is immediate, as, by integrating by parts on the surface $\Sigma$ (which is valid, as $X,Y$ are holomorphic), the dual Liouvillian $\mathcal L^*$ can then be shown to be symmetric with respect to the bilinear form 
\begin{align}
    \langle X,Y\rangle \equiv \int_\Sigma d^n \alpha d^n \beta \,P_{ss}XY.
\end{align}
From this and the expressions
\begin{align}
    \overline{X(t)Y(0)}& = \langle e^{-t\mathcal L^*}(X),Y\rangle,\\
    \overline{Y(t)X(0)}& = \langle X,e^{-t\mathcal L^*}(Y)\rangle,
\end{align}
we have detailed balance. For the converse direction, the proof follows analogously to the classical case as well: one starts by proving the following (assuming $P_{ss}$ is nonvanishing on $\Sigma$):
\begin{align}
    \langle -\mathcal L^*(X),Y\rangle = \langle D^{\mu\nu} \partial_\mu X,\partial_\nu Y\rangle + \langle X, P_{ss}^{-1}\partial_\mu Y J^\mu_{ss}\rangle
\end{align}
The above is a generalization of the lemma in Section 4.6 of \cite{pavliotis_stochastic_2014}, to pseudo Fokker-Planck equations, and may be proven by invoking the holomorphicity of $X,Y$. With the above, one can compute the asymmetry of the Liouvillian:
\begin{align}
    &\langle \mathcal L^*(X),Y\rangle -\langle X,\mathcal L^*(Y)\rangle\\
    &=2\int_\Sigma d^n\alpha d^n \beta \,Y(\vec{\alpha},\vec{\beta})\partial_\mu X(\vec{\alpha},\vec{\beta}) J^\mu_{ss}(\vec{\alpha},\vec{\beta})\nonumber
\end{align}
Now, if detailed balance holds with respect to a trivial TRS, then the Liouvillian $\mathcal L^*$ is symmetric, and thus the right-hand side must vanish for all pairs of holomorphic functions $X,Y$. The only way that this can happen is if the stationary current vanishes everywhere on the integration surface, i.e. 
\begin{align}
    J^\mu_{ss}(\vec{\alpha},\vec{\beta})\equiv 0~~~~\forall (\vec{\alpha},\vec{\beta})\in \Sigma.
\end{align}
However, the above is precisely the statement of the potential conditions.

\newpage

\bibliographystyle{apsrev4-1}
\bibliography{CQA}

\end{document}